\documentclass[a4paper,fleqn,usenatbib]{mnras} 

\usepackage{latexsym}
\usepackage{amssymb}
\usepackage{amsfonts}
\usepackage{amsmath}
\usepackage{bm}
\usepackage{graphicx}
\usepackage{subfigure}
\usepackage{times}
\usepackage{units}
\usepackage{hyperref}
\usepackage{multirow}
\usepackage{comment}
\usepackage{ulem}
\usepackage{hyperref}
\usepackage{graphicx}
\usepackage{acronym}
\usepackage{xcolor}
\usepackage{booktabs}
\usepackage[utf8]{inputenc}
\usepackage[T1]{fontenc}

\usepackage{pifont} 

\usepackage{enumitem}
\setlist[itemize]{noitemsep, topsep=0pt}

\usepackage{multirow}
\usepackage{morefloats}
\usepackage{mathrsfs}
\usepackage{latexsym}
\usepackage{dcolumn}
\usepackage{lipsum}
\usepackage{mathtools}
\usepackage{cuted}
\renewcommand{\cite}[1]{\citep{#1}}
\renewcommand\eqref[1]{(\ref{#1})}

\newcommand{\GRB}{GRB170817A}
\newcommand{\swind}{spiral-wave wind}

\newcommand{\surname}[1]{\text{#1}}

\newcommand{\ie}{\textit{i.e.}}
\newcommand{\eg}{\textit{e.g.}}

\newcommand{\be}{\begin{equation}}
    \newcommand{\ee}{\end{equation}}
\newcommand{\bea}{\begin{eqnarray}}
    \newcommand{\eea}{\end{eqnarray}}
\newcommand{\bel}{\begin{align}}
    \newcommand{\eel}{\end{align}}

\newcommand{\pba}{\texttt{PyBlastAfterglow}}

\def\GMc2{{\rm G M_{\odot} c^{-2}}}

\def\ccm{\,\text{cm}^{-3}}

\newacro{MRI}{magnetorotational instability}

\newacro{BH}{black hole}
\newacro{BNS}{binary neutron star}
\newacro{EM}{electromagnetic}
\newacro{EOS}{equation of state}
\newacroplural{EOS}[EOSs]{equations of state}
\newacro{GR}{general relativity}
\newacro{HD}{hydrodynamics}
\newacro{MHD}{magnetohydrodynamics}
\newacro{PIC}{particle-in-cell}
\newacro{BPL}{broken power law}
\newacro{SNR}{supernova remnant}
\newacro{BM}{Blandford \& McKee}
\newacro{ST}{Taylor-von Neumann-Sedov}
\newacro{FBOT}{fast blue optical transient}
\newacro{SSA}{synchrotron self-absorption}
\newacro{SED}{spectral energy distribution}
\newacro{GRHD}{general-relativistic hydrodynamics}
\newacro{GRMHD}{general-relativistic magnetohydrodynamics}
\newacro{GW}{gravitational wave}
\newacro{BZ}{Blandford-Znajek}
\newacroplural{GW}[GWs]{gravitational waves}
\newacro{LES}{large-eddy simulation}
\newacroplural{LES}[LES]{large-eddy simulations}
\newacro{GRLES}{general-relativistic large-eddy simulation}
\newacro{NR}{numerical relativity}
\newacro{NS}{neutron star}
\newacroplural{NS}[NSs]{neutron stars}
\newacro{GRB}{gamma-ray burst}
\newacroplural{GRB}[GRBs]{gamma-ray bursts}
\newacro{kN}{kilonova}
\newacroplural{kN}[kNe]{kilonovae}
\newacro{SGRB}{short \ac{GRB}}
\newacro{ISM}{interstellar medium}
\newacro{MNS}{massive neutron star}
\newacro{NSBH}{neutron star-black hole}
\newacro{NSE}{nuclear statistical equilibrium}
\newacro{SN}{supernova}
\newacroplural{SN}[SNe]{supernovae}
\newacro{CCSN}{core-collapse supernova}
\newacroplural{CCSN}[CCSNe]{Core-collapse supernovae}
\newacro{MP}{metal-poor}
\newacro{UFG}{ultra-faint dwarf galaxy}
\newacroplural{UFG}[UFGs]{ultra-faint dwarf galaxies}
\newacro{NRN}{nuclear reaction network}
\newacroplural{NRN}[NRNs]{nuclear reaction networks}
\newacro{RR}{reaction rate}
\newacroplural{RR}[RRs]{reaction rates}
\newacro{ODE}{ordinary differential equation}
\newacroplural{ODE}[ODEs]{ordinary differential equations}
\newacro{IR}{infrared}
\newacro{NIR}{near-infrared}
\newacro{FIR}{far infrared}
\newacro{LC}{light curve}
\newacroplural{LC}[LCs]{light curves}
\newacro{UV}{ultraviolet}
\newacro{LTE}{local thermodynamic equilibrium}
\newacro{TOV}{Tolman-Oppenheimer-Volkoff}
\newacro{RMS}{root mean square}
\newacro{MM}{multi-messenger}
\newacro{LK}{light curve}
\newacro{CBM}{circumburst medium}
\newacro{LF}{Lorentz factor}
\newacro{SSC}{synchrotron self-Compton}
\newacro{IC}{inverse Compton}
\newacro{SSA}{synchrotron self-absorption}
\newacro{PP}{pair production}
\newacro{EBL}{extragalactic background light}
\newacro{FS}{forward shock}
\newacro{RS}{reverse shock}
\newacro{VHE}{very high energy}
\newacro{EFE}{Einstein’s field equations}
\newacro{ADM}{Arnowitt, Deser and Misner}
\newacro{IVP}{initial value problem}
\newacro{RHS}{right hand side}
\newacro{RMF}{relativistic mean-field}
\newacro{PDE}{partial differential equation}
\newacroplural{PDE}[PDEs]{partial differential equations}
\newacro{EM}{electromagnetic}
\newacro{HMNS}{hyper-massive neutron star}
\newacro{SMNS}{supra-massive neutron star}
\newacro{EATS}{equal time arrival surface}
\newacroplural{EATS}[EATSs]{equal time arrival surfaces}
\newacro{JWST}{James Webb Space Telescope}
\newacro{VLA}{Very Large Array}
\newacro{SKA}{Square Kilometre Array}
\newacro{DSA}{Diffusive Shock Acceleration}
\newacro{PN}{post-Newtonian}
\newacro{EOB}{effective-one-body}
\newacro{PC}{prompt collapse}
\newacro{LIGO}{Laser Interferometer Gravitational-Wave Observatory}
\newacro{MF}{magentic field}
\newacroplural{MF}[MFs]{magnetic fields}
\newacro{AGN}{active galactic nucleus}
\newacroplural{AGN}[AGNs]{active galactic nuclei}
\newacro{LOS}{line of sight}
\newacro{DE}{dynamical ejecta}
\newacro{SWW}{\swind{}}

\newacro{BW}{blast wave}
\newacroplural{BW}[BWs]{blast waves}

\newacro{VLBI}{Very-long-baseline interferometry}
\newacro{ISS}{interstellar scintillation}
\newacro{mas}{milliarcsecond}
\newacro{FWHM}{full width at half maximum}

\renewcommand{\cite}[1]{\citep{#1}}
\defcitealias{Nava:2013}{\scshape N13} 
\defcitealias{Miceli:2022efx}{\scshape M22} 

\def\PBA{\texttt{PyBlastAfterglow}}

\title[BNS afterglow skymaps]{
    Multi-physics framework for fast modeling of gamma-ray burst afterglows
}

\author[V.\ Nedora et al.]{
    Vsevolod \surname{Nedora}$^{1,2}$, 
    Ludovica \surname{Crosato Menegazzi}$^{1}$, 
	Enrico \surname{Peretti}$^{3}$, 
    Tim \surname{Dietrich}$^{1,2}$,
    Masaru \surname{Shibata}$^{1,4}$ 
    \\
    ${}^1$Max Planck Institute for Gravitational Physics (Albert Einstein Institute), Am M{\"u}hlenberg 1, Potsdam 14476, Germany\\
    ${}^2$Institute for Physics and Astronomy, University of Potsdam, Potsdam 14476, Germany\\
    ${}^3$Université Paris Cité, CNRS, Astroparticule et Cosmologie, 10 Rue Alice Domon et Léonie Duquet, 75013 Paris, France\\
    ${}^4$Center for Gravitational Physics and Quantum Information, Yukawa Institute for Theoretical Physics, Kyoto University, Kyoto, 606-8502, Japan
}

\date{
    Accepted XXX. Received YYY; in original form ZZZ
}

\begin{document}
\label{firstpage}

\maketitle

\date{\today}

\begin{abstract}
    In this paper, we present \PBA{}, a modular C++ code with a Python interface to model light curves and sky maps of gamma-ray burst afterglows. 
	The code is open-source, modular, and sufficiently fast to perform parameter grid studies. \PBA{} is designed to be easily extendable and used as a testing bed for new physics and methods related to gamma-ray burst afterglows.
	For the dynamical evolution of relativistic ejecta, a thin-shell approximation is adopted, where both forward and reverse shocks are included self-consistently, as well as lateral structure, lateral spreading, and radiation losses. 
	Several models of the shock microphysics are implemented, including fully 
	numerical model of the downstream electron distribution evolution, synchrotron emission, self-absorption, and synchrotron self-Compton emission under 
	the one-zone approximation. Thus, the code is designed to be able to model 
	complex afterglows that include emission from reverse shock, very high energy 
	emission, structured jets, and off-axis observations. 
\end{abstract}


\section{Introduction}\label{sec:intro}

A \ac{GRB} is an intense, transient, cosmological source of electromagnetic 
radiation characterized by two distinct emission phases. The initial phase, 
known as prompt emission, typically lasts from a fraction of a second to 
several minutes. This prompt emission is highly variable; the spectrum 
usually peaks in keV - MeV range; peak isotropic luminosity typically lies in 
$10^{49-53}$erg~s$^{-1}$ range.
This phase is followed by the afterglow during which the emission shows a 
broad spectrum ranging from gamma-rays to radio waves, and its temporal 
behavior that can be generally described by a smooth power-law 
(see \citet{Zhang:2018book} for a textbook discussion).

\Acp{GRB} are generally categorized into short and long types based on whether 
the prompt emission lasts less than or more than $2$ seconds\footnote{
	This categorization may suffer from the uncertainties in measuring 
	burst duration \cite{Moss:2022}
} and are thought to originate from ultra-relativistic jets. The fluid and 
energy dynamics within these jets are complex, involving both internal and 
external dissipation processes via mechanisms like matter shell interactions and 
magnetic reconnections 
\cite{Thompson:1994zh,Rees:1994nw,Spruit:2000zm,Rees:1992ek,Chevalier:1999jy}. 

Various models have been proposed to explain the nature and mechanics of 
\acp{GRB}. For instance, long \acp{GRB} are often associated with jets forming 
during collapse of rapidly rotating massive stars whereas short \acp{GRB} 
are thought to be connected to mergers of compact objects at least one of which is 
a neutron star \cite{Woosley:1993,Paczynski:1997yg,LIGOScientific:2017vwq}. 
However, the composition and driving mechanisms within these jets -- whether magnetic 
or kinetic -- are still areas of active research, particularly in how these mechanisms 
affect the efficiency and nature of prompt radiation  
\cite{Preece:1998jy,Lazzati:1999zc,Ghisellini:1999wu,Giannios:2004di,Fan:2006sx,Zhang:2011,Sironi:2015eoa,Beniamini:2016hzc,Oganesyan:2017ork,Oganesyan:2017eds,Ravasio:2017vak,Lazarian:2019}.

The afterglow emission is generally better understood. 
It encompasses all broad-band radiation from a \ac{GRB} observed 
over long periods -- from minutes to years -- following the initial burst of prompt 
radiation. 
Generally it can be described by decaying power-laws 
\cite{Nousek:2005fm,Nardini:2005mb,Zaninoni:2013hca,Bernardini:2012}. This 
suggests that the afterglow radiation originates at larger radii -- beyond 
$10^{15}\,$cm -- and is produced through interactions between the jet and the surrounding 
circumburst medium. This interaction drives two types of hydrodynamic shocks: a \ac{FS} 
that moves into the surrounding medium and a \ac{RS} that travels back into the 
jet \cite{Blandford:1976,Sari:1995}. The system that includes these shocks and 
contact discontinuity between them is commonly referred to as a \ac{BW}. 

Both \ac{FS} and \ac{RS} drive respective collisionless shocks, where interactions between 
particles are mediated via electromagnetic forces \cite{Sironi:2015oza} instead of direct 
particle collisions. While the microphysics of these shocks is very complex, two main effects 
can be identified: amplification of random magnetic fields present in the shock upstream 
and acceleration of the inbound particles as those repeatedly cross the shock. 
The result of these processes is a non-thermal, synchrotron radiation produced by particles 
gyrating around magnetic field lines 
\cite{Meszaros:1993,Meszaros:1996sv,Sari:1999b,Sari:1997qe,Kobayashi:1999ei}. 
Notably, many open question remain regarding physics conditions at relativistic collisionless 
shocks and \acp{GRB} offer a unique opportunity to study them 
\cite{Vanthieghem:2020nvr,Sironi:2011,Sironi:2013ri}.

Most of the observed radiation was shown to come from particles accelerated at the \ac{FS}. 
There, environmental conditions -- where \acp{GRB} occur -- and the nature of their progenitors 
affect electron and radiation spectra in addition to the details of shock microphysics. Notably, 
jet conditions at the start of the afterglow phase are defined not only by the overall energy 
budget of the burst but also by the dissipation processes that occurred during the prompt phase 
\cite{Kumar:2014upa}.

While not being the main contributor to the observed emission, the \ac{RS} was found to be 
important in interpreting observations of some \acp{GRB} 
\cite{Laskar:2019xfo,Lamb:2019lao,Salafia:2021njb}. 
A \ac{RS} forms when \ac{GRB} ejecta collides with external medium. It travels back through 
the ejected matter, compressing, heating and decelerating it \cite{Blandford:1976,Ayache:2020akr}. 
A \ac{RS} is generally slower than a \ac{FS} and moves through a significantly more dense medium. 
Thus, afterglow emission from it peaks at lower frequencies (radio-to-optical) and at early times
-- before the \ac{RS} crosses the ejecta. 
The impact of the \ac{RS} emission on the total afterglow emission is highly dependent on the 
initial jet \ac{LF}, density of the circumburst medium, and jet structure and composition -- 
whether it is baryonic or Poynting flux dominated 
\cite{Zhang:2004ie,McMahon:2005je,Giannios:2007tu,Lyutikov:2011,Gao:2015lga}. 
One prominent example of a \ac{RS} contribution to afterglow is the \ac{GRB}~180418A where 
a bright peak observed between $28$ and $90$ seconds after the burst in the optical band 
was attributed to it \cite{Becerra:2019tvi}.

Recently it was demonstrated that long \acp{GRB} can emit significant fraction of their energy in 
TeV range during the afterglow phase, ($20$--$50$\% of the total emitted energy)  
\cite{MAGIC:2019irs,HESS:2021dbz,LHAASO:2023lkv}. Specifically, in $2019$ MAGIC and H.E.S.S. 
collaborations detected \ac{VHE} emissions -- above $100$~GeV -- from 
long \acp{GRB}~190114C during the afterglow phase \cite{Abdalla:2019dlr,MAGIC:2019lau}. 
These observations confirmed the theoretically expected \ac{VHE} long after the initial 
burst of gamma rays. The previous lack of such detections is believed to be at least in 
part attributed to technological limitations of earlier Cherenkov telescopes
\cite{Vurm:2016qqi,Gilmore:2012,Dominguez:2011,Franceschini:2017iwq}. 
Current theoretical models suggest that \ac{VHE} emission in the afterglow phase comes primarily 
from inverse Compton scattering, -- 
a process in which lower energy photons are upscattered to gamma-ray energies 
through high energy electrons.
While the details of the upscattering mechanism are still uncertain, 
recent observations hinted at \ac{SSC} as a likely candidate. Notably, it was also noted 
that single-component synchrotron models might extend into the \ac{VHE} spectrum under 
certain conditions 
\cite{Pilla:1997jm,Dermer:1999eh,Wang:2001fu,Wang:2006eq,Fan:2007tqx}.

\GRB{}, observed in September $2017$, is one of the best sampled and closest ($z=0.0099$) \acp{GRB} 
ever detected 
\cite{Arcavi:2017xiz,Coulter:2017wya,Drout:2017ijr,Evans:2017mmy,Hallinan:2017woc,Kasliwal:2017ngb,Nicholl:2017ahq,Smartt:2017fuw,Soares-santos:2017lru,Tanvir:2017pws,Troja:2017nqp,Mooley:2018dlz,Ruan:2017bha,Lyman:2018qjg}.  
It remains the only \ac{GRB} associated with a burst of gravitational waves and quasi-thermal emission, 
called kilonova  \cite{Metzger:2010,Savchenko:2017ffs,Alexander:2017aly,Troja:2017nqp,Monitor:2017mdv,Nynka:2018vup,Hajela:2019mjy}.
The leading interpretation of this event is the merger of two neutron stars \cite{LIGOScientific:2017vwq}. 
A detailed analysis of multi-frequency \ac{GRB} afterglow data collected over several years 
suggested a non-trivial lateral structure within the jet 
\cite{Fong:2017ekk,Troja:2017nqp,Margutti:2018xqd,Lamb:2017ych,Lamb:2018ohw,Ryan:2019fhz,Alexander:2018dcl,Mooley:2018dlz,Ghirlanda:2018uyx}, 
created, at least in part, during jet propagation through dense kilonova ejecta \cite{Lamb:2022pvr}.  
Furthermore, radio and optical imaging of the burst region showed a motion of the image flux centroid 
\cite{Mooley:2022uqa}. It further confirmed the jetted nature of the \ac{GRB} ejecta and provided 
additional constraints on jet properties \cite{Mooley:2018dlz,Mooley:2022uqa}.

Most observed \ac{GRB} afterglows can be explained by radiation produced 
within relativistic shocks formed during ejecta propagation through the circumburst medium. 
However, there exist \acp{GRB} afterglow signatures that cannot be easily explained with this 
model. 
For instance, X-ray flares observed in 
approximately 33\% of \acp{GRB}, exhibiting a broad range of timings and characteristics 
hinting towards an ongoing activity from the \acp{GRB} central engine 
\cite{Chincarini:2010,Margutti:2011,Bernardini:2011,Yi:2016uxj,Yi:2017trl} or an 
unaccounted geometrical effects \cite{Duque:2021xis}.  
Additionally, analysis of long \acp{GRB} often favor a constant density circumburst 
environment in contradiction to the expected wind profile, that should be produced by a 
massive star before it undergoes supernova explosion. Furthermore, degeneracies in model 
parameters further complicate inferring the conditions at burst and the nature of the 
progenitor, especially when observational data is limited 
\cite{Panaitescu:2001bx,Schulze:2011,Li:2015cua,Gompertz:2018anr}. 
Furthermore, model parameters associated with shock microphysics that are commonly used to 
relate the shock conditions to the emission properties in afterglow models have large 
discrepancies. Specifically, it was found that in order to explain some observations, certain 
parameter values tend to lay outside of ranges predicted by first-principle microphysics 
simulations 
\cite{Kumar:2009,Kumar:2010,BarniolDuran:2011,He:2011}.

Overall, extensive examinations of \ac{GRB} afterglows provide insights into the 
progenitor and its environment, jet properties and the microphysics of collisionless  
shocks. It is also worth mentioning that \acp{GRB} are used extensively in cosmology 
\cite{Dainotti:2016yxl,Schady:2017pwx,Bulla:2022ppy}. 
However, \ac{GRB} jets are complex physical systems with spatial and temporal scales 
ranging from electron gyroradii to parsecs and from milliseconds to years respectively. 
Modeling these systems requires relativistic magnetohydrodynamics, plasma and collisionless 
shocks microphsyics and radiation transport.
Thus, it is practically impossible to construct a first-principle model of a \ac{GRB} 
jet and all its observables. The state-of-the-art in this filed is a numerical-relativistic 
\ac{MHD} simulation combined with simplified treatment of shock 
microphysics and radiation transport  \cite{Daigne:2000xg,vanEerten:2009pa,Ayache:2020akr}. 
These simulations are computationally very expensive and thus are ill-suited for parameter 
inference from observational data. Thus, in most cases, simplified analytic 
\cite{Blandford:1976,Sari:1997qe,Panaitescu:2000bk,Granot:2001ge,Yamasaki:2021klt,Fraija:2023zco}  
or semi-analytic  \cite{Huang:1999di,Uhm:2006qk,Peer:2012,Nava:2013,Zhang:2018book,Ryan:2019fhz,Guarini:2021gwh,Miceli:2022efx,Wang:2024wbt} 
\ac{GRB} afterglow models are employed. 

The computational efficiency of this models allows them to be used in multimessenger 
studies, where a combined ``hyper''-model that includes models of individual signatures 
is applied to diverse set of observational data that include \ac{GRB} afterglow, 
supernova/kilonva, gravitational waves and neutrinos 
\cite{Dietrich:2020efo,Pang:2022,Kunert:2023vqd,Sarin:2023khf}.
The main disadvantage of these models, however, is limited generalization due to limited 
and simplified physics input and non-trivial extendability that lead to methodological 
biases and inaccuracies. 

In this work we present a \ac{GRB}\footnote{Source code: \url{https://github.com/vsevolodnedora/PyBlastAfterglowMag}} afterglow model that attempts to achieve a balance between 
the computational efficiency and accuracy of physics input. The model is implemented in the 
open source numerical code \PBA{} designed to be (i) easily extendable and (ii) sufficiently  
fast to allow for construction of simulation grids that can in turn be used to train a surrogate 
model for direct Bayesian inference. 

The paper is organized as follows. 
In Sec.~\ref{sec:method}, we introduce methods used to model \ac{GRB} afterglow, starting with 
\ac{BW} dynamics (Sec.~\ref{sec:method:dynamics}), then discussing shock 
microphysics and comoving radiation (Sec.~\ref{sec:method:mphis_comovrad}) 
and subsequently showing how the observed radiation is obtained (Sec.~\ref{sec:method:obs_rad}). 
In Sec.~\ref{sec:method}, we elaborate on the details of numerical implementation 
of the aforementioned physics. 
Then, in Sec.~\ref{sec:result}, we discuss in detail two \ac{GRB} afterglow simulations 
for a unstractured (top-hat) jet and a structured (Gaussian) jet. 
To validate the code we perform extensive comparisons with existing codes and published 
resungs in Sec.~\ref{sec:comparison}. 
After that, in Sec.~\ref{sec:applications}, we illustrate code application to \ac{VHE} afterglow 
of GRB~190411C and complex afterglow signatures of \GRB{}. 
Finally, in Sec.~\ref{sec:conclusion}, we summarize the presented work and provide and 
outlook.

\section{Methods}\label{sec:method}

In this section we discuss the physics that is implemented in \PBA{}. 
For the sake of completeness and future referencing we discuss some derivation 
in detail. 

\subsection{Blastwave dynamics}\label{sec:method:dynamics}

The interaction between an ejecta shell and an ambient medium can be 
considered as a relativistic Riemann problem, in which shocks form 
when required conditions for velocities densities and pressures are met 
(cf.~\citet{Rezzolla:2013} for a textbook discussion). 

There are several semi-analytic formulations of \ac{BW} dynamics with a \ac{FS} and a 
\ac{RS} in the literature. They can be broadly divided into pressure 
balance formulations and mechanical models. Pressure balance formulations 
assume that the pressure in \ac{FS} and \ac{RS} downstreams are equal 
\cite{Nava:2013,Chen:2021rbe,Zhang:2022phg}. 
The mechanical model was proposed by \cite{Beloborodov:2006qt,Uhm:2006qk} 
and later improved by \cite{Ai:2021dbw,Ai:2022pig}. It does not rely 
on the pressure equilibrium and has a better energy conserving property. 
In \PBA{} we implement the formulation motivated by \cite{Nava:2013} 
(hereafter \citetalias{Nava:2013}; see their appendix~B).

\subsubsection{Evolution equations}

Consider the stress-energy tensor for perfect fluid,  
\begin{equation} \label{eq:method:tmunu}
    T^{\mu\nu} = (\rho' c^2 + e' + p')u^{\mu}u^{\nu} + p'\eta^{\mu\nu}\, ,
\end{equation}
where $u^{\mu} = \Gamma(1,\beta)$ is the fluid four-velocity with 
$\Gamma$ being the bulk \acf{LF} and $\beta=\sqrt{1-\Gamma^{-2}}$ is the  
dimensionless velocity (in units of the speed of light, $c$), 
$p' = (\hat{\gamma}-1)e'$
is the pressure,  
$e'$ 
is the internal energy density,  
$\hat{\gamma}$ is the adiabatic index (also 
called the ratio of specific heats), and 
$\eta^{\mu\nu}$ is the metric with signature 
$\{-1,1,1,1\}$. 
Hereafter, we define prime quantities in the comoving frame.

If the fluid is ultra-relativistic, $\hat{\gamma}=4/3$, and if it is 
non-relativistic, $\hat{\gamma}=5/3$. These limits can be recovered 
by assuming the following form of the $\hat{\gamma}$ 
dependency on the fluid \ac{LF} \citep[\eg][]{Kumar:2003yt} 
\begin{equation}\label{eq:method:eos}
    \hat{\gamma} \approxeq \frac{4 + \Gamma^{-1}}{3} \, .
\end{equation}
A more accurate prescription can be inferred from numerical simulations 
\cite{Mignone:2005ns}. 

The $\mu=\nu=0$ component of the stress-energy tensor 
Eq.~\eqref{eq:method:tmunu}, then reads 
\begin{equation}
    T^{00} = \Gamma^2(\rho'c^2 + e' + p') - p' = \Gamma^2 \rho' c^2 + (\hat{\gamma}\Gamma^2-\hat{\gamma}+1)e' \, .
\end{equation}
Under the thin-shell approximation (\eg, assuming that an ejecta shell is uniform 
in its properties) the total energy of the shell reads,  
\begin{equation} 
    E_{\rm tot} = \int T^{00} dV = \Gamma c^2 \rho' V' + \Gamma_{\rm eff} e' V' = \Gamma c^2 m + \Gamma_{\rm eff}E_{\rm int}' \, ,
\end{equation}
where $\Gamma_{\rm eff} = (\hat{\gamma}\Gamma^2 - \hat{\gamma} + 1)/\Gamma$  
is the effective \ac{LF} (\eg, see \citet{Nava:2013,Zhang:2018book,Guarini:2021gwh}),
the enclosed mass $m=\rho'V'$ with $V'$ is the comoving volume, 
and the co-moving internal energy is $E_{\rm int}' = e'V'$.

As mentioned before, a \ac{BW} is comprised of \ac{FS}, \ac{RS}, and a contact discontinuity between them. Following standard notation, we label the unshocked ejecta (\ac{ISM}) 
as regions $4$ ($1$) and shocked ejecta (\ac{ISM}) 
as regions $3$ ($2$) (see figure~B1 in \citetalias{Nava:2013}). The region to 
which a quantity belongs to will be indicated with the subscript. 
   
The total energy of the \ac{BW} with \ac{FS} propagating into the circumburst 
medium, contact discontinuity and the \ac{RS} propagating into the ejecta itself 
reads \citepalias{Nava:2013}, 
%
\begin{equation}\label{eq:method:Etot}
    \begin{aligned}
        E_{\rm tot} &= 
        \Gamma_0 m_{0;4}c^2 + \Gamma m_{0;3} c^2 + \Gamma_{\rm eff;3} E_{\rm int;3}' \\ &+ \Gamma_0 m_{0;2} c^2 + \Gamma_{\rm eff;2} E_{\rm int;2}'\, .
    \end{aligned}
\end{equation}
Notably, most semi-analytic afterglow models in the literature only consider the 
\ac{FS}. This is equivalent to assuming that at the beginning of a simulation, the \ac{RS} 
has already crossed ejecta and that the entire ejecta shell moves with the bulk \ac{LF} 
$\Gamma$. Without this assumption, an ejecta shell is initially not shocked and 
moves with the \ac{LF}  $\Gamma_{0}$ as the first two terms in Eq.~\eqref{eq:method:Etot} 
indicate.  

As the \ac{BW} moves through the circumburst medium, it accretes mass, $dm$, and 
loses energy due to radiation from both shocked regions, $dE_{\rm rad;3}'$, and, 
$dE_{\rm rad;2}'$, so that the total change in \ac{BW} energy is given as, 
\begin{equation}
    dE_{\rm tot} = \Gamma dm c^2 + \Gamma_{\rm eff;2}dE_{\rm rad;2}' + \Gamma_{\rm eff;3}dE_{\rm rad;3}'\, ,
\end{equation}
where $\Gamma_{\rm eff;3} = (\hat{\gamma}_3\Gamma^2 - \hat{\gamma}_3 + 1)/\Gamma$, 
$\hat{\gamma}_3$ is given by the \ac{EOS}, Eq.~\eqref{eq:method:eos},  
using the relative \ac{LF} between upstream and downstream, 
$\Gamma_{43}$.  
Combining the equations and re-arranging the terms, we obtain, 
\begin{equation}\label{eq:method:dEtot}
    \begin{aligned}
        dE_{\rm tot} = & (\Gamma_0 dm_{0;4} + d\Gamma m_{0;3} + \Gamma dm_{0;3} \\
                     & + d\Gamma dm + \Gamma dm - dm) c^2 \\
        + & d\Gamma_{\rm eff;3}E_{\rm int;3}' + \Gamma_{\rm eff;3}(dE_{\rm int;3}' - dE_{\rm rad;3}') \\
        + & d\Gamma_{\rm eff;2}E_{\rm int;2}' + \Gamma_{\rm eff;2}(dE_{\rm int;2}' -  dE_{\rm rad;2}') \\
    \end{aligned}
\end{equation}

Here $dm_{0;3}$ is the amount of mass that crosses the \ac{RS} over the time period that the \ac{FS} takes to attain mass $dm$. The change in mass in the unshocked part of the shell then is $dm_{0;4}=-dm_{0;3}$. 

The change in the internal energy behind the \ac{FS} reads, 
\begin{equation} \label{eq:method:dEdEdE2}
    dE_{\rm int;2}' = dE_{\rm sh;2}' + dE_{\rm ad;2}' + dE_{\rm rad;2}'\, ,
\end{equation}
where $dE_{\rm sh}'$ is the random kinetic energy produced at the shock due to inelastic 
collisions \cite{Blandford:1976} with element $dm$ accreted from circumburst medium, 
$dE_{\rm ad}'$ is the energy lost to adiabatic expansion, and $dE_{\rm rad}'$ is the energy 
lost to radiation. The latter we discuss separately in Sec.~\ref{sec:method:dErad}. 

We further assume that for the \ac{FS} the upstream medium can be viewed as cold and static. 
Then in the post-shock frame the average kinetic energy per unit mass is  $(\Gamma-1)c^2$ 
and is constant across the shock. 
Thus, 
\begin{equation} \label{eq:method:dEsh2}
    dE_{\rm sh;2}' = (\Gamma - 1)c^2 dm \, . 
\end{equation}

Adiabatic losses, $dE_{\rm ad;2}'$, can be obtain in two main formulations: 
microscopic and macroscopic. The former relies on integrating the momenta of hadrons and 
leptons \citep{Dermer:2000gu,Miceli:2022efx} (see equation~B(15) in \citetalias{Nava:2013}).  
The latter formulation can be easily derived from the first law of thermodynamics, 
$dE_{\rm int;2}' = T'dS' - p'_{2}dV_{2}'$, for an adiabatic process, \ie, $T'dS'=0$. 
Then, 
\begin{equation}\label{eq:method:dEad2}
    dE_{\rm ad;2}' = -(\hat{\gamma}_2-1)E_{\rm int;2}' d \ln V_{2}' \, . 
\end{equation}
where we used $p_2' = (\hat{\gamma}-1)E_{\rm int;2}'/V_2'$. 
For a static and cold upstream, the comoving volume is $V'\propto R^3 / \Gamma$, 
so that the derivative $d \ln V_2'$ reads 
\begin{equation}\label{eq:method:dlnV2}
    d\ln V_{2}' = d\ln m - d\ln \rho - d\ln\Gamma \, .
\end{equation}

For the \ac{RS} the energy change in the downstream reads, 
\begin{equation} \label{eq:method:dEdEdE3}
    dE_{\rm int;3}' = dE_{\rm sh;3}' + dE_{\rm ad;3}' + dE_{\rm rad;3}'\, .
\end{equation}
Notably, the upstream with respect to the \ac{RS} is not static,  
so that the average kinetic energy per unit mass in the post-shock frame equals to 
$(\Gamma_{\rm rel}-1)c^2$, where $\Gamma_{\rm rel} = \Gamma_{43}$ 
is the relative \ac{LF} between upstream and downstream. 
Then, the energy generated at the shock is,  
\begin{equation} \label{eq:method:dEsh3}
    dE_{\rm sh;3}' = (\Gamma_{43} - 1)c^2 dm_{0;3} \, ,
\end{equation}
and the adiabatic losses $dE_{\rm ad;3}'$ in macroscopic formulation reads,  
\begin{equation}\label{eq:method:dEad3}
    dE_{\rm ad;3}' = -(\hat{\gamma}_3-1)E_{\rm int;3}' d \ln V_{3}' \, . 
\end{equation}
As $V'\propto R^3 / \Gamma_{43}$, its derivative reads, 
\begin{equation}\label{eq:method:dlnV3}
    d\ln V_{2}' = d\ln m_{0;3} - d\ln \rho_{4} - d\ln\Gamma_{43} \, ,
\end{equation}
where $d\ln\Gamma_{43} = \Gamma_{43}^{-1}(d\Gamma_{43}/d\Gamma)d\Gamma$.

In Eq.~\eqref{eq:method:dEtot}, the expression for internal energy for both shocks, 
$dE_{\rm int;i}' - dE_{\rm rad;i}'$ can be replaced with 
$dE_{\rm sh;i}' - dE_{\rm ad;i}'$, where $i\in\{2,3\}$. After doing this 
and rearranging the terms we arrive at the evolution equation for the 
\ac{BW} bulk \ac{LF} 
\begin{equation}\label{eq:method:dG_fsrs}
    \begin{aligned}
        d\Gamma = & -\frac{N}{D}\, , \text{where} \\
        N = & (\Gamma-1)(\Gamma_{\rm eff;2}+1)dm \\
        + & (\Gamma-\Gamma_0+\Gamma_{\rm eff;3}(\Gamma_{43}-1))dm_{0;3} \\
        + & \Gamma_{\rm eff;2}(\hat{\gamma}_2-1)E_{\rm int;2}'(d\ln m - d\ln\rho) \\
        - & \Gamma_{\rm eff;3}(\hat{\gamma}_3-1)E_{\rm int;3}'(d\ln m_{0;3} - d\ln\rho_{4}) \\
        D = & (m + m_{0;3}) + E_{\rm int;2}'\frac{d\Gamma_{\rm eff;2}}{d\Gamma} + E_{\rm int;3}'\frac{d\Gamma_{\rm eff;3}}{d\Gamma} \\
        + & \frac{\Gamma_{\rm eff;2}}{\Gamma}(\hat{\gamma}_2-1)E_{\rm int;2}' 
        + \frac{\Gamma_{\rm eff;3}}{\Gamma_{43}}(\hat{\gamma}_3-1)E_{\rm int;3}'\frac{d\Gamma_{43}}{d\Gamma}\, .
    \end{aligned}
\end{equation}
The advantage of choosing the analytical expression for \ac{EOS}, 
Eq.~\eqref{eq:method:eos}, is that we can compute the derivatives in 
Eq.~\eqref{eq:method:dG_fsrs} analytically as, 
\begin{equation}
    \frac{d\Gamma_{\rm eff;2}}{d\Gamma} = \frac{4}{3} + \frac{1}{3\Gamma^2} + \frac{2}{3 \Gamma^3}
\end{equation}
and 
\begin{equation}\label{eq:method:dGeff3dG}
\begin{aligned}
    \frac{d\Gamma_{\rm eff;3}}{d\Gamma} 
    & = \frac{d}{d\Gamma}\Big[\big(\hat{\gamma}_3(\Gamma^2-1)+1\big)\frac{1}{\Gamma}\Big] \\
    & = \hat{\gamma}_3 \Big(1 + \frac{1}{\Gamma^2}\Big) + \frac{d\hat{\gamma}_3}{d\Gamma_{43}}\frac{d\Gamma_{43}}{d\Gamma}\Gamma\Big(1 - \frac{1}{\Gamma}\Big) - \frac{1}{\Gamma^2} \\
    & = \hat{\gamma}_3 \Big(1 + \frac{1}{\Gamma^{2}}\Big) - \frac{1}{3}\frac{d\Gamma_{43}}{d\Gamma}\frac{1}{\Gamma_{43}^2} \Gamma\Big(1 - \frac{1}{\Gamma^2}\Big) - \frac{1}{\Gamma^{2}}\, .
\end{aligned}
\end{equation}
For numerical reasons it is convenient to express $\Gamma_{43}$ as 
\begin{equation}
	\begin{aligned}
	    \Gamma_{43} &= \Gamma\Gamma_0(1-\beta\beta_0) = \Gamma\Gamma_0 \frac{1+\beta\beta_0}{1+\beta\beta_0}  (1-\beta\beta_0) \\
	    & = \Gamma \Gamma_0 \Big(\frac{1}{\Gamma_0^2} + \frac{1}{\Gamma^2} - \frac{1}{\Gamma^2 \Gamma_0^2}\Big)  \frac{1}{(1 + \beta \beta_0)}\, ,
	\end{aligned}
\end{equation}
and its derivative, present in Eq.~\eqref{eq:method:dG_fsrs} and Eq.~\eqref{eq:method:dGeff3dG}, as
\begin{equation}
    \frac{d\Gamma_{43}}{d\Gamma} = \Gamma_0 + \frac{\Gamma (1 - \Gamma_0^2)}{\sqrt{\Gamma^2\Gamma_0^2 - \Gamma_0^2 - \Gamma^2 + 1}}\, .
\end{equation}

As the \ac{RS} propagates through the ejecta shell, it accretes matter 
at a rate given by $dm_{0;3}$. Following \citetalias{Nava:2013} we write, 
\begin{equation}\label{eq:method:dm3_0}
	dm_{0;3} = 2\pi R_{\rm rsh}^2 (1-\cos(\omega)) (\beta_{34} + \beta_{3})\rho_{4}' cdt' ,
\end{equation}
where $\omega$ is the half-opening angle of the \ac{BW}, 
$R_{\rm rsh}$ is the radius of the \ac{RS}, $\beta_{34}$ is the dimensionless 
velocity of the unshocked ejecta relative to the \ac{BW}, $\beta_{3}$ is the shocked 
ejecta velocity in its own frame and $\rho_{4}'$ is the mass density of the 
unshocked ejecta measured in the frame of the shocked ejecta. The latter can be 
related to the ejecta density in the progenitor frame as follows, 
\begin{equation}\label{eq:method:rho4ptimeTorho4}
	\rho_{4}'=\Gamma_{43}\rho_{4}''=\frac{\Gamma_{43}}{\Gamma}\rho_{4}=\Gamma(1-\beta\beta_0)\rho_{4}\, ,
\end{equation}
where in the last step we used that $\Gamma_{43}=\Gamma\Gamma_{0}(1-\beta\beta_{0})$. 
In Eq.~\eqref{eq:method:rho4ptimeTorho4}, $\rho_{4}''$ is the ejecta density measured in its 
own frame, \ie, the proper density. 
The quantity $\beta_{3}$ in Eq.~\eqref{eq:method:dm3_0} can be obtained from  
shock jump conditions (assuming strong shock and assuming that upstream is cold, \ie, 
that the energy density is $e_{4} = \rho_{4}''c^2$ and pressure $p_4=0$) as follows,  
\begin{equation}
	\beta_{3}=\frac{\beta_{34}}{3} = \frac{\beta_{0}-\beta}{3(1-\beta\beta_{0})}\, .
\end{equation}

The distance that a shock travels over the comoving time $dt'$ can be obtained 
from the shock velocity in the progenitor frame as follows 
$dR = \beta_{\rm sh} / (\Gamma (1-\beta\beta_{\rm sh}))$. 
In the case of \ac{RS}, the $\beta_{\rm sh} = \beta_{3}$ and after some algebra 
we can write $dt'=3/(4\beta\Gamma c)dR$. Substituting this into Eq.~\eqref{eq:method:dm3_0}, 
together with Eq.~\eqref{eq:method:rho4ptimeTorho4}, we obtain 
\begin{equation}\label{eq:method:dm3dr_1}
	\frac{dm_{0;3}}{dR} = 2\pi R_{\rm rsh}^2 (1-\cos(\omega)) \rho_4 \frac{d\Delta_{4}'}{dR} ,\,
\end{equation}
where we introduced the rate of change of the thickness of shocked region behind the 
\ac{RS}, $d\Delta_4'$, that reads,   
\begin{equation}\label{eq:method:ddelta4dr}
	\begin{aligned}
		\frac{d\Delta_{4}}{dR} & = \frac{d}{dR}(t_{\rm burst}\beta_{0} c - R_{\rm sh}) = \frac{\beta_0 -\beta}{\beta} \\
		& = \beta_0 \frac{\beta^{-4} - \beta_{0}^{-4}}{(\beta^{-1} + 	\beta_0^{-1})(\beta^{-2}+\beta^{-2}_0)}\, ,
	\end{aligned}
\end{equation}
where in the last stage we expanded it for numerical reasons. 
Since we are working within the thin-shell approximation to \ac{BW} radial structure, 
we can write $R\sim R_{\rm rsh} \sim R_{\rm shell}$. 

The initial width of the ejecta shell in the progenitor frame can be taken as
$\Delta_0 = c t_{\rm prompt} \beta_{0}$, where $t_{\rm prompt}$ is the 
duration of the mass ejection in the progenitor frame. 
In our model $t_{\rm prompt}$ is a free parameter. 
Additionally, we assume that the ejecta density in the observer frame, $\rho_4$, 
scales with its initial value at follows
\begin{equation}\label{eq:method:rho4}
	\rho_4 = \frac{M_0}{2\pi(1-\cos(\omega))R_{\rm rsh} \Delta_0 }
	\exp\Bigg( -\frac{\Delta_{4}}{\Delta_0} \Bigg)\, .
\end{equation}
Its derivative, needed for Eq.~\eqref{eq:method:dG_fsrs}, reads 
\begin{equation}
	\frac{d\ln\rho_{4}}{dR} = -\frac{2}{R} - \frac{1}{\Delta_0}\frac{d\Delta_{4}}{dR}\, .
\end{equation}
Substituting Eq.~\eqref{eq:method:rho4} into Eq.~\eqref{eq:method:dm3dr_1}, 
we obtain the equation for the mass accreted by the \ac{RS}, 
\begin{equation}\label{eq:method:dm3dr}
	\frac{dm_{0;3}}{dR} = \frac{M_0}{\Delta_0}\frac{d\Delta_{4}}{dR}
	\exp\Bigg( -\frac{\Delta_{4}}{\Delta_0} \Bigg)\, .
\end{equation}

It is commonly assumed that at the beginning of the simulation, the \ac{RS} has already 
crossed the ejecta. In this can we can rewrite   Eq.~\eqref{eq:method:dG_fsrs} as 
following, 
\begin{equation} \label{eq:method:dGdRold}
	d\Gamma = \frac{ -(1+\Gamma_{\rm eff; 2})(\Gamma-1) + \Gamma_{\rm eff; 2}(\hat{\gamma}_{2}-1)E_{\rm int; 2}' \frac{dm}{m}   }{ (M_0 + m) c^2 + \frac{d\Gamma_{\rm eff; 2}}{d\Gamma}E_{\rm int; 2}' + \Gamma_{\rm eff; 2}(\hat{\gamma}_{2}-1)E_{\rm int; 2}' 
		\frac{1}{\Gamma} }\, .
\end{equation}
Eq.~\eqref{eq:method:dGdRold} corresponds to the Eq.~7 in \citetalias{Nava:2013} after inserting the adiabatic term 
term given by Eq.~\eqref{eq:method:dEad2}.

\subsubsection{Lateral spreading}\label{sec:method:lat_spreading}

When a \ac{GRB} jet decelerates and different regions become  casually  
connected with each other, the transverse pressure gradient will lead to motion along 
the tangent to the surface. Consequently, the jet starts to spread laterally 
\citep[\eg][]{vanEerten:2009pa,Granot:2012,Duffell:2018iig}. 

In most semi-analytic \ac{GRB} afterglow models that include the jet lateral 
structure and employ the thin-shell approximation of the \ac{BW}, lateral spreading 
cannot be incorporated in a self-consistent way.  
Recently, however, new models were proposed where \ac{GRB} ejecta is modeled 
as a $2$D thin-shell including a pressure gradient along its surface. 
In such models lateral spreading occurs naturally \cite{Lu:2020hka,Wang:2024wbt}. 

In a model that approximates ejecta as a set of independent \acp{BW}, each of which 
is evolved under thin-shell approximation, the exact form of the lateral expansion 
prescription -- functional form of \ac{BW} opening angle evolution -- is one of the 
main free parameters. 
We consider the following prescription motivated by \citet{Huang:1999qa} and 
\citet{Ryan:2019fhz}.  
We assume that an expanding fluid element interacts only with matter in its immediate 
vicinity. There, the lateral and radial components of the velocity are related as 
$\beta_{r}/\beta_{\omega}=\partial \omega / \partial\ln R $ \cite{Huang:2000ni}. 
The co-moving sound speed in the shocked region reads $c_s^2 = dp'/de'|_{\rm shock}$ 
\cite{Kirk:1999km} that can be written as,  
\begin{equation}\label{eq:method:cs_shock}
	c_{s}^2 = \frac{\hat{\gamma}p'}{\rho'}\Big[ \frac{(\hat{\gamma}-1)\rho'}{(\hat{\gamma}-1)\rho' + \hat{\gamma}\rho'} \Big] c^2 = 
	\frac{\hat{\gamma} (\hat{\gamma} - 1) (\Gamma - 1)}{1 + \hat{\gamma} (\Gamma - 1)} c^2\, ,
\end{equation}
where we inserted $\hat{\gamma}$ using Eq.~\eqref{eq:method:eos}. 
Then, assuming that the spreading proceeds at the sound speed, 
$\upsilon_{\omega} = c_s$, the lateral expansion can be written as 
\cite{Huang:1999qa} 
\begin{equation} \label{eq:method:dthetadr_huang}
	\frac{d\omega}{dR} = \frac{c_s}{R \Gamma_{\rm sh} \beta_{\rm sh} c}\, .
\end{equation}
This formulation has been used in the early semi-analytic \ac{GRB} 
afterglow models \citep[\eg][]{Rossi:2004dz}.  

For a structured jet, however, the so-called ``conical'' spreading model 
was shown to yield a better agreement with numerical simulations \cite{Ryan:2019fhz}. 
There, all the material that has been swept up in the past affects spreading at a 
given time. 
However, in deriving their prescription, \citet{Ryan:2019fhz} implicitly 
assumed the ``TM'' \ac{EOS} \cite{Mignone:2005ns} for transrelativistic fluid. 
In order to remain consistent with the rest of our model, we, instead, continue 
to use Eq.~\eqref{eq:method:cs_shock}, while adopting structure-related terms 
from the aforementioned study to obtain the final form of the prescription 
as follows, 
\begin{equation}\label{eq:method:dthetadr}
	\begin{aligned}
		\frac{d\omega}{dR} & = \frac{c_s}{R \Gamma_{\rm sh} \beta_{\rm sh} c}\,  
	\\
		& \times 
		\begin{cases}
			0 & \text{ if } Q_0 \Gamma_{\rm sh}\beta_{\rm sh} \omega_c < 1 \\
			\frac{Q(1-Q_0\omega_c\Gamma_{\rm sh}\beta_{\rm sh})}{Q-Q_0} & \text{ if } Q \Gamma_{\rm sh}\beta_{\rm sh} \omega_c > 1 \\
			1 & \text{ otherwise } 
		\end{cases}
	\, ,
	\\
		& \times 
		\begin{cases}
			\tan(\omega_0/2) / \tan(\omega_c/2) & \text{ if } \omega_0 < \omega_c \\
			1 & \text{ otherwise } \\
		\end{cases}
	\, ,
	\end{aligned}
\end{equation} 
where $Q=3\sqrt{2}$, $Q_0=2$, $\omega_0$ is the initial half-opening angle 
of the \ac{BW}, and $\omega_c$ is the half-opening angle of the jet core, 
given as a free parameter when setting the Gaussian jet structure.

Numerical simulations of jet spreading show that only when 
a jet has decelerated, the spreading can commence 
\cite{Xie:2018vya,vanEerten:2009pa,Granot:2012,Duffell:2018iig}. 
Authors of the aforementioned prescriptions account for this by 
setting $d\omega/dt = 0$ if $Q_0 \Gamma_{\rm sh}\beta_{\rm sh} \omega_c \geq 1$ 
(see first case braces in Eq.~\eqref{eq:method:dthetadr}).

As the \ac{BW} laterally spreads, the amount of mass 
it sweeps increases. 
Following \citet{Granot:2012}, we write the equation for the mass 
accreted by the \ac{BW} as follows,
\begin{equation}\label{eq:method:dmdr_grb}
    dm = 2 \pi \rho \Big[ \big(1-\cos(\omega)\big) + \frac{1}{3}\sin(\omega)R d\omega \Big] R^2 \, .
\end{equation}

In the comoving frame, the shock downstream densities read
\begin{equation}\label{eq:method:rho_prime2}
	\begin{aligned}
		\rho'_{2} &= (\hat{\gamma}_2 \Gamma + 1) / (\hat{\gamma}_2 - 1) \rho_{\rm ISM}\, , \\
		\rho'_{3} &= (\hat{\gamma}_3 \Gamma + 1) / (\hat{\gamma}_3 - 1) \rho_{4}\, ,
	\end{aligned}
\end{equation}
for the \ac{FS} and \ac{RS} respectively. 

While we employ the thin-shell approximation to the \ac{BW} structure, 
we also need actual thicknesses of shocked regions in order to compute 
radiation transport. Under the general assumption of a homogeneous shell, 
but relaxing the assumption of the uniform upstream medium, the shocked 
region thicknesses in the burster frame read \cite{Johannesson:2006zs}, 
\begin{equation}\label{eq:method:thickness}
	\begin{aligned}
		\Delta R_{\rm fsh} &= \frac{m_2}{ 2 \pi R^2 ( 1 - \cos(\omega) ) \Gamma \rho'_{2} } \, , \\
		\Delta R_{\rm rsh} &= \frac{m_3}{ 2 \pi R^2 ( 1 - \cos(\omega) ) \Gamma \rho'_{3} } \, .
	\end{aligned}
\end{equation}
%
It is worth noting that for the \ac{FS}, if $1-\cos(\omega)=2$ and the swept-up 
mass $m_2 = 4 \pi R^3 n' m_p / 3$, we obtain the commonly used \ac{BM} shock 
thickness,  $\Delta R'=R/12\Gamma^2$ \citep[\eg][]{Johannesson:2006zs,vanEerten:2009pa}. 

\subsection{Microphysics \& Comoving Radiation}\label{sec:method:mphis_comovrad}

As a \ac{BW} moves through the medium with small but present magnetization, 
the seed magnetic field becomes amplified through a range of instabilities such as  
the current-driven instability \cite{Reville:2006px}, 
the Kelvin-Helmholtz shear instability \cite{Zhang:2011}, the
Weibel (filamentation) instability \cite{Medvedev:1999tu,Lemoine:2010,Tomita:2016yib},
the \v{C}erenkov resonant instability \cite{Lemoine:2010}, 
the Rayleigh-Taylor instability \cite{Duffell:2013tha}, 
the magneto-rotational instability \cite{Cerda-Duran:2011}, 
or the pile-up effect \cite{RochadaSilva:2014ehi}.  

Inbound charged particles gain energy, reflecting off and  scattering on \ac{MHD} 
instabilities, and acquire a spectrum, a certain distribution in energy. 
In order to model this process, \ac{PIC} simulations are commonly employed 
to study particle dynamics at electron's gyro-radius scale \citep[\eg][]{Sironi:2015oza}. 
However, even sophisticated \ac{MHD}-\ac{PIC} simulations used to model larger 
spatiotemporal scales are still limited to a few $10^3$ of proton gyro-scales and few 
milliseconds \cite{Bai:2014kca,Mignone:2018per}.  
These first-principles studies showed that the main process responsible for electron 
acceleration at collisionless shocks is the first-order Fermi acceleration 
\cite{Spitkovsky:2008fi,Sironi:2009,Sironi:2011,Park:2014lqa}.  

As mentioned in the introduction, the shock microphysics is 
practically impossible to model on spatial and temporal scales relevant for 
\ac{GRB} afterglows. It is thus common to make simplifying assumptions regarding shock 
microphsyics and employ approximate models or prescriptions. These can be generally 
divided into two groups, depending on whether the electron (and photon) distributions 
are evolved self-consistently or assumed to be fixed and given. The example of the 
latter is a very common model proposed by \citet{Sari:1997qe} and further extended by 
\citet{Granot:2001ge}. 
The underlying assumption of this model is that the instantaneous electron 
spectrum can always be approximated with a \ac{BPL} separated by critical 
electron \acp{LF}. Then, the synchrotron spectrum, computed as an integral over 
the electron distribution, is also a \ac{BPL}. Moreover, it is possible 
to include \ac{SSA} and high-energy \ac{SSC} spectrum analytically as well 
\cite{Nakar:2009,Joshi:2019opd,Yamasaki:2021klt,Pellouin:2024gqj}. However, by 
construction, these formulations, while allowing for very computationally 
efficient afterglow models (\eg, \texttt{afterglowpy} and \texttt{jetsimpy}), 
are not flexible or generalizable. 

Another approach to modeling microphysics is to evolve electron distribution explicitly 
accounting for heating and cooling and obtain instantaneous emission spectra 
numerically by convolving emission kernels with current electron distribution 
\cite{Dermer:2000gu,Petropoulou:2009,Miceli:2022efx,Huang:2022fnv,Zhang:2023uei}. 
In these models, only the injection electron spectrum -- the electron distribution 
that emerges after the acceleration processes at collisionless shock -- is usually 
fixed, while the electron distribution in the shock downstream is allowed to evolve. 
This approach is also quite common in modeling quasi-stationary systems, such as 
active galactic nuclei jets \cite{Kino:2002fc,Jamil:2012} and it allows for 
larger flexibility, \eg, employing more physically-motivated injection spectra 
\cite{Warren:2021whb,Gao:2023kjw}, different heating \cite{Warren:2015lsa} and 
cooling processes and non-trivial interactions between photons and electrons 
\cite{Wang:2006rp,He:2009}. 

Notably, in both approaches, the microphysics of particle acceleration and magnetic 
field amplification is hidden behind free parameters. Specifically, it is common 
to assume that a certain fraction of downstream shock energy goes into particle 
acceleration, $\epsilon_{\rm e}$, and magnetic field amplification, $\epsilon_{\rm b}$. 
Together with the slope of the injection electron spectrum (assuming it is a single 
power-law) $p$, they from a set of free parameters that most afterglow models consider. 

Since shock microphysics is the same for \ac{FS} and \ac{RS} and depends on the 
properties of these shocks' downstream, we omit here subscripts $2$ and $3$ that 
were used in the previous section for clarity. For the same reason we omit 
apostrophe, $'$, for the electron \acp{LF} since we only work with the comoving electron 
spectrum here. 


Recall that comoving energy density in shock downstream reads $e' = E_{\rm int}'/ V'$ 
where the comoving volume $V' = m / \rho'$, $m$ is the shocked region mass and 
$\rho'$ is the comoving mass density. Then, the magnetic field strength 
in the shock downstream reads 
\begin{equation}\label{eq:method:B}
	B' =  \sqrt{8 \pi \epsilon_{\rm b} e'}\, .
\end{equation}
Notably, $e'$ can also be evaluated directly as 
$e' = 4\Gamma(\Gamma-1)n_{\rm ISM} m_p c^2$ in the relativistic regime. 
Then, the magnetic field strength can simply be expressed as 
$B' = \sqrt{32\pi \epsilon_{\rm b} m_p c^2 n_{\rm ISM} (\Gamma - 1)\Gamma}$.


Consider a power-law electron spectrum in injected electrons, 
$dN_{e}/d\gamma_{e}\propto\gamma_{e}^{-p}$ 
for $\gamma_{e}\in(\gamma_{e;\,\rm m},\,\gamma_{e;\,\rm M})$, 
where $\gamma_{e}$ is the 
electron \ac{LF}, $p$ is the spectral index and
$\gamma_{e;\,\rm m}$ and $\gamma_{e;\,\rm M}$ are minimum and 
maximum \acp{LF} \cite{Dermer:1997pv,Sari:1997qe}.

The maximum \ac{LF} can be obtained by balancing the acceleration 
timescales with the minimum cooling timescale for an electron. 
For electrons, the dynamical timescale, 
$t'_{\rm dyn} \simeq R/c\Gamma$, would generally be significantly larger 
than the radiation cooling timescale that reads, 
\begin{equation}\label{eq:method:tcool}
	t_{\rm cool}' 
	= \frac{6\pi m_e c}{\gamma_e \sigma_T B^{'2}(1 + \tilde{Y})} \, ,
\end{equation}
where $\tilde{Y}$ stands for the corrections due to \ac{IC} scattering. 
Thus, for an electron with $\gamma_{e}=10^3$ in magnetic field $B'=100\,$G, $t_{\rm cool}'\simeq 77\,$sec., assuming $\tilde{Y}=1$. 

From Eq.~\eqref{eq:method:tcool} it is possible to derive the 
characteristic cooling \ac{LF} above which injected electrons rapidly 
cool, 
\begin{equation}\label{eq:method:gc}
	\gamma_{e; \, \rm c} = \frac{6\pi m_e c}{\sigma_T t' B^{'2}(1+\tilde{Y})}\, .
\end{equation}
Equation~\eqref{eq:method:gc} is written in the comoving frame. Commonly, 
in the literature $\gamma_{e; \, \rm c}$ is evaluated in the observer frame,  
then, an additional factor $1/\Gamma$ is added. 

The comoving acceleration time for an electron can be written as, 
\begin{equation}\label{eq:method:tacc}
	t_{\rm acc}' \simeq \zeta\frac{r_{\rm B}}{c} = \zeta\frac{\gamma_{e}m_e c}{q_e B'}\, ,
\end{equation}
where $r_{\rm B} = \gamma_{e} m_e c^2 / (q_e B')$ is the guration radius and 
$\zeta\simeq1$ is a free parameter that depends on the acceleration mechanism. 
Equating Eq.~\eqref{eq:method:tcool} and Eq.~\eqref{eq:method:tacc}, 
we obtain,  
\begin{equation}\label{eq:method:gM}
	\gamma_{e;\,\rm M} = \sqrt{\frac{6\pi q_e}{\sigma_T B' \zeta (1 + \tilde{Y})}}\, .
\end{equation}
It is important to note that $\gamma_{\rm e;\, M}$ may depend on the 
non-uniformity in the shock downstream magnetic field. 
If $B'$ is not constant, but decays with distance from the shock front, 
electrons will lose most of their energy further downstream, where the 
Larmor radius, $r_{\rm B}$, is larger \cite{Kumar:2012}. To account for this, 
the weakest magnetic field, $B_{\rm w}'$, should be used in 
Eq.~\eqref{eq:method:gM} instead of $B'$ \citep{Miceli:2022efx}.

The lower limit of the electron injection spectrum $\gamma_{e;\, \rm m}$, 
can be obtained from the following considerations. Within an assumed
power-law distribution, $N_e \propto \gamma_{e}^{-p}$, the average 
electron \ac{LF} reads, 
\begin{equation}
	\langle \gamma_{e} \rangle =
	\frac{\int_{\gamma_{e;\,\rm m}}^{\gamma_{e;\,\rm M}} \gamma_{e}N_{e}d\gamma_{e}}{\int_{\gamma_{e;\,\rm m}}^{\gamma_{e;\,\rm M}} N_{e}d\gamma_{e}} = 
	\frac{\int_{\gamma_{e;\,\rm m}}^{\gamma_{e;\,\rm M}}{\gamma_{e}^{-p+1}d\gamma_{e}}}{\int_{\gamma_{e;\,\rm m}}^{\gamma_{e;\,\rm M}}{\gamma_{e}^{-p}d\gamma_{e}}}\, .
\end{equation}
This integral can be solved analytically. If $p \neq 1$ and 
$p \neq 2$, we can write, 
\begin{equation}\label{eq:method:gm_1}
	\langle \gamma_{e} \rangle = \Bigg( \frac{p-1}{p-2} \Bigg)\Bigg(\frac{\gamma_{e;\,\rm M}^{-p+2}-\gamma_{e;\,\rm m}^{-p+2}}{\gamma_{e;\,\rm M}^{-p+1}-\gamma_{e;\,\rm m}^{-p+1}} \Bigg) \, .
\end{equation}
Otherwise, the solution is \cite{Zhang:2018book}, 
\begin{equation}\label{eq:method:gm_2}
	\langle \gamma_{e} \rangle = 
	\begin{cases}
		\frac{\ln(\gamma_{e;\,\rm M}) - \ln(\gamma_{e;\,\rm m})}{-\gamma_{e;\,\rm M}^{-1} + \gamma_{e;\,\rm m}^{-1}} \text{ if } p = 2\, , \\
		\frac{\gamma_{e;\,\rm M} - \gamma_{e;\,\rm m}}{\ln(\gamma_{e;\,\rm M}) - \ln(\gamma_{e;\,\rm m})} 
		\text{ if } p = 1 \, . 
	\end{cases}
\end{equation}

On the other hand, under the assumption that only a fraction of shock 
downstream energy is used in electron acceleration, we can write the 
following expression for the average electron \ac{LF}, 
\begin{equation}\label{eq:method:gm_3}
	\langle \gamma_{e} \rangle = 
	\epsilon_{\rm e}\frac{e'}{\rho'}\frac{m_p}{m_e c^2}\, ,
\end{equation}
where $m_{e}$ and $m_{p}$ are the electron and proton mass, respectively. 

Notably, in the presence of electron-positron pairs, the ratio of proton to electron 
densities has to be included in aforementioned equations \cite{Beloborodov:2005nd,Nava:2013}. 
Here we implicitly assume that the shock downstream is not pair-rich.

Eq.~\eqref{eq:method:gm_1} and Eq.~\eqref{eq:method:gm_3} can be solved analytically 
for $\gamma_{e;\,\rm m}$ assuming that $p>2$ and $\gamma_{e;\,\rm M} \gg \gamma_{e;\,\rm m}$, 
to obtain $\gamma_{e;\,\rm m} = ((p - 2) / (p - 1)) e' m_p / (\rho' m_e c^2)$.  
These limits are usually respected in the context of \ac{GRB} afterglows, \citep[\eg][]{Kumar:2014upa}.   
Also, the theory of particle acceleration at relativistic shocks predicts $p \gtrsim 2$ 
\cite{Sironi:2015oza,Marcowith:2020vho}. 

In Eq.~\eqref{eq:method:gm_3} we chose not to include explicitly a quantity that represents a 
fraction of the shocked electrons that participate in the acceleration process. This parameter, 
usually denoted as $\xi_{e}$ in \ac{GRB} afterglow literature, is degenerated with 
$\epsilon_{\rm e}$ \cite{Eichler:2005ug}. 
As such, we implicitly assume that $\xi_{e}=1$, \ie, all inbound electrons are accelerated, 
and only $\epsilon_{\rm e}$ regulates with what efficiency.

As a shock decelerates and $\gamma_{e; \, \rm m} \rightarrow 1$, 
electron acceleration enters the so-called deep-Newtonian regime \cite{Sironi:2013tva}, 
that starts when $\beta_{\rm sh} \lesssim 8\sqrt{m_p/m_e}\bar{\epsilon}_{\rm e}$, 
where $\bar{\epsilon}_{\rm e} = 4\epsilon_{\rm e}(p-2)/(p-1)$ \cite{Margalit:2020bdk}. 
Notably, synchrotron emission from electrons accelerated at lower shock velocity is 
dominated by electrons with \ac{LF} ${\simeq}2$, instead of those with $\gamma_{e;\, \rm m}$. 
Thus, when $\gamma_{e;\,\rm m}$ gets close to $1$, additional adjustments to our model are 
needed. Specifically, we set that only a fraction of injected electrons, $\xi_{\rm DN}$, 
can contribute to the observed emission. 
The $\xi_{\rm DN}$ is computed according as follows \cite{Sironi:2013tva},
\begin{equation} \label{eq:method:gm_lim}
	\xi_{\rm DN} = \frac{p-2}{p-1}\epsilon_{\rm e}\frac{e'}{\rho'}\frac{m_p}{m_e c^2}\, .
\end{equation}
The inclusion of $\xi_{\rm DN}$ was shown to yield improved fits for late-time \GRB{} 
data \cite{Ryan:2023pzk,Wang:2024wbt}.

\subsubsection{Electron distribution evolution}\label{sec:method:electrons}

In order to calculate the time-dependent synchrotron and \ac{SSC} radiation, we 
self-consistently evolve the electron \ac{LF} distribution solving numerically 
the 1D continuity -- Fokker-Plank-type -- equation 
\cite{Chandrasekhar:1943,Zhang:2018book},  
\begin{equation}\label{eq:method:dNedt}
	\begin{aligned}
	\frac{\partial N_e(\gamma_{e},t')}{\partial t'} = & 
	-\frac{\partial }{\partial \gamma_{e}}\Big[ \dot{\gamma} N_{e}(\gamma_{e},t') \Big] \\
	& +  Q(\gamma_{e},t') + Q_{\rm pp}(\gamma_{e},t')\, , 
	\end{aligned}
\end{equation}
where $ N_e(\gamma_{e},t') $ is the number of electrons in the energy interval 
$[\gamma_{e}, \gamma_{e}+d\gamma_{e}]$ at time $t'$, $\dot{\gamma}$ is the rate at 
which electrons with \ac{LF} $\gamma_{e}$ gain (if $>0$) or lose (if $<0$) energy,  
$Q(\gamma_{e},t)'$ is the injection term due to acceleration at a shock, 
and $Q_{\rm pp}(\gamma_{e},t')$ is the injection term due to \ac{PP}.

For deriving Eq.~\eqref{eq:method:dNedt} several approximations were made. 
We neglect the diffusion in energy space that would otherwise give 
a term $\partial/\partial\gamma_{e} [D_e(\gamma_e) \partial N_{e}/\partial\gamma_{e}]$; 
we neglect the electron escape term $-N_{e}(\gamma_{e},t')/t_{e;\rm \, esc}\,$,  
where $t_{e;\rm\, esc}$ is the typical escape time scale of an electron with 
\ac{LF} $\gamma_{e}$.

The source term, $Q(\gamma_{e},t')$ in Eq.~\eqref{eq:method:dNedt} describes the 
distribution with which electrons are injected into the system, which in our case 
is the power-law, $Q(\gamma_{e},t') = Q_{0} \gamma_{e}^{-p}$, where $Q_{0}$ is the 
normalization coefficient, that can be obtained by integrating the injection electron 
distribution from $\gamma_{e; \, \rm m}$ to $\gamma_{e;\,\rm  M}$ as follows, 
\begin{equation}
	\begin{aligned}
		N_{e;\,\rm inj}(t') &= \int_{\gamma_{e;\,\rm  m}}^{\gamma_{e;\,\rm  M}} Q(\gamma_{e},t')d\gamma_{e} = \int_{\gamma_{e;\,\rm  m}}^{\gamma_{e;\,\rm  M}} Q_{0}\gamma_{e}^{-p} d\gamma_{e} \\
		&= Q_{0} \frac{\gamma_{e;\,\rm  M}^{-p+1}-\gamma_{e; \, \rm m}^{-p+1}}{1-p}\, .
	\end{aligned}
\end{equation}
Since the total number of injected electrons is equal to the number of protons crossing the 
shock front, \ie, $m/m_p$, the final form of the source term reads
\begin{equation}\label{eq:method:Qinj}
	Q(\gamma_{e},t') = \frac{m}{m_p} 
	\Bigg( \frac{1-p}{\gamma_{e;\,\rm  M}^{-p+1}-\gamma_{e; \, \rm m}^{-p+1}} \Bigg) \gamma_{e}^{-p} \, 
\end{equation}
for $\gamma_{e}\in(\gamma_{e;\,\rm  m},\gamma_{e;\,\rm  M})$ and $0$ otherwise.

The $\dot{\gamma}_{e}$ term in Eq.~\eqref{eq:method:dNedt} describes heating 
and cooling processes that take place in system. Here, we consider 
the following contributions to this term,  
\begin{equation}\label{eq:method:gamma_dot}
	\dot{\gamma} = \dot{\gamma}_{\rm syn} + \dot{\gamma}_{\rm adi} + \dot{\gamma}_{\rm SSC}\, ,
\end{equation}
as described in detail below.

As \ac{BW} evolves, the shocked region volume, $V'=m/\rho'$ increases, 
and electrons lose their energy to adiabatic expansion. Recalling that  
Eq.~\eqref{eq:method:dEad2} can be written as 
$d(\gamma_{e}-1)/dR = -(\hat{\gamma}-1)(\gamma_{e}-1)d\ln(V')/dR$, 
we can write the adiabatic cooling term for electrons as
\begin{equation}
	\dot{\gamma}_{\rm adi} =\frac{\gamma_{e}}{3}\Big(\frac{\gamma_{e}^2-1}{\gamma_{e}}\Big)d\ln(V')\, .      
\end{equation}

In the context of \ac{GRB} afterglows, the shock magnetic fields are expected to 
be randomized. Then, the synchrotron emission power of a single electron 
is $P_{\gamma}' = (4/3)\sigma_T c\gamma_{e}^2\beta_e^2 u_{\rm B}'$, 
where $\beta_e$ is the dimensionless velocity of an electron, obtained 
by averaging over the pitch angles \cite{RybickiLightman:1985}.
The synchrotron cooling rate, $\dot{\gamma}_{\rm syn}$, of an 
electron then reads 
\begin{equation}\label{eq:method:gamma_dot_synch}
	\dot{\gamma}_{\rm syn} = - \Big(\frac{\sigma_T B^{'2} }{6 \pi m_e c }\Big)\gamma_{e}^2\, .
\end{equation}

Before considering the last term in Eq.~\eqref{eq:method:gamma_dot}, it is 
worth noting that generally, only synchrotron cooling is considered in most \ac{GRB} 
afterglow models, \ie, $\dot{\gamma}_{e} = \dot{\gamma}_{\rm syn}$ in 
Eq.~\eqref{eq:method:gamma_dot}. In this case, and if the injection function is a 
power-law, $Q(\gamma_{e},t')\propto\gamma_{e}^{-p}$, there exists an analytic solution 
to Eq.~\eqref{eq:method:dNedt} \citep[see, \eg,][]{Sari:1997qe}. In fact, there exist 
two solutions, depending on the order of $\gamma_{e; \, \rm m}$ (Eq.~\eqref{eq:method:gm_1}, Eq.~\eqref{eq:method:gm_3}) and 
$\gamma_{e; \, \rm c}$ (Eq.~\eqref{eq:method:gc}). If the $\gamma_{e; \, \rm m} < \gamma_{e; \, \rm c}$, the 
resulting spectrum is generally referred to as being in a ``slow cooling'' regime, 
while the case of $\gamma_{e; \, \rm m} > \gamma_{e; \, \rm c}$ corresponds to the 
``fast cooling'' regime. 
The analytic spectra for both regimes is 
\begin{equation}\label{eq:method:Ne_f}
	\begin{aligned}
		N_{e} (\gamma_{e}) &= K_{\rm f} 
		\begin{cases}
			\gamma_{e;\,\rm  m} ^{1-p} \gamma_{e}^{-2}\, 
			& \text{ when } 
			\gamma_{e; \, \rm c} < \gamma_{e} < \gamma_{e;\,\rm  m} \, ,
			\\
			\gamma_{e} ^{-p-1} 
			& \text{ when }
			\gamma_{e;\,\rm  m} < \gamma_{e} < \gamma_{e;\,\rm  M}\, ,
		\end{cases} 
		\\
		K_{\rm f} &= \frac{m}{m_{p}} 
		\Bigg(
		\frac{\gamma_{e; \, \rm c}^{-1} - \gamma_{e;\, \rm m}^{-1}}{\gamma_{e;\,\rm  m}^{p-1}} 
		+ \frac{\gamma_{e;\,\rm  m}^{-p} - \gamma_{e;\, \rm M}^{-p}}{p}
		\Bigg)^{-1}\, ,
	\end{aligned}
\end{equation}
for the fast cooling and 
\begin{equation}\label{eq:method:Ne_s}
	\begin{aligned}
		N_{e} (\gamma_{e}) &=   K_{\rm s} 
		\begin{cases}
			\gamma_{e}^{-p} 
			& \text{ when }
			\gamma_{\rm e;\, m} < \gamma_{e} < \gamma_{e; \, \rm c} \, ,
			\\
			\gamma_{e; \, \rm c}\gamma_{e}^{-p-1}
			& \text{ when }
			\gamma_{e; \, \rm c} < \gamma_{e} < \gamma_{\rm e;\, M}\, ,
		\end{cases}
		\\
		K_{\rm s} & = \frac{m}{m_{p}} 
		\Bigg(
		\frac{\gamma_{e; \, \rm c}^{1-p} - \gamma_{e;\,\rm  m}^{1-p}}{1-p} - 
		\frac{\gamma_{e; \, \rm c}(\gamma_{e;\,\rm  M}^{-p} - \gamma_{e; \, \rm c}^{-p})}{p} 
		\Bigg)^{-1}\, .
	\end{aligned}
\end{equation}
for the slow cooling.

Normalization factors $K_{\rm f}$ and $K_{\rm s}$ in Eq.~\eqref{eq:method:Ne_f} and 
Eq.~\eqref{eq:method:Ne_s} are derived from the electron spectrum by 
first dividing the spectra (for both regimes) into two segments for both regimes as,  
\begin{equation}\label{eq:method:dNe_dgamma_f}
	\begin{aligned}
		dN_{e} (\gamma_{e}) &=  
		\begin{cases}
			K_{f,1} \gamma_{e}^{-2}\, 
			& \text{ when } 
			\gamma_{e; \, \rm c} < \gamma_{e} < \gamma_{e;\,\rm  m} \, ,
			\\
			K_{f,2}\gamma_{e} ^{-p-1} 
			& \text{ when }
			\gamma_{e;\,\rm  m} < \gamma_{e} < \gamma_{e;\,\rm  M}\, .
		\end{cases} 
  \end{aligned}
\end{equation}
for the fast cooling 
\begin{equation}\label{eq:method:dNe_dgamma_s}
	\begin{aligned}
		dN_{e} (\gamma_{e}) &=  
		\begin{cases}
			K_{s,1} \gamma_{e}^{-p}\, 
			& \text{ when } 
			\gamma_{e; \, \rm m} < \gamma_{e} < \gamma_{e;\,\rm  c} \, ,
			\\
			K_{s,2}\gamma_{e} ^{-p-1} 
			& \text{ when }
			\gamma_{e;\,\rm  c} < \gamma_{e} < \gamma_{e;\,\rm  M}\, .
		\end{cases} 
  \end{aligned}
\end{equation}
for the slow cooling, and then integrating the resulting \acp{BPL} as, 
\begin{equation}\label{eq:method:Ne_inj_f}
    \begin{aligned}
        N_{e} = \int_{\gamma_{e; \, \rm c}}^{\gamma_{e; \, \rm m}}{K_{f;1}\gamma_e^{1-p}\gamma_e^{-2}d\gamma_e + \int_{\gamma_{e; \, \rm m}}^{\gamma_{e;\,\rm  M}}{K_{f;2}\gamma_e^{-p-1}}}\, ,
    \end{aligned}
\end{equation}
in the fast cooling regime, and
\begin{equation}\label{eq:method:Ne_inj_s}
    \begin{aligned}
        N_{e} = \int_{\gamma_{e; \, \rm m}}^{\gamma_{e; \, \rm c}}{K_{s;1}\gamma_e^{-p}d\gamma_e + \int_{\gamma_{e; \, \rm c}}^{\gamma_{e;\,\rm  M}}{K_{s;2}\gamma_e^{-p-1}}}\, ,
    \end{aligned}
\end{equation}
in the slow cooling regime.  
The final result is then obtained by combining the normalization factors 
of these \ac{BPL} segments as, 
\begin{equation}\label{eq:method:Ne_inj_f_integrated}
    \begin{aligned}
     K_{f;2} &= \frac{N_{e}}{\gamma_{e; \, \rm m}^{1-p}\big(\gamma_{e; \, \rm c}^{-1}-\gamma_{e; \, \rm m}^{-1} \big)-p^{-1}\big(\gamma_{e; \, \rm M}^{-p}-\gamma_{e; \, \rm m}^{-p} \big)}  \\
     K_{f;1} &= K_{f;2}\gamma_{e; \, \rm m}^{1-p} \, ,
    \end{aligned}
\end{equation}
for the fast cooling regime and 
\begin{equation}\label{eq:method:Ne_inj_s_integrated}
    \begin{aligned}
     K_{s;1} &= \frac{N_{e}}{(1-p)^{-1}\big(\gamma_{e; \, \rm c}^{1-p}-\gamma_{e; \, \rm m}^{1-p} \big)-\gamma_{e; \, \rm c}p^{-1}\big(\gamma_{e; \, \rm M}^{-p}-\gamma_{e; \, \rm c}^{-p} \big)} \\
     K_{s,2} &= K_{s,1}\gamma_{e; \, \rm c}   \, ,
    \end{aligned}
\end{equation}
for the slow cooling regime. Substituting coefficients from 
Eqs.~\eqref{eq:method:Ne_inj_f_integrated} \eqref{eq:method:Ne_inj_s_integrated} 
into Eq.~\eqref{eq:method:Ne_inj_f} and \eqref{eq:method:Ne_inj_s} we obtained 
Eqs.~\eqref{eq:method:Ne_f} and \eqref{eq:method:Ne_s}. 

This analytic solution is a foundation of most analytic synchrotron emission 
prescriptions commonly used in afterglow modeling. We implement it in 
\PBA{} for comparison with full numerical method.


Synchrotron photons produced by a population of electrons may also 
scatter on the same electrons. This is an \ac{SSC} process \cite{RybickiLightman:1985}. 
Notably, photons that have already been upscattered can interact with electrons 
again. However, this high orders scatterings are hampered by the  Klein-Nishina 
effect, \ie, the linear decrease of the scattering cross-section with the incident 
electron \ac{LF} \cite{RybickiLightman:1985}.  Moreover, computing higher order 
scatterings is numerically expensive. Thus, we limit the implementation to only the 
first-order \ac{IC} component. 
In order to obtain the energy loss rate due to \ac{SSC} scatterings we need to convolve the scattering kernel with the seed photon spectrum, $n_{\tilde{\nu}'}$, 
estimation of which we discuss later as it requires computing comoving emissivities 
first. 
The \ac{SSC} cooling rate that includes Klein-Nishina correction is given by   
\cite{Blumenthal:1970,Fan:2008,Geng:2017aku},
\begin{equation}\label{eq:method:gamma_dot_ssc}
	\dot{\gamma}_{\rm ssc} = \frac{3}{4}\frac{h\sigma_T}{m_e c \gamma_{e}^2}
	\int_{\tilde{\nu}'_{0}}^{\tilde{\nu}'_{1}} \frac{n_{\tilde{\nu}'}}{\tilde{\nu}'}
	\Bigg[ \int_{\nu'_0}^{\nu'_1} h\nu' K(\tilde{\nu}',\gamma_{e},\nu') d\nu'\Bigg] d\tilde{\nu}'\, ,
\end{equation}
where $K(\cdots)$ is the scattering kernel defined below, 
$\tilde{\nu}$ is the frequency of seed photons (before scattering) and 
$\nu$ is the photon frequency after the scatting, \ie, the \ac{SSC} photons.

The \ac{SSC} kernel can be expressed as a function of normalized energy of 
incoming photons $\varepsilon = h\nu/(m_e c^2)$, energy of the electron, 
$\tilde{\varepsilon} = h\tilde{\nu}/m_e c^2$ and electron \ac{LF} as 
follows \cite{Jones:1968,Miceli:2022efx}, 
\begin{equation}\label{eq:method:ssc_k1}
	K(\tilde{\varepsilon}',\gamma_{e},\varepsilon') = 
	\frac{\varepsilon}{\tilde{\varepsilon}} - \frac{1}{4\gamma_{e}} 
	\text{ for }
	\frac{\varepsilon}{4\gamma_{e}} < \varepsilon < \tilde{\varepsilon}\, 
\end{equation}
and 
\begin{equation}\label{eq:method:ssc_k2}
	\begin{aligned}
	K(\tilde{\varepsilon}',\gamma_{e},\varepsilon) 
		& = 2q\ln(q) + (1+2q)(1-q)  \\
		& + 0.5(1-q) \frac{(4\gamma_{e}\tilde{\varepsilon}q)^2 }{(1+4\gamma_{e}\tilde{\varepsilon})}\, , \\
		& \text{ for } 
		\tilde{\varepsilon} < \varepsilon < \frac{4\gamma_e^2\tilde{\varepsilon}}{1+4\gamma_{e}\tilde{\varepsilon}}
		\, ,
	\end{aligned}
\end{equation}
where $q = \varepsilon / (4\gamma_{e}\varepsilon) / (\gamma_{e} - \varepsilon)$. 
Eq.~\eqref{eq:method:ssc_k1} represents the down-scattering on an 
incoming photon and Eq.~\eqref{eq:method:ssc_k2} stands for the 
up-scattering. 
%

When low energy electrons re-absorb newly produced synchrotron photons in free-free 
transitions, they can gain energy. This process can be quantified with an \ac{SSA} 
cross section \cite{Ghisellini:1991}, and is referred to as \ac{SSA} heating. 
At present, we do not consider this process for the sake of simplicity. 
However, it can be added as an additional heating $\dot{\gamma}_{\rm ssa}^+$ 
term in Eq.~\eqref{eq:method:gamma_dot} computed following \cite{Gao:2013}. 
\ac{SSA} becomes important when its charactersitic frequency is larger than the 
cooling frequency, associated with $\gamma_{e; \, \rm c}$. In this regime, 
called strong absorption regime, electrons may pile up at a $\gamma_{e; \, \rm pile-up}$, 
that can be derived by balancing the cooling rate due to \ac{SSC} losses and heating 
due to \ac{SSA}. In the context of \ac{GRB} afterglows, this regime was shown to 
occur in a \ac{RS} when ejecta is moving through a dense wind medium 
\cite{Kobayashi:2003zk,Gao:2013}.

\subsubsection{Comoving synchrotron spectrum}

While in most cases of interest \ac{FS} remains relativistic and 
$\langle\gamma_{e}\rangle \gg 1$, \ac{RS} becomes relativistic only under 
certain conditions and generally $\Gamma_{43}\sim1$ \cite{Uhm:2012}. 
Thus, in computing synchrotron emissivity from an electron population we 
need to account for small electron \acp{LF}. 
This can be achieved by adding cyclo-synchrotron emissivity. An example of such an 
approach is given in \citet{Petrosian:1981}. For the sake of computational efficiently 
we adopt here the following phenomenological expression for the comoving 
synchrotron emissivity from a single electron \ac{LF}, 
$j'_{\rm syn}({\nu',\gamma_{e}})$ \cite{Ghisellini:1997gu}, 
\begin{equation}\label{eq:method:cycl}
	j'_{\rm syn}({\nu',\gamma_{e}}) = \frac{3 q_{e}^3 B'}{m_e c^2} \frac{2 p^2}{1+3p^2}\exp\Bigg(\frac{2(1-x)}{1+3p^2}\Bigg)\, ,
\end{equation}
where $\gamma_{e} \leq 2$ and $p=\gamma^2 _{e} - 1$ is the electron dimensionless 
momentum, $x=\nu'/\nu_{\rm L}'$ with $\nu'_{\rm L}$ is the Larmor frequency, 
$\nu_{\rm L}' = q_e B' / (2\pi m_e c)$.

Eq.~\eqref{eq:method:cycl} was shown to have the correct cooling rate (\ie, 
Eq.~\eqref{eq:method:gamma_dot_synch}) when integrated over frequency; the correct 
frequency dependence, $\propto\exp(-2 x)$, at large harmonics ($x\gg1$ in both 
non-relativistic and ultrarelativistic regimes). It was also shown to have 
better than $~40\,\%$ agreement with exact expressions at $\gamma_{e} = 2$ 
\cite{Ghisellini:1988}.

For electrons with $\gamma_{e} > 2$ we can consider the standard synchrotron radiation 
formula averaged over isotopically distributed pitch angles \cite{Aharonian:2010}, 
\begin{equation}\label{eq:method:synch_bessel}
	\begin{aligned}
	j'_{\rm syn}(\nu',\gamma_{e}) = 
	\frac{3 q_{e}^3 B'}{m_e c^2}
	2 y^2 \Big( & K_{4/3}(y)K_{1/3}(y) \\
	& - \frac{3}{5}y\big( K_{4/3}^2 (y) - K_{1/3}^2 (y) \big) \Big)\, ,
	\end{aligned}
\end{equation}
where $K_{z}(y)$ is the modified Bessel function of the order $z$, 
and $y=\nu'/\nu_{\rm crit}'$ with 
\begin{equation}\label{eq:method:nu_crit}
	\nu_{\rm crit}' = \frac{3}{4}\frac{q_e B'}{\pi m_e c} 
\end{equation}
being the critical frequency.

However, it is numerically expensive to evaluate modified Bessel functions 
for all evolution time steps, all electron \acp{LF} and all frequencies, 
which would be required to compute the comoving spectra with Eq.~\eqref{eq:method:synch_bessel}.   
Thus, we follow \citet{Crusius:1986}, who derived an exact expression for Eq.~\eqref{eq:method:synch_bessel} in terms of Whittaker's function and 
\citet{Zirakashvili:2006pv} and \citet{Aharonian:2010} who presented a  
simple analytical form that does not contain special functions. The 
synchrotron emissivity at a given frequency and electron \ac{LF} 
then can be evaluated as,   
\begin{equation}\label{eq:method:synch}
	\begin{aligned}
		j'_{\rm syn}(\nu',\gamma_{e}) & = 
		\frac{3 q_{e}^3 B'}{m_e c^2}
		\frac{1.808 y^{1/3}}{\sqrt{1+3.4 y^{2/3}}}  \\
		& \times \frac{ 1 + 2.21 x^{2/3} + 0.347 y^{4/3} }{ 1 + 1.353y^{2/3} + 0.217 y^{4/3} } e^{-y}\, ,
	\end{aligned}
\end{equation}
which was shown to agree with Eq.~\eqref{eq:method:synch_bessel} 
within $0.2\,\%$ accuracy.

The comoving synchrotron emissivity from a population of electrons at a 
given comoving time $t'$ is evaluated by convolving the emissivity of one 
electron, $j'_{\rm syn}(\nu',\gamma_{e})$, with the electron distribution, 
$N_{e} (\gamma_{e}, t')$, as  
\begin{equation}\label{eq:method:j_syn}
	j'_{\rm syn}(\nu') = \int_{\gamma_{e; 0}} ^{\gamma_{e; 1}} N_{e} (\gamma_{e})
	j'_{\rm syn}(\nu',\gamma_{e}) d\gamma_{e}\, ,
\end{equation} 
where $\gamma_{e; \rm 0}$ and $\gamma_{e; \rm 1}$ are the lower an upper 
limits of the electron \ac{LF} grid.

While we do not compute the \ac{SSA} heating that affects primarily the electron 
distribution, we do account for \ac{SSA} effects on the synchrotron radiation.
The classical treatment for the \ac{SSA} for the power-law electron distribution in
the context of \ac{GRB} afterglows was derived, \eg, by \citet{Sari:1997qe} 
and \citet{Granot:1998ek}. 
For an electron distribution that does not have a simple analytical form but 
under the one-zone approximation to the radiation isotropy assumption we 
can compute the \ac{SSA} coefficient at a given radiation frequency as 
follows \cite{RybickiLightman:1985}, 
\begin{equation}\label{eq:method:a_syn}
	\begin{aligned}
	\alpha'_{\rm syn}(\nu') & = -\frac{1}{8\pi m_e \nu^{'2}} \\
	& \times \int_{\gamma_{e; 0}} ^{\gamma_{e; 1}}
	j'_{\rm syn}(\nu',\gamma_{e}) 
	\gamma_{e}^2 
	\frac{\partial}{\partial\gamma_{e}}\Big[\frac{N_{e}(\gamma_{e})}{\gamma^2_{e}}\Big] d\gamma_{e}\, .
	\end{aligned}
\end{equation}

\subsubsection{Comoving \ac{SSC} spectrum}

To compute the \ac{SSC} radiation spectrum we follow 
\citet{Jones:1968,Blumenthal:1970,Miceli:2022efx}. The 
comoving emissivity at a given frequency and electron \ac{LF} is, 
\begin{equation}\label{eq:method:j_ssc_i}
	j'_{\rm ssc}(\nu',\gamma_{e}) = \frac{3}{4} h\sigma_T c\frac{\nu'}{\gamma_{e}^2}
	\int_{\nu'_{0}} ^{\nu'_{1}} \frac{n_{\tilde{\nu}'}}{\tilde{\nu}'} F(\tilde{\varepsilon}',\gamma_{e},\varepsilon') d\tilde{\nu}'\, ,
\end{equation}
where $n_{\tilde{\nu}'}$ is the seed photon spectrum. 
It is evaluated as,  
\begin{equation}\label{eq:method:n_seed}
	n_{\tilde{\nu}'} = n_{\tilde{\nu}';\,\rm syn} + n_{\tilde{\nu}';\,\rm ssc} 
	= \Bigg( \frac{j_{\rm syn}'(\tilde{\nu}')}{h \tilde{\nu}'} + \frac{j_{\rm ssc}'(\tilde{\nu}')}{h \tilde{\nu}'} \Bigg) \frac{\Delta t'}{V'} \, , 
\end{equation}
%
where $\Delta t' = \Delta R' /c = \Delta R \Gamma_{\rm sh} / c$ 
is the comoving time that photons stay within the shocked region 
\cite{Granot:1998ek,Huang:2022fnv}, and $\Delta R$ is the shock thickness 
in the burster frame obtained with Eq.~\eqref{eq:method:thickness}.  

As stated before, we limit the number of scatterings to $1$, \ie, we compute 
Eq.~\eqref{eq:method:j_ssc_i} and Eq.~\eqref{eq:method:n_seed} only once. 

The \ac{SSC} emissivity from a population of electrons is obtained as, 
\begin{equation}\label{eq:method:j_ssc}
	j'_{\rm ssc}(\nu') = \int_{\gamma_{e; 0}} ^{\gamma_{e; 1}} N_{e}(\gamma_{e})
	j'_{\rm ssc}(\nu',\gamma_{e})  d\gamma_{e}\, .
\end{equation}

\subsubsection{Pair-Production}

High-energy photons may interact with other photons inside the source, producing an electron-positron pair before they are able to 
escape. This \acf{PP} process can be characterized with a cross section 
\cite{Vernetto:2016alq,Murase:2010fq}, 
\begin{equation}\label{eq:method:pp_crosssec}
	\begin{aligned}
	\sigma_{\gamma\gamma}(\beta') & = \frac{3}{16}\sigma_{\rm T} (1-\beta_{\rm cm}^2) \\
	& \times \Bigg[(3-\beta_{\rm cm}'^4)\ln{\Bigg(\frac{1+\beta_{\rm cm}'}{1-\beta_{\rm cm}'}\Bigg)}-2\beta_{\rm cm}'(2-\beta_{\rm cm}'^2) \Bigg]\, ,
	\end{aligned}
\end{equation}
where $\beta_{\rm cm}$ is the center-of-mass speed of an electron produced 
in this process, that can be expressed as \cite{Coppi:1990}, 
\begin{equation}\label{eq:method:pp_beta}
	\beta_{\rm cm}' = \sqrt{1-\frac{2}{\tilde{\varepsilon}'\varepsilon'(1-\mu_{\gamma\gamma})}}\, ,
\end{equation}
where $\tilde{\varepsilon}=h\tilde{\nu}/(m_e c^2)$ is the normalized energy of the target 
photon, $\varepsilon=h\nu/(m_e c^2)$ is the normalized energy of the incoming energetic photon, 
and $\mu_{\gamma\gamma}=\cos( \theta_{\gamma\gamma} )$ is the angle between the directions of 
motion of two photons. \ie, the scattering angle. 

The annihilation rate then reads, 
\begin{equation}\label{eq:method:pp_rate}
	\mathcal{R}(\tilde{\varepsilon},\varepsilon)= \frac{c}{2}\int_{-1}^{\mu_{\gamma\gamma;\,\rm max}}{\sigma_{\gamma\gamma}(\tilde{\varepsilon},\varepsilon\,\mu_{\gamma\gamma}) (1-\mu_{\gamma\gamma}) d\mu_{\gamma\gamma}}\, .
\end{equation}

It was shown that Eq.~\eqref{eq:method:pp_rate} can be approximated with a simplified 
formula that is significantly cheaper to compute  \cite{Coppi:1990}, 
\begin{equation}\label{eq:method:pp_rate_approx}
	\mathcal{R}(x)\approx 0.652c\sigma_{\rm T}\frac{x^2-1}{x^3}\ln{(x)}\Theta(x-1),
\end{equation}
where $\Theta$ is the Heaviside function and 
$x=\tilde{\varepsilon}\varepsilon$. This approximation 
was shown to agree with the exact formula within $7\%$ accuracy \cite{Miceli:2022efx}. 

The effect of the \ac{PP} is twofold. A fraction of photons is lost, and a certain 
number of electrons is produced. Additional source of electrons enters 
Eq.~\eqref{eq:method:dNedt} as an extra source term on the right-hand-side, 
$Q_{\rm pp}(\gamma_{e},t')$, that following \citet{Miceli:2022efx} we can write as, 
\begin{equation}\label{eq:method:q_pp}
		Q_{\rm pp}(\gamma_{e},t')=4\frac{m_c c^2}{h} n_{\tilde{\nu}_{\gamma}'}
		\int_{\nu_0'}^{\nu_1'} d\tilde{\nu}' n_{\tilde{\nu}'} \mathcal{R} (\tilde{\nu}_{\gamma}')\, ,
\end{equation} 
where $\tilde{\nu}_{\gamma}'$ is the frequency that corresponds to the 
electron with \ac{LF} $\gamma_{e}$ and is computed as 
$\tilde{\nu}_{\gamma}' = 2\gamma_e m_e c^2 / h$. 

The \ac{PP} effect on the photon spectrum is difficult to estimate since we do 
not evolve the comoving photon spectrum in the same way we evolve electron 
distribution. Thus, we resort to a common approximation where an attenuation of 
the spectrum that leaves the emitting region can be computed using optical depth 
\cite{Huang:2020uti}, 
\begin{equation}\label{eq:method:tau_pp}
	\tau_{\gamma\gamma} = \frac{\Delta R'}{c}\int_{\nu'_1}^{\nu'_2} \mathcal{R}(\tilde{\nu},'\nu') n_{\tilde{\nu}'}d \tilde{\nu}'
	\,.
\end{equation}
Then, the attenuation can be included by multiplying the total emissivity 
by $(1-\exp(-\tau_{\gamma\gamma})) / \tau_{\gamma\gamma}$.

\subsection{Radiative \ac{BW} evolution}\label{sec:method:dErad}

Deriving the \ac{BW} evolution equations in Sec.~\ref{sec:method:dynamics}, we introduced 
$dE_{\rm rad}'$ term -- energy lost to radiation. Furthermore, in 
Sec.~\ref{sec:method:mphis_comovrad} we assumed that a fraction $\epsilon_{\rm e}$ 
of shock downstream energy is used to accelerating electrons. These electrons can 
radiate a fraction $\epsilon_{\rm rad}$ of their internal energy, affecting the 
underlying dynamics of the system 
\cite{Piran:1999kx,Huang:1999di,Chiang:1998wh,Dermer:1997pv,Dermer:2000gu}. 

Given an electron distribution $N_{e}$ and the corresponding comoving synchrotron 
spectrum $j'_{\rm syn}$, the total energy loss rate reads,   
$u'_{\rm syn;\,tot} = \int j'_{\rm syn} d\nu'$. 
If the size of the emitting region is $\Delta R'$, then the photons escape 
time is $dt_{\rm esc}'=\Delta R' / c$, and the total energy lost from the system is 
$dE_{\rm rad}' = dt_{\rm esc}' u'_{\rm syn;\, tot}$. 

Since we evolve the dowstream electron distribution numerically we can in principle 
omit the commonly used assumption that electrons emit all their energy instantaneously. 
However, numerically it is to compute $u'_{\rm syn;\, tot}$ at each sub-step 
of the adaptive step-size \acs{ODE} solver when computing the \ac{BW} dynamics. 

A more efficient formulation of radiative losses can be derived by assuming 
an analytic \ac{BPL} synchrotron spectrum that can be integrated analytically. 
Specifically, we employ classical fast and slow cooling spectra from 
\citet{Sari:1997qe} (see their Eqs.~(7) and (8)) and analytically integrate them 
from $\nu_0=10^6\,$Hz to $\nu_{\rm M}$. The result reads, 
\begin{equation}\label{eq:method:u_rad_an_0}
	\begin{aligned}
		u'_{\rm syn;\,tot} 
		&= \frac{3}{4} \left(\frac{1}{\nu_{\rm m}}\right)^{\frac{1}{3}} \left(\nu_{\rm m}^{\frac{4}{3}} - \nu_0^{\frac{4}{3}}\right) \\
		& -\frac{1}{3-p} \left(2 \left(\frac{1}{\nu_{\rm m}}\right)^{\frac{1}{2}-\frac{p}{2}}\right) \left(\nu_{\rm m}^{\frac{3}{2}-\frac{p}{2}} - \nu_{\rm c}^{\frac{3}{2}-\frac{p}{2}}\right) \\
		&+ \left(\frac{\nu_{\rm c}}{\nu_{\rm m}}\right)^{-\frac{p-1}{2}} \frac{2}{p-2} \left(\nu_{\rm c} \nu_{\rm M}^{\frac{p}{2}} - \nu_{\rm M} \nu_{\rm c}^{\frac{p}{2}}\right) \nu_{\rm M}^{-\frac{p}{2}}
	\end{aligned}
\end{equation}
if $\nu_{\rm m} < \nu_{\rm c}$ and
\begin{equation}\label{eq:method:u_rad_an_1}
	\begin{aligned}
		u'_{\rm syn;\,tot} 
		&= \frac{3}{4} \left(\frac{1}{\nu_{\rm c}}\right)^{\frac{1}{3}} \left(\nu_{\rm c}^{\frac{4}{3}} - \nu_0^{\frac{4}{3}}\right) \\
		&- \frac{2}{\sqrt{\nu_{\rm c}^{-1}}} \left(\sqrt{\nu_{\rm c}} - \sqrt{\nu_{\rm m}}\right) \\
		&+ \sqrt{\frac{\nu_c}{\nu_{\rm m}}} \frac{2}{p-2} \left(\nu_{\rm m} \nu_{\rm M}^{\frac{p}{2}} - \nu_{\rm M} \nu_{\rm m}^{\frac{p}{2}}\right) \nu_{\rm M}^{-\frac{p}{2}}\, ,
	\end{aligned}
\end{equation}
otherwise. 
In Eq.~\eqref{eq:method:u_rad_an_0} and Eq.~\eqref{eq:method:u_rad_an_1} 
all frequencies are in the comoving frame but the hyphen ' is omitted for clarity.

Characteristic frequencies, $\nu_{\rm m}'$, $\nu_{\rm c}'$, and $\nu_{\rm M}'$, are 
obtained from corresponding characteristic \acp{LF}, $\gamma_{e;\, i}$, 
as follows \cite{Johannesson:2006zs}, 
\begin{equation}\label{eq:method:nu_char}
	\nu'_{i} = \gamma_{e;\, i}\nu_{\rm crit}' \, 
	\begin{cases}
		0.06 + 0.28 p & \text{ if } \gamma_{e; \, \rm c} < \gamma_{e; \, \rm m} \\
		0.455 + 0.08 p & \text{ if } \gamma_{e; \, \rm m} < \gamma_{e;\, \rm c}\, ,
	\end{cases}
\end{equation}
where $\nu_{\rm crit}'$ is defined in Eq.~\eqref{eq:method:nu_crit}. 
We compare this analytical approximant to the full numerical calculation 
of $dE_{\rm rad}'$ in Sec.~\ref{sec:result}. 
 
More generally, the energy lost due to radiation can be expressed as 
$dE_{\rm rad}'=-\epsilon_{\rm rad}\epsilon_{\rm e}dE_{\rm sh}'$, where $dE_{\rm sh}'$ 
is the energy generated at the \ac{FS} or \ac{RS} 
(see Eq.~\eqref{eq:method:dEsh2} and Eq.~\eqref{eq:method:dEsh3}). 
If $\epsilon_{\rm rad} \ll 1$ \ac{BW} evolution is referred to 
as adiabatic, while if $\epsilon_{\rm e} \epsilon_{\rm rad} \simeq 1$ 
it is called radiative.

\subsection{Emitting Region in Observer Frame}\label{sec:method:obs_rad}

After the comoving emissivities and absorption coefficients are computed, 
we evaluate the observed radiation via \ac{EATS} integration 
\citep[\eg,][]{Granot:1998ek,Granot:2007gn,Gill:2018kcw,vanEerten:2009pa}. 

In order to properly account for the light abberation, we compute the radiation 
transport through the thin shell that represents the emitting region in the 
observer frame. 
The conversions of comoving emissivity and absorption coefficients into the observer 
frame are \cite{vanEerten:2009pa},  
$j_{\nu} = j_{\nu}' / ( \Gamma (1 - \beta\mu) )^2$,  
$\alpha_{\nu} = \alpha_{\nu}' ( \Gamma ( 1-\beta\mu ) )$, 
where $\mu = \cos(\theta_{ij,\rm LOS})$ is the angle between the \ac{BW} angle 
(defined in the next section) and the \ac{LOS}. 
The transformation for the frequency reads $\nu'=\nu (1+Z) \Gamma ( 1-\beta\mu )$, 
where $Z$ is the source redshift. 

For the uniform plane-parallel emitting region the radiation transport equation 
has an analytic solution, 
\begin{equation} \label{eq:method:intensity}
	I_{\nu} = \frac{j_{\nu}}{\alpha_{\nu}}( 1 - e^{-\tau_{\nu}}) \approxeq 
	j_{\nu} \times \begin{cases*}
		- (e^{-\tau_{\nu}} - 1)	/ \tau \text{ if } \tau_{\nu} > 0 \\	
		(e^{-\tau_{\nu}} - 1)	/ \tau_{\nu} \text{ otherwise } 
	\end{cases*}
\end{equation}
where $I_{\nu}$ is the intensity, the two cases correspond to the forward 
facing and back facing sides of a shock \cite{Ryan:2019fhz} and 
$\tau_{\nu} \approxeq -\alpha_{\nu} \Delta R / \mu'$ is the optical depth in the 
observer frame with  
\begin{equation}
	\mu' = \frac{\mu - \beta}{1 - \beta \mu}
\end{equation} 
being the parameter relating the angle of emission in local frame to that in the observer 
frame \cite{Granot:1998ek}. This parameter accounts for cases when light rays cross the 
ejecta shell along directions different from radial. The shock thickness in the observer 
frame is $\Delta R = \Delta R'/(1 - \mu \beta_{\rm sh})$.

The observed flux density from a shock is then estimated by integrating over \ac{EATS} 
using a spherical coordinate system that we discuss in Sec.~\ref{sec:method:numerics}, as 
\begin{equation}\label{eq:method:Fnu_1}
	F_{\nu} = \frac{1+Z}{2\pi d_L^2} 
	\int
	\int I_{\nu}(\theta,\phi) d\theta d\phi \, ,
\end{equation}
where $d_L$ is the luminosity distance to the source. 

High-energy photons propagating through the \ac{ISM} may be absorbed by \ac{EBL}. This 
effect can be accounted for by introducing additional optical depth following 
\citet{Franceschini:2017iwq}. The authors compute a table of photon-photon optical depths,  
$\tau_{\rm EBL}$ as a function of photon energy and redshift, $Z$. We use an updated table 
given in \cite{Franceschini:2017iwq} and attenuate the observed flux as 
$F_{\nu} = F_{\nu} \exp(-\tau_{\rm EBL})$.

\subsection{Numerical Implementation}\label{sec:method:numerics}

In this section, we detail numerical methods and techniques implemented in 
\PBA{} to solve the equations described in the previous subsections. 

\subsubsection{\ac{BW} dynamics}

After the initial conditions and settings are specified, the first stage of a 
simulation is the dynamical evolution of the \acp{BW}. Depending on the jet structure, 
the number of \acp{BW} can vary from $1$ for a top-hat jet (\ie, no lateral structure) 
to $N$ for a structured (\eg, Gaussian jet), specified by a user. 
For each \ac{BW}, a set of \acsp{ODE} is assembled. If the \ac{RS} option 
is selected, the following equations are used:
Eq.~\eqref{eq:method:dG_fsrs}, Eq.~\eqref{eq:method:dmdr_grb},
Eq.~\eqref{eq:method:dm3dr_1}, Eq.~\eqref{eq:method:ddelta4dr}, 
Eq.~\eqref{eq:method:dEdEdE2}, Eq.~\eqref{eq:method:dEdEdE3}, 
and
Eq.~\eqref{eq:method:dthetadr}. 
Otherwise, for a simulation with \ac{FS} only the following equations 
are used:
Eq.~\eqref{eq:method:dGdRold}, Eq.~\eqref{eq:method:dmdr_grb},
Eq.~\eqref{eq:method:dEdEdE2}, 
and
Eq.~\eqref{eq:method:dthetadr}. 

Sets of \acsp{ODE} are combined for all \acp{BW} and solved simultaneously 
for a specified grid of time steps. Notably, when the \ac{RS} is included 
\acs{ODE} system can become stiff at the onset of the \ac{RS} and the end of 
\ac{RS} crossing the ejecta. We do not employ a special solver for stiff 
\acsp{ODE}. Instead we implement an adaptive step-size explicit Runge-Kutta method 
of order $8(5,3)$  \cite{Prince:1981,Hairer:1993,Suresh:1997}\footnote{
	For the numerical implementation of coefficients we followed 
	\url{https://www.unige.ch/~hairer/software.html}
}. 
Generally, we limit the number of time steps to $1000$.  

Note, if radiation losses are included, microphsyics parameters, 
\ie, $\epsilon_{e;\,\rm fs}$, $\epsilon_{b;\,\rm fs}$, $p_{\rm fs}$
$\epsilon_{e;\,\rm rs}$, $\epsilon_{b;\,\rm rs}$, $p_{\rm rs}$
are used to compute $dE_{\rm rad; 2}'$ and $dE_{\rm rad; 3}'$.

\subsubsection{Comoving spectra}

After \acp{BW} are evolved and the downstream properties for all shocks at all time 
steps are obtained, electron distribution is evolved, as well as comoving radiation 
spectra. For this purpose, we determine the injection electron spectrum by 
solving the system of equations given by Eqs.~\eqref{eq:method:gm_1}-\eqref{eq:method:gm_3} using the bisect method. This allows us to compute the 
source term, Eq.~\eqref{eq:method:Qinj}, in the evolution equation.  
At the initial time step, we assume an analytic electron spectrum, given by Eq.~\eqref{eq:method:Ne_f} 
and Eq.~\eqref{eq:method:Ne_s}. The seed photon spectrum is initialized with zeroes and is 
populated during the subsequent evolution step. 

The electron distribution is evolved by solving the kinetic equation, 
Eq.~\eqref{eq:method:dNedt}. For this, we employ an unconditionally stable and particle number 
preserving fully implicit scheme \cite{Chang:1970,Chiaberge:1998cv}, which is a fully 
implicit difference scheme recently used in similar context \cite{Huang:2022fnv}. 
Electron \ac{LF} grid is initialized using equal logarithmic resolution from $1$ to $10^8$. 
The grid intervals are $\Delta \gamma_{e;j}=\gamma_{j+1/2}-\gamma_{j-1/2}$, where quantities 
with the subscript $j \pm 1/2$ are calculated at half-grid points. In order to discretize the 
continuity equation, we define the number of electrons per grid cell at a given time step as 
\begin{equation}\label{eq:method:CC_N_j}
	N_j^i=N(\gamma_{e,j},i\times\Delta t)\, ,
\end{equation}
where $\Delta t$ is the time step that we will be discuss later.  
The particle flux between the grid cells is, 
\begin{equation}\label{eq:method:CC_F_j}
	F_{j\pm 1/2}^{i+1} = \dot{\gamma}_{j\pm 1/2}^i N_{j\pm 1/2}^{i+1}\, . 
\end{equation}
The discretized form of Eq.~\eqref{eq:method:dNedt} is given by 
\begin{equation}\label{eq:method_CCdiscretised}
        \frac{N_j^{i+1}-N_j^i}{\Delta t} = \frac{\dot{\gamma}_{e,j+1/2}^iN^{i+1}_{j+1/2}-\dot{\gamma}_{e,j-1/2}^iN^{i+1}_{j-1/2}}{\Delta\gamma_e} + Q_j^i \, ,
\end{equation}
where $\dot{\gamma}_e$ is the total cooling rate specified in Eq.~\eqref{eq:method:gamma_dot}. 
The discretized from of continuity equation, Eq.~\eqref{eq:method_CCdiscretised}, 
then becomes \citep{Chang:1970}, 
\begin{equation}\label{eq:method_CCdiscretised2}
        V3_jN_{j+1}^{i+1}+V2_jN_j^{i+1}+V1_jN_{j-1}^{i+1}=S_j^i\, ,
\end{equation}
where the $V$ coefficients are: 
\begin{equation}\label{eq:method:CC_Vcoeff}
    \begin{aligned}
        &V1_j = 0\, , \\
        &V2_j = 1+\frac{\Delta t \dot{\gamma}_{e,j-1/2}}{\Delta\gamma_{e,j}} \, ,\\
        &V3_j = -\frac{\Delta t \dot{\gamma}_{e,j+1/2}}{\Delta\gamma_{e,j}}\, ,
    \end{aligned}
\end{equation}
and the source term is, 
\begin{equation}\label{eq:method:source_CCscheme}
    \begin{aligned}
        S_j^i = N_j^i + Q_j^i\Delta t \, .
    \end{aligned}
\end{equation}
Equation~\eqref{eq:method_CCdiscretised2} forms a tridiagonal matrix equation that is solved 
via standard backwards substitution method.

Notably, the time step for the evolution has to allow electrons to cool, \ie, 
it has to respect synchrotron cooling timescale, 
\begin{equation}
	\Delta t'_{\rm syn} = \frac{\sigma_T \gamma_{e; \, \rm M} B^{'2}}{6\pi m_e c}\, ,
\end{equation}
and adiabatic cooling timescale (due to the expansion of the emitting region),  
\begin{equation}
	\Delta t' _{\rm adi} = \frac{\gamma_{e; \, \rm M}^2 - 1}{3\gamma_{e; \, \rm M}}\frac{1}{V'}dV' ,
\end{equation}
where $V'$ is the comoving volume. 
The maximum allowed time step, $\Delta t'$, then is determined by the Courant-Friedrichs-Lewy (CFL)
condition as 
$\Delta t' = \Delta \ln( \gamma_{e} )/ (\Delta t'_{\rm syn} + \Delta t' _{\rm adi})$. 
Notably, this $\Delta t'$ is not equal to the time step used for the \ac{BW} evolution. 
If it is smaller, we perform a sub-stepping procedure between the main, \ac{BW} 
evolution time steps. During sub-steps we fix the shock downstream properties and do not 
update radiation field for the sake of computational efficiency. 

After the next evolution step is reached, we compute the comoving synchrotron emissivity 
(Eq.~\eqref{eq:method:j_syn}) and the \ac{SSA} coefficient (Eq.~\eqref{eq:method:a_syn} 
with the derivative obtained via $1$st order finite-differencing) as well as the 
\ac{SSC} emissivity (Eq.~\eqref{eq:method:j_ssc}) using the photon field, $n_{\tilde{\nu}}'$, 
from the previous time step. Then, we update the photon field (Eq.~\eqref{eq:method:n_seed}) 
that is used later in computing \ac{SSC} cooling (Eq.~\eqref{eq:method:gamma_dot_ssc}) 
as well as the pair-production source term (Eq.~\eqref{eq:method:q_pp}) during the next step. 
Notably, the innermost integral in Eq.~\eqref{eq:method:gamma_dot_ssc} depends only 
on values that are known at the beginning of a simulation, -- parameters of comoving 
electron \ac{LF} and radiation frequency grids: $\nu'$, $\tilde{\nu}'$ and $\gamma_{e}$. 
Thus, we compute it once at the beginning of the simulation.
The updated photon field is also used to compute the absorption due to the \ac{PP} 
(Eq.~\eqref{eq:method:tau_pp}). 

Thus, for each \ac{BW} evolution time step we build comoving emissivity and absorption 
spectra.

\subsubsection{Jet Discretization and Observed Spectrum}\label{sec:method:discret}

Emission that an observer sees from a system of \acp{BW}, which represents a \ac{GRB} 
jet, depends on the geometry of the system. We employ a spherical coordinate system 
$(r,\,\theta,\,\phi)$ where $r$ is the distance from the coordinate 
origin, and $\theta$ and $\phi$ are the latitudinal and azimuthal 
angles respectively. The central engine (post-merger remnant) is located at the 
coordinate origin, and the system's symmetry axis ($z$-axis) lies along $\theta=0$. 
The observer is located in the $\phi=\pi/2$ plane and  
$\theta_{\rm obs}$ is the angle between the \ac{LOS} and the $z$-axis.
Thus, the unit vector of the observer is given by 
$\vec{n}_{\rm obs} = \big( 0,\, \sin(\theta_{\rm obs})\vec{y},\, \cos(\theta_{\rm obs}\big)\vec{z} )$.

To construct a structured jet model we follow the approach suggested by 
\citet{Ryan:2019fhz}, namely, we divide the into a set of independent \acp{BW}, each of 
which has a progressively larger initial half-opening angle.  
At the same time each hemisphere is discretized uniformly in terms of $\theta$  into rings assigned to each of the \acp{BW}, 
so that hemisphere is split into $k=\{0,1,2,...n-1\}$ rings centered on the 
symmetry axis with boundaries $\theta_{i;\,\rm l}$ and $\theta_{i;\,\rm h}$, and with the ring center located at 
$\theta_{i;\,\rm  c}=(\theta_{i;\,\rm  h}-\theta_{i;\,\rm  l})/2$. 

Observed radiation is obtained by summing contributions from each \ac{BW}, using 
Eq.~\eqref{eq:method:Fnu_1}, that in turn is computed by integrating intensity 
over a given ring segment as, 
\begin{equation}\label{eq:method:Fnu}
    F_{\nu} = \frac{1+Z}{2\pi d_L^2} \sum_{i}^{\text{layers}} 
    \int_{\theta_{i;\,\rm l}}^{\theta_{i;\,\rm h}} 
    \int_{\phi_0=0}^{\phi_1=\pi/2} I_{i,\nu}(\theta,\phi) d\theta d\phi \, .
\end{equation}

Afterwards, we account for the \ac{EBL} absorption as discussed before.

\subsubsection{Sky map calculation}\label{sec:method:skymap}

A sky map is an intensity distribution projected onto a plane orthogonal to the 
\acf{LOS}. To compute it, we further discretize each ring (associated with the \ac{BW}) 
into $S\in{1,2,3...}$ $\theta$-subrings using an additional $\theta$-grid uniform 
in $\cos{\theta}$. 
Then, we split each subring along the $\phi$-axis into a set of $\Omega$-cells in 
such a way that the resulted $\Omega$-cells have the same solid angle. For example, the 
$i$-th subring is comprised of $2i + 1$ $\Omega$-cells. Then, each ring has  
$\sum_{i=0}^{i=S-1}(2i+1))$ cells in total. 

For each cell that lies within the visible part of the ring at a given observer 
time and frequency, we compute the intensity  
$I_{i;\nu}(\theta,\phi)$. Numerically, we also make sure that at least $3$ 
subrings fall between $\theta_{i;\, \rm l}$ and $\theta_{i;\, \rm h}$ and that 
each of those subrings has at least $3$ non-zero $\Omega$-cells. This is accomplished 
via an iterative algorithm that re-discretizes each ring into progressively more 
sub-rings and cells until the required conditions are met. 

After the discretization, the coordinate vector of the $i$-cell 
in both is given by
$\vec{v}_{i} = r_{i}\big( \sin{(\theta_{i})}\cos{(\phi_{i})}\vec{x},\, \sin{(\theta_{i})}\cos{(\phi_{i})}\vec{y},\, \cos{(\theta_{i})}\vec{z} \big)$,
The cosine of the angle between the \ac{LOS} and $\vec{v}_{i}$ reads,  
\begin{equation}\label{eq:method:mu_obs}
    \mu_{i} = 
    \sin{(\theta_{i})}\sin{(\phi_{i})}\sin(\theta_{\rm obs}) +  
    \cos{(\theta_{i})}\cos(\theta_{\rm obs}) \, .
\end{equation}

A sky map is computed by projecting the specific intensity on the $x$-$z$ plane that 
is perpendicular to the \ac{LOS}. We chose the basis with which the principal jet 
moves in the positive $\tilde{x}$-direction. The basis vectors of the plane then read, 
\begin{equation}
    \begin{aligned}
        \tilde{\vec{x}}_{i} &= \sin(\theta_{\rm obs})\vec{z}_{i}-\cos(\theta_{\rm obs})\vec{x}_{i} \, , \\
        \tilde{\vec{z}}_{i} &= \vec{x}_{i} \, ,
    \end{aligned}
\end{equation}
and the coordinates of the $ij$ cell on the image plane (for the principle jet) 
are given by 
\begin{equation}\label{eq:method:xccoord}
    \begin{aligned}
        \tilde{x}_{i} & = -r_{i}[\cos(\theta_{\rm obs})\sin(\theta_i)\sin(\phi_{i})+ \sin(\theta_{\rm obs})\cos(\theta_{i}))], \\
        \tilde{z}_{i} & = r_{i}\sin(\theta_{i})\cos(\phi_{i})\, . 
    \end{aligned}
\end{equation}
In the following, we omit the use of tildes for simplicity.

In order to characterize sky maps we consider the surface brightness-weighted 
center of a sky map, also called image or a flux centroid, defined as 
\begin{equation}\label{eq:method:xc}
    x_{\rm c} = \frac{1}{\int I_{\nu} dx dz}\int x I_{\nu} dx dz,
\end{equation}
and the $x$ and $z$-averaged brightness distributions, 
\begin{equation}\label{eq:method:int_dist}
    \begin{aligned}
        I_{\nu; \rm m}(x) &= \frac{1}{\Delta z} \int I_{\nu}(x,z) dz\, , \\
        I_{\nu; \rm m}(z) &= \frac{1}{\Delta x} \int I_{\nu}(x,z) dx\, .
    \end{aligned}
\end{equation}

\subsubsection{Sky Map calculation}\label{sec:method:numerics:skymap}

From the \pba{} simulations we obtain the intensity distribution 
on the projection plane separately from each \ac{BW}. These  
large unstructured arrays require post-processing to generate physically interpretable 
sky maps. We implement the following pipeline in a 
separate python code that is included in a python package that is released 
alongside the main code.  
First, ``raw'' sky maps are interpolated using the Delaunay triangulation 
for each \ac{BW} onto a grid unifrom in $x$ and $z$. This is done separately 
for principle and counter jets for numerical reasons. Then, interpolated sky 
maps are used to integrate them along a given axes to compute the $x$- and 
$z$-averaged brightness distributions, that in turn allow us to compute the 
sky map size as a \ac{FWHM} of the distributions.  
In order to obtain a sky maps for plotting, we consider a $2$D histogram with 
edges given by the same uniform grid in $x$ and $z$. Then we bin ``raw'' sky 
maps with this histogram to obtain final intensity distribution $I(x,z)$. 
This procedure mimics how an observing instrument would collect photons from an 
extended source and is also used in other \ac{GRB} afterglow models 
(\eg, \citet{Fernandez:2021xce}).

\section{Simulation Examples} \label{sec:result}

In this section we provide a comprehensive overview of several \ac{GRB} afterglow 
simulations performed with \PBA. We focus on two main geometries of the jet: 
top-hat, where the jet is comprised of a singular \ac{BW}, and commonly considered 
Gaussian jet, where distributions of \acp{BW}' initial energy and \ac{LF} follow 
a Gaussian. For both structures we perform two simulations: with \ac{FS} only and 
with \ac{FS}-\ac{RS} system. We label these simulations as ``top-hat-\ac{FS}'' and 
``top-hat-\ac{FS}-\ac{RS}'' for the top-hat structure and similarly for the Gaussian 
one.

\subsection{Top-hat-\ac{FS} simulation}\label{sec:3.1}

In this subsection we discuss the simulation with top-hat jet structure 
with initial isotropic equivalent energy of the burst being, 
$E_{\rm iso;\,\rm c}=10^{53}\,$ergs, initial jet \ac{LF} given as, 
$\Gamma_{0;\, \rm c}=400$, and the initial jet half-opening angle being, 
$\theta_{\rm w}=\theta_{\rm c}=0.1\,$rad, where the half-angle of the jet wings 
$\theta_{\rm w}$ is the same as of the jet core, $\theta_{\rm c}$. 
The jet is expanding into the constant density \ac{ISM} with $n_{\rm ISM} = 1\,\ccm$. 
Since the simulation is performed including \ac{FS}, we implicitly assume that 
at the beginning of the simulation the \ac{RS} has already crossed the ejecta. 
In this case, the region behind contact continuity is fully shocked and 
moves with \ac{LF} $\Gamma$ from the beginning. 
Microphysics parameters for \ac{FS} are set as  
$\epsilon_{e;\,\rm fs}=0.1$, $\epsilon_{b;\, \rm fs}=0.001$, $p_{\rm fs}=2.2$, 
Unless stated otherwise, these parameters remain fixed for the discussion. 

From given jet properties, the initial conditions for the \ac{BW} evolution 
are, 
\begin{equation}\label{eq:method:tophat_id}
	\begin{aligned}
		E_{0} &= E_{\rm iso;\,\rm c} \sin^2(\omega_0 / 2)\, , \\
		\Gamma_{0} &= \Gamma_{0;\, \rm c}\, , \\
		M_{0} &= \frac{E_{\rm iso;\,\rm c}}{(\Gamma_0 - 1)c^2} \sin^2(\omega_0 / 2)\, ,
	\end{aligned}
\end{equation}
where $\Gamma_{0}$, $\omega_0$, $E_{0}$ and $M_{0}$ are initial 
\ac{LF}, half-opening angle, energy and mass, respectively.

\subsubsection{\ac{BW} dynamics and energy conservation}\label{sec:res:energy_tophat_fs}

\begin{figure}
	\centering 
	\includegraphics[width=0.49\textwidth]{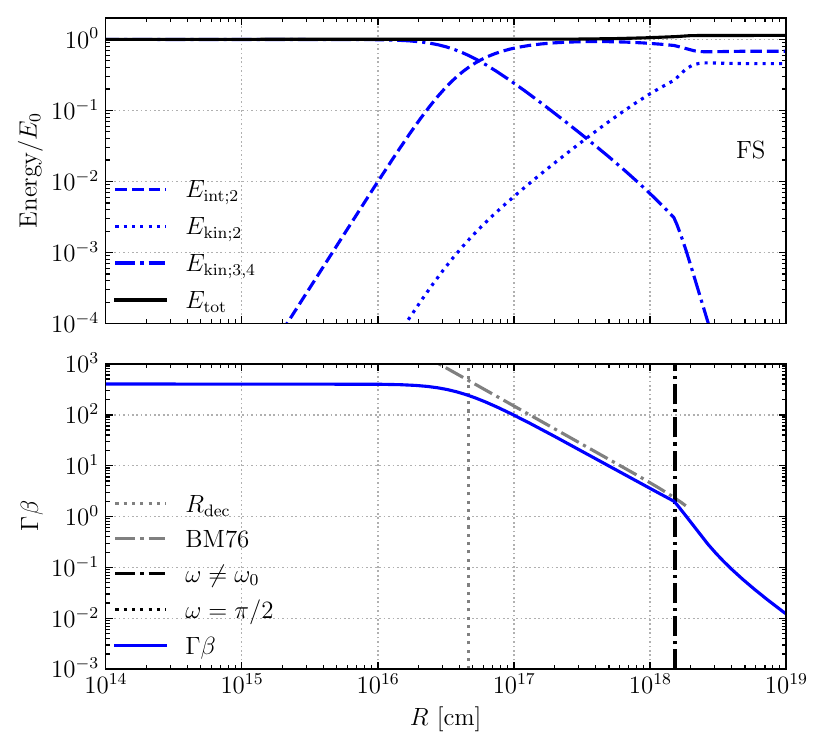}
	\caption{
		Energy (\textit{top panel}) and momentum (\textit{bottom panel})
		for the \ac{BW} from top-hat-\ac{FS} simulation
		as a function of the \ac{BW} radius $R$, 
		computed as $R=\int t_{\rm burst} \beta c$, where $\beta$ is 
		the \ac{BW} dimensionless velocity.
		Here $R_{\rm dec}$ is the deceleration radius,  
		$\omega\neq\omega_0$ indicates the onset of the lateral spreading
		$\omega=\pi/2$ marks the end of spreading; 
		BM76 indicates the analytic \ac{BM} solution.  
		}
	\label{fig:method:energy_fs}
\end{figure}

In Fig.~\ref{fig:method:energy_fs} we show the example of the \ac{BW} evolution.  
The evolution begins with the free-coasting phase, where \ac{BW} momentum, 
$\Gamma\beta=$const. 
It continues until the amount of matter swept-up from the \ac{ISM} becomes 
comparable to the \ac{BW} mass at 
$R_{\rm d} = (3 E_{\rm iso, c}/(4\pi c^2 m_p n_{\rm ISM}\Gamma^2))^{1/3}$.
After that the kinetic energy of the shocked ejecta, 
$E_{\rm kin;3,4}=(\Gamma - 1) M_{0} c^2$ starts to decrease as the energy is 
converted into internal energy of the shock downstream, 
$E_{\rm int;2} = \Gamma_{\rm eff;2} E_{\int;2}'$. Kinetic energy in this stage, $E_{\rm kin;2}=(\Gamma - 1) m c^2$, also increases following the growing 
amount of mass swept up from the \ac{ISM}. 

During the deceleration phase, the \ac{BW} \ac{LF} asymptotically follows the 
\ac{BM} solution, given as 
$\Gamma_{\rm BM} = \sqrt{17 E_{\rm iso;c} / (16\pi c^2 R ^3)}$. 
We note that generally, a model based on a homogeneous shell does not fully 
agree with the \ac{BW} solution for adiabatic evolution which is obtained by 
integrating the energy density of the shocked fluid over the extended 
(not a thin) shell. A possible solution is to add a normalization factor 
\citepalias{Nava:2013}. Here we omit it for simplicity. 

Summing all the energy components we obtain total energy of the \ac{BW},   
$E_{\rm tot} = E_{\rm kin;4} + E_{\rm kin;2} + E_{\rm int;2}$. 
Its evolution is shown with black line in Fig.~\ref{fig:method:energy_fs}. 
For most of the evolution the energy is conserved within ${\sim}1\,\%$. 
However, when \ac{BW} decelerates to $\Gamma\beta\sim1$, the 
energy conservation violation reaches ${\sim10}\,\%$, due to the limitations of our simple \ac{EOS} (Eq.~\eqref{eq:method:eos}), and the treatment of adiabatic losses. 
Notably, we find that the energy conservation at late times is improved if more accurate 
\ac{EOS} for transrealtivistc fluid is used, \eg, one 
presented in \citet{Peer:2012}.

Once the conditions for the lateral spreading are satisfied 
(see Sec.~\ref{sec:method:lat_spreading}) and $d\omega/dR$ becomes non-zero, 
the \ac{BW} evolution starts to deviate from \ac{BM} solution as 
$dm/dR \propto \omega$ in a non-linear way 
(see Eq.~\eqref{eq:method:dthetadr} and Eq.~\eqref{eq:method:dmdr_grb}). 
This behavior is expected and agrees with simplified numerical hydrodynamic 
solution \cite{Granot:2012}. Furthermore, we compare our dynamics during lateral 
spreading with a more sophisticated, $2$D thin-shell \ac{HD} code in 
Sec.~\ref{sec:comparison}. After the \ac{BW} half-opening angle reaches $\pi/2$, 
and when $\Gamma \sim 1$ the \ac{BW} evolution enters the \ac{ST} regime and 
the following evolution proceeds with $\Gamma \propto R^{-3/2}$.

\begin{figure}
	\centering 
	\includegraphics[width=0.49\textwidth]{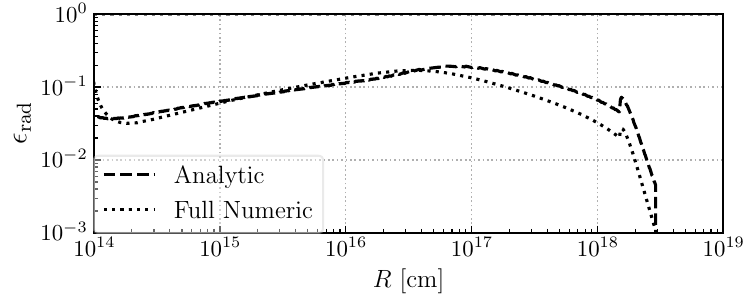}
	\includegraphics[width=0.49\textwidth]{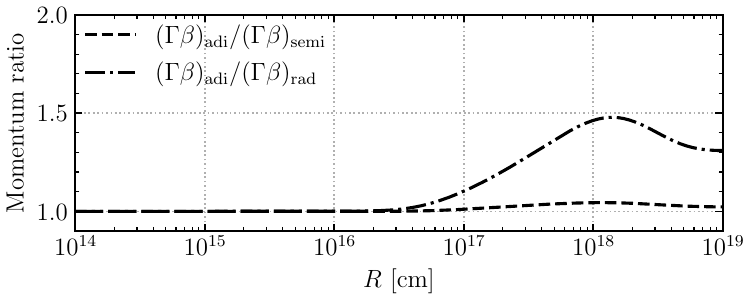}
	\caption{
		Effect of the radiative losses on the \ac{BW} evolution in the 
		top-hat-\ac{FS} simulation. 
		\textit{Top panel} shows the evolution of the fraction of the 
		shocked energy ($E'_{\rm rad; 2} / E_{\rm sh; 2}'$) lost to synchrotron 
		radiation, \ie, $\epsilon_{\rm rad}$. The dotted line corresponds to 
		this quantity computed using the synchrotron spectrum 
		from self-consistently evolved electron distribution. 
		The dashed line corresponds to $\epsilon_{\rm rad}$ computed using an 
		analytic integral over \ac{BPL} approximation of synchrotron emission
		(Eq.~\eqref{eq:method:u_rad_an_0} and Eq.~\eqref{eq:method:u_rad_an_1}). 
		\textit{Bottom panel} shows the effect of radiative losses on the 
		\ac{BW} momenta, $\Gamma\beta$, evolution. For the sake of clarity, 
		ratio of momenta is plotted. Dashed line corresponds to the case where 
		adiabatic evolution is compared with semi-radiative evolution.  
		Dash-dotted line corresponds to comparison of adiabatic evolution 
		with fully radiative evolution, $\epsilon_{\rm rad}=1$. 
	}
	\label{fig:method:radiative_bw}
\end{figure}

The effect of radiative losses term, $dE'_{\rm rad; 2}$, in 
Eq.~\eqref{eq:method:dEdEdE2} on the \ac{BW} evolution is shown in 
Fig.~\ref{fig:method:radiative_bw}. When a fraction of the \ac{BW} internal energy 
is lost to radiation, the \ac{BW} decelerates faster. The strongest effect thus 
is achieved when all energy generated at the shock is lost to radiation, \ie, 
$\epsilon_{\rm rad}=1$, the so-called, fully radiative evolution. In the figure 
it is shown with dot-dashed line on the bottom subplot of the figure. 

In semi-radiative regime, where $\epsilon_{\rm rad}$ is obtained by integrating the 
synchrotron spectrum (Eq.~\eqref{eq:method:u_rad_an_0} and 
Eq.~\eqref{eq:method:u_rad_an_1}) the effect is the strongest when 
$\epsilon_{\rm e}\rightarrow 1$ and $\epsilon_{\rm b}\rightarrow 1$. 
For the values chosen for this simulation, however, $\epsilon_{\rm rad}\simeq 0.1$, 
which implies that 
$E_{\rm rad}'/E_{\rm sh}' = \epsilon_{\rm e}\epsilon_{\rm rad} \simeq 0.01$, 
and \ac{BW} evolves almost adiabatically as the bottom subplot of the figure shows. 

In the top subplot of the figure we compare the analytic approximation to radiative 
losses (Eq.~\eqref{eq:method:u_rad_an_0} and Eq.~\eqref{eq:method:u_rad_an_1}) 
to the full numerical integration of the synchrotron spectrum 
(using Eq.~\eqref{eq:method:j_syn}). While there is a small difference at late times, 
we do not find this difference to be noticeable in the \ac{BW} evolution. 
Thus, we set the analytic method as a default option for the sake of computational 
speed.

\begin{figure}
	\centering 
	\includegraphics[width=0.49\textwidth]{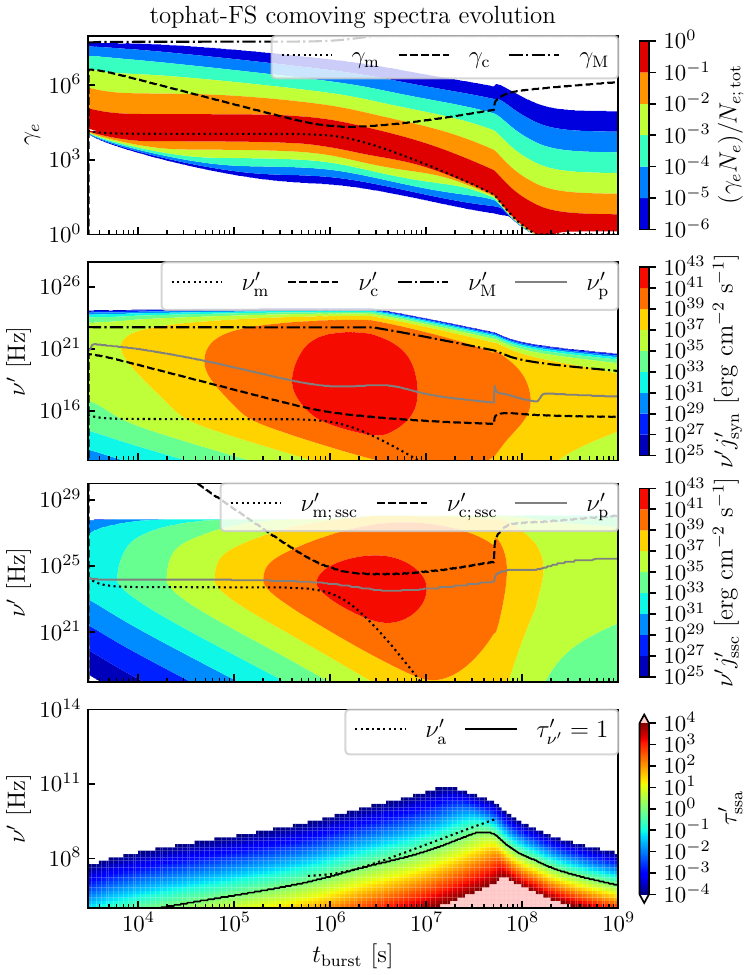}
	\caption{
		Comoving spectra evolution for the top-hat-\ac{FS} simulation 
		(without \ac{PP} effects). 
		\textit{The top panel} corresponds to the electron spectrum. 
		There, black dotted, dashed and dash-dotted lines indicate critical \acp{LF},  
		$\gamma_{e;\, \rm m}$ (Eq.~\eqref{eq:method:gm_1}-\eqref{eq:method:gm_3}), 
		$\gamma_{e; \, \rm c}$ (Eq.~\eqref{eq:method:gc}) 
		and $\gamma_{e;\,\rm M}$ (Eq.~\eqref{eq:method:gM})
		respectively. 
		\textit{The second panel} corresponds to the synchrotron spectrum 
		(Eq.~\eqref{eq:method:j_syn}) with black dotted, dashed and dash-dotted 
		lines indicating critical \acp{LF}, $\nu_{\rm m}$, $\nu_{\rm c}$, $\nu_{\rm M}$
		respectively, computed using Eq.~\eqref{eq:method:nu_crit} and 
		Eq.~\eqref{eq:method:nu_char}. The solid gray line traces the 
		frequency of the spectrum maximum. 
		\textit{Third panel} corresponds to the \ac{SSC} spectrum 
		(Eq.~\eqref{eq:method:j_ssc}), with black and gray lines computed as above but 
		for the \ac{SSC} process. 
		\textit{Fourth panel} corresponds to the \ac{SSA} spectrum 
		(Eq.~\eqref{eq:method:a_syn}) shown in terms of the optical depth, $\tau_{\nu'}'$, 
		computed using comoving thickness of the shocked region. 
		The black line marking there marks the location of $\tau_{\nu'}'=1$, where the 
		spectrum transition from optically thin to optically thick. 
		Dotted line corresponds to the analytic estimate, Eq.~\eqref{eq:method:nu_a}.  
	}
	\label{fig:method:comov_spectra_fs}
\end{figure}

\begin{figure}
	\centering 
	\includegraphics[width=0.49\textwidth]{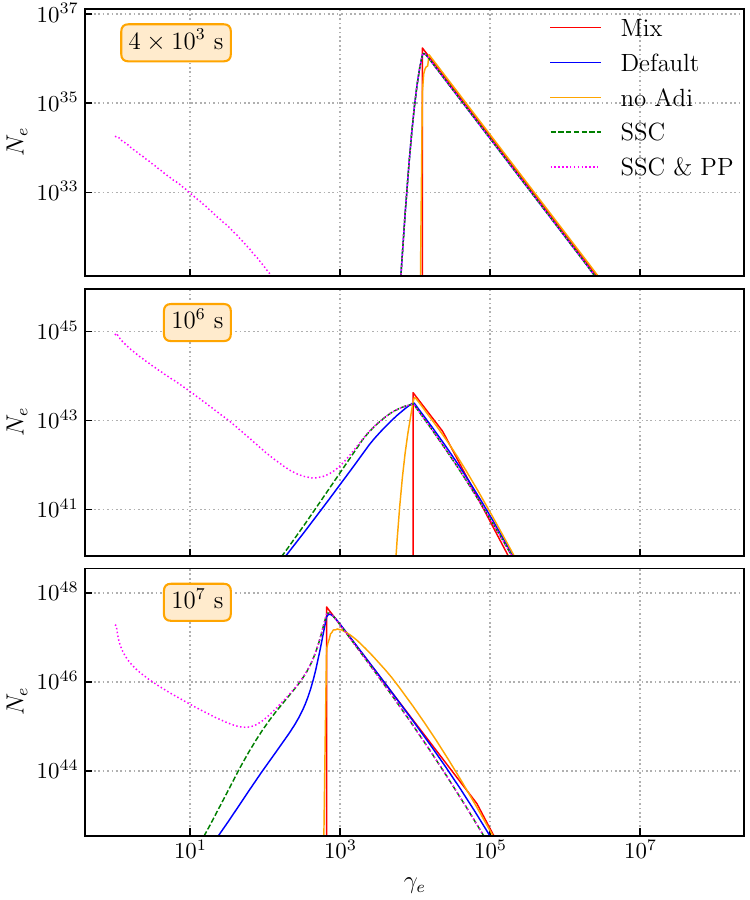}
	\caption{
		Electron distribution at three different time steps (one per panel) for three 
		runs with different physics setups all related to the top-hat-\ac{FS} 
		simulations. ``\ac{SSC} \& \ac{PP}'' run includes both \ac{SSC} and \ac{PP} processes; 
		the ``\ac{SSC}'' run accounts only for \ac{SSC} process; the ``Default'' run 
		does not include any of the above; and the ``Mix'' run corresponds to the simulation 
		where electron distribution is not evolved (analytic \ac{BPL} is assumed at each time step). 
	}
	\label{fig:method:n_ele_num_mix_fs}
\end{figure}

\begin{figure}
	\centering 
	\includegraphics[width=0.49\textwidth]{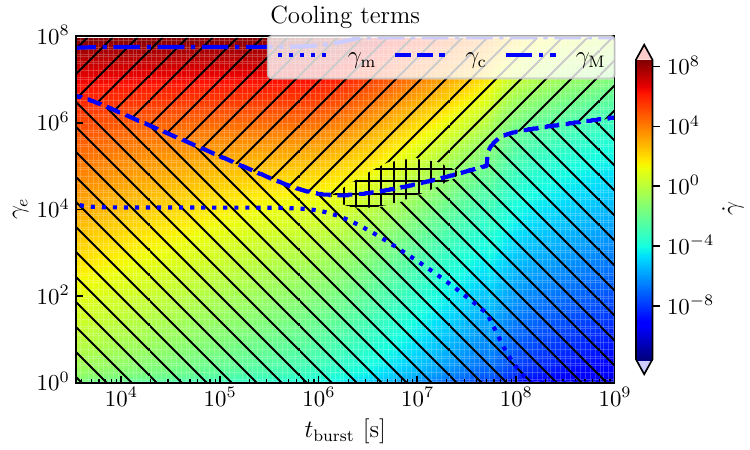}
	\caption{
		The total electron cooling rate (Eq.~\eqref{eq:method:gamma_dot}) for the 
		top-hat-\ac{FS} simulation. Hatching indicates which cooling
		term dominates, where $//$ hatches are used when
		$\dot{\gamma}_{\rm syn} > \max( \dot{\gamma}_{\rm adi},\,\dot{\gamma}_{\rm ssc} )$, 
		$\backslash\backslash$ hatches are used when
		$\dot{\gamma}_{\rm adi} > \max( \dot{\gamma}_{\rm syn},\,\dot{\gamma}_{\rm ssc} )$, 
		and $+$ hatches are used when 
		$\dot{\gamma}_{\rm ssc} > \max( \dot{\gamma}_{\rm syn},\,\dot{\gamma}_{\rm adi} )$.
		Also, characteristic \acp{LF} computed with analytic expressions are shown with 
		blue lines same as in the top panel of Fig.~\ref{fig:method:comov_spectra_fs}. 
	}
	\label{fig:method:gamm_dot_fs}
\end{figure}

\subsubsection{\ac{FS} comoving spectra evolution}\label{sec:method:spec_fs}

In Fig.~\ref{fig:method:comov_spectra_fs}, the evolution of comoving spectra 
is shown. 
The electron spectrum evolution is shown in the top panel of the figure 
alongside the characteristic \acp{LF}. The spectrum in normalized such that at 
every time step the electron distribution $N_e$ is divided by the total number 
of electrons obtained as $N_{e\,\rm tot} = \int N_e d\gamma_{e}$ and multiplied 
by electron \ac{LF}, $\gamma_{e}$. This allows for an overall clarity but 
sacrifices some detail, \eg, the cooling \ac{LF} $\gamma_{e; \, \rm c}$ does not 
visibly follow contours of the plot. The same procedure is employed in, \eg, 
\citet[][]{Bosnjak:2008bd}. 

The subplot shows that at the beginning of the simulation the electron spectrum 
is confined between $\gamma_{e; \, \rm m}$ and $\gamma_{e; \, \rm M}$ following 
the analytical initial conditions for slow cooling regime 
(Eq.~\eqref{eq:method:Ne_s}) since $\gamma_{e; \, \rm c} > \gamma_{e; \, \rm m}$. 
During the evolution, a population of cooled electrons with 
$\gamma_{e} < \gamma_{e; \, \rm m}$ grows. This is one of the key differences between 
an analytical and numerical electron spectra, shown also in  
Fig.~\eqref{fig:method:n_ele_num_mix_fs} at three time steps. Thus, at any given time, 
the electron distribution is comprised of the freshely injected electrons with 
$\gamma_e>\min(\gamma_{e; \, \rm m},\, \gamma_{e; \, \rm c})$ and old, cooled 
electrons that occupy $\gamma_e < \min(\gamma_{e; \, \rm m},\, \gamma_{e; \, \rm c})$. 

To illustrate the effect of adiabatic cooling, \ie, the $\dot{\gamma}_{\rm adi}$ 
term in Eq.~\eqref{eq:method:gamma_dot}, on the electron distribution evolution 
we perform the same simulation setting $\dot{\gamma}_{\rm adi}=0$ and label it 
``no Adi''. The comparison with a simulations that includes adiabatic cooling  
(labeled as ``Default'') is shown in Fig.~\ref{fig:method:n_ele_num_mix_fs}. 
The figure highlights the importance 
of this term in developing low-$\gamma_{e}$ part of the spectrum  
(see, \eg, \citet{Geng:2017aku} for similar analysis). 
Figure also shows the effects of \ac{SSC} term (Eq.~\eqref{eq:method:gamma_dot_ssc}) 
and the \ac{PP} source term (Eq.~\eqref{eq:method:q_pp}) on the electron spectrum. 
Both contribute to the formation of more extended low-energy tail of the spectrum.

A more detailed analysis of various contributors to the overall electron 
cooling throughout the \ac{BW} evolution is shown in Fig.~\ref{fig:method:gamm_dot_fs}. 
The figure is divided into three areas marked by distinct hatching styles depending 
on which $\dot{\gamma_{e}}$ term is the largest. 
Additionally, analytically computed characteristic \acp{LF} are also shown for 
comparison. As expected synchrotron cooling dominates the overall cooling rate 
for $\gamma_{e} > \gamma_{e;\, \rm c}$, at almost all times except for when 
$\gamma_{e; \, \rm c}$ approaches $\gamma_{e;\, \rm m}$ and \ac{SSC} scattering 
becomes a dominate cooling process for electrons with 
$\gamma_{e} \simeq \gamma_{e;\, \rm c}$, which is computed 
using Eq.~\eqref{eq:method:gc} without the $\tilde{Y}$ term.
Electrons with $\gamma_{e}<\gamma_{e; \, \rm c}$ cool primarily via adiabatic losses. 
Overall, the cooling rate, $\dot{\gamma}$, decreases with time and with electron 
\ac{LF} as the color-coding in this figure indicates. Thus, at late times, when the injection 
\ac{LF} is $\gamma_{e; \, \rm m} \sim 1$, the electron spectrum becomes quasi-stationary. 
This can also be seen in Fig.~\ref{fig:method:comov_spectra_fs}, at the end of the 
\ac{BW} evolution. Notably, since we evolve electron distribution in terms 
of electron \ac{LF}, our implementation becomes increasing less accurate as 
$\gamma_{e}\rightarrow 1$.

\begin{figure}
	\centering 
	\includegraphics[width=0.49\textwidth]{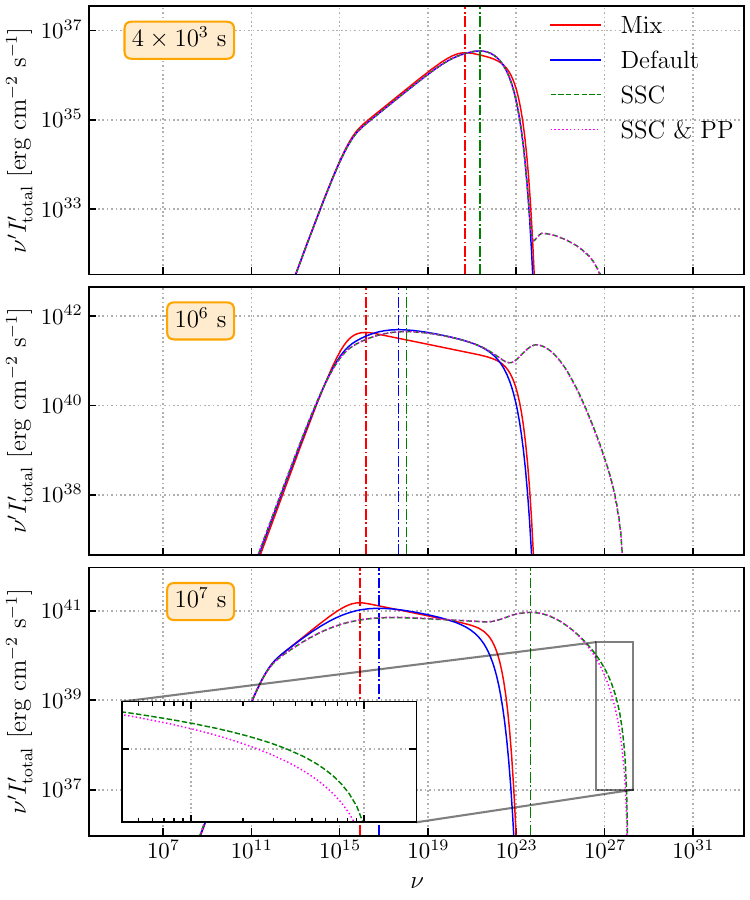}
	\caption{
		Total comoving intensity at three different time steps 
		(one per panel) for three runs with different physics setup of the top-hat-\ac{FS} 
		simulations. The ``\ac{SSC} \& \ac{PP}'' run includes \ac{SSC} and \ac{PP}; 
		the ``\ac{SSC}'' run includes only \ac{SSC}, and the ``Default'' run does not 
		include any of these two processes, and in the ``Mix'' run 
		electron distribution is not evolved (analytic \ac{BPL} is assumed at each time step). 
		The enlarged subplot on the bottom panel shows the effect of the \ac{PP} 
		attenuation term (Eq.~\eqref{eq:method:tau_pp}) on the intensity. 
		Here \ac{EBL} absorption effect is included. 
	}
	\label{fig:method:j_syn_methods}
\end{figure}

The comoving synchrotron emissivity spectrum is shown in the second panel of 
Fig.~\ref{fig:method:comov_spectra_fs}. At the beginning of the simulation, 
the peak of the spectrum, $\nu_{\rm p}'$, coincides with the $\nu_{\rm m}'$.  
During the evolution, however, $\nu_{\rm p}'$ shifts to higher frequencies as a 
fraction of electrons with 
$\gamma_{e}\lesssim\min(\gamma_{e; \, \rm m},\gamma_{e; \, \rm c})$ grows. This can 
also be seen in Fig.~\ref{fig:method:j_syn_methods}, where comoving intensity spectrum,  
$\nu' I'_{\rm totoal}$, is shown for three time steps for several physics setups. 
In particular, the figure shows that the synchrotron spectrum computed using an analytic 
electron spectrum (Eq.~\eqref{eq:method:Ne_f} and Eq.~\eqref{eq:method:Ne_s})  
has lower $\nu_p'$ than the spectra computed with numerically evolved electron 
spectrum. Both the smoothness of the latter spectra and additional electron population 
with $\gamma_{e}$ close to the lower boundary of the injection electron spectrum 
contribute to this outcome. 

As the \ac{BW} freely coasts, comoving magnetic field strength and injection spectrum do not 
evolve, but the number of particles radiating at every time step increases and  
so does the maximum synchrotron emissivity. It reaches its peak at the 
onset of \ac{BW} deceleration. Throughout the entire \ac{BW} evolution, the spectrum 
remains in the slow-cooling regime. 

The last panel in Fig.~\ref{fig:method:comov_spectra_fs} shows the \ac{SSA} spectrum 
evolution in terms of the comoving optical depth, 
$\tau'_{\rm ssa}(\nu') = \alpha_{\rm syn}'(\nu') \Delta R'$. As the amount of 
matter swept-up by the \ac{FS} increases so does the \ac{SSA} optical depth. 
The frequency at which $\tau'_{\rm ssa}(\nu') = 1$, \ie,  $\nu'_{a}$, can be used to 
separate the optically thin ($< 1$) from optically thick ($>1$) parts of the 
synchrotron spectrum. It is commonly referred to as \ac{SSA} frequency,  
\eg, \citet{Granot:1998ek}. The \ac{SSA} frequency can be derived analytically, 
by additionally assuming that the \ac{BW} follows the \ac{BM} solution 
\cite{Warren:2018lyx}, 
\begin{equation}\label{eq:method:nu_a}
	\nu'_{a} = 3.41\times10^9 \Big(\frac{p+2}{3p+2}\Big)^{3/5} \frac{(p-1)^{8/5}}{p-2}\, .
\end{equation}
Equation~\eqref{eq:method:nu_a} shows that $\nu'_{\rm a}$ increases continuously 
with time during this stage of evolution. 
In the last panel of Fig.~\ref{fig:method:comov_spectra_fs} we  compare Eq.~\eqref{eq:method:nu_a} with full numerical calculation of the $\nu'_{a}$. Overall, the difference does not exceed a factor of $2$ until \ac{BW} starts to spread laterally and deviate from \ac{BM} solution.

The third panel in Fig.~\ref{fig:method:comov_spectra_fs} shows the evolution of 
\ac{SSC} emissivity spectrum, $j_{\rm ssc}$. 
The \ac{SSC} emissivity corresponding to a given electron \ac{LF} depends on the seed 
photon distribution ( Eq.~\eqref{eq:method:j_ssc_i}). 
Thus, the integrated \ac{SSC} spectrum (Eq.~\eqref{eq:method:j_ssc}),   
has a complex dependency on \ac{BW} properties, comoving electron, and radiation spectra. 

Characteristic frequencies of the spectrum are computed as
\begin{equation}
	\nu_{i;\, \rm ssc}' = \frac{4\sqrt{2}}{3} \nu'_{i;\,\rm syn} \gamma_{e; \, i}^2\, ,
\end{equation}
where $\nu'_{i;\,\rm syn}\in\{\nu'_{\rm m}, \nu'_{\rm c}\}$ and 
$\gamma_{e; \, i}\in\{\gamma_{\rm m},\gamma_{\rm c}\}$, respectively.

It can be shown that the number of electrons that contribute to the \ac{SSC} process 
around the peak is proportional to the number of electrons in the system, 
$j_{\rm ssc;\, max} \propto N_e$. This is reflected in the 
figure, where the maximum of the \ac{SSC} qualitatively follows the evolution 
of the maximum of the synchrotron spectrum. 

Notably, the Thomson optical depth for electron scattering, \ie, 
the optical depth of electrons seen by a photon, defined as \cite{Zhang:2018book}, 
\begin{equation}
	\tau_{\rm es} = \sigma_T \frac{m}{m_p V'} \Delta R' 
\end{equation}
remains always $\ll 1$.
Thus, when computing the \ac{SSC} process, in is indeed sufficient consider only one 
scattering process.

The effect of the \ac{SSC} cooling term, Eq.~\eqref{eq:method:gamma_dot_ssc},  
on the synchrotron spectrum is shown in Fig.~\ref{fig:method:j_syn_methods} 
as a slight shift of the $\nu_p'$ to higher frequencies when \ac{SSC} is included 
in the simulation. 

The effect of the \ac{PP} attenuation (Eq.~\eqref{eq:method:tau_pp}) is shown 
in the bottom panel of Fig.~\ref{fig:method:j_syn_methods}, in the enlarged subplot. 
As expected, \ac{SSC} emissivity is reduced at highest frequencies. However, the effect 
is weak for a \ac{GRB} jet settings chosen for the analysis. 

\begin{figure}
	\centering 
	\includegraphics[width=0.49\textwidth]{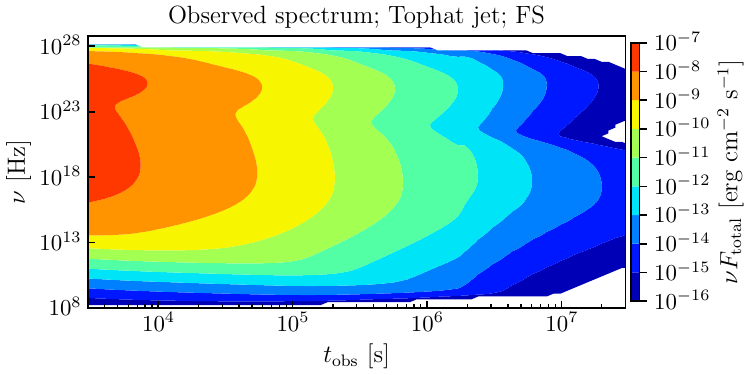}
	\caption{
		Time evolution of the observer spectrum from the top-hat-\ac{FS}-\ac{RS} 
		simulation with \ac{SSC}, \ac{PP} \ac{SSA} and \ac{EBL} absorption 
		turned on. 
	}
	\label{fig:method:tophat_spec_fs}
\end{figure}

\subsubsection{Observed spectrum}

In Fig.~\ref{fig:method:tophat_spec_fs} the observed spectrum evolution computed
via \ac{EATS} integration from the simulation is shown. The spectrum has a distinct 
double-peak shape at all times with the high (low) energy peak corresponding to 
\ac{SSC} (synchrotron) emission. At a given time, the \ac{SSC} emission decreases 
sharply towards the highest frequencies primarily due to the \ac{EBL} absorption.

\subsection{Top-hat-\ac{FS}-\ac{RS} simulation}\label{sec:res:tophat_fsrs}

In this section, we review the top-hat-\ac{FS}-\ac{RS} simulation. This simulation 
is performed using the extended set of \acp{ODE} for \ac{BW} dynamics based on  
Eq.~\eqref{eq:method:dG_fsrs} (see Sec.~\ref{sec:method:numerics}). 
Microphysics parameters for the \ac{RS} are set as follows, 
$\epsilon_{e;\,\rm rs}=0.1$, $\epsilon_{b;\, \rm rs}=0.001$, $p_{\rm rs}=2.2$ (\ie, the same as those for the FS: see \S\ref{sec:3.1}), 
while the burst duration that defines initial ejecta shell width is 
$t_{\rm prmpt}=10^3\,$s.
 
As before, we first discuss the \ac{BW} dynamics and then the electron 
and radiation spectra. However, as we are going to show, the presence of 
the \ac{RS} alters the dynamics of the \ac{FS}. However, for the sake of 
brevity, we focus only on the \ac{RS} itself in this section.

\subsubsection{\ac{BW} dynamics and energy conservation}

\begin{figure}
	\centering 
	\includegraphics[width=0.49\textwidth]{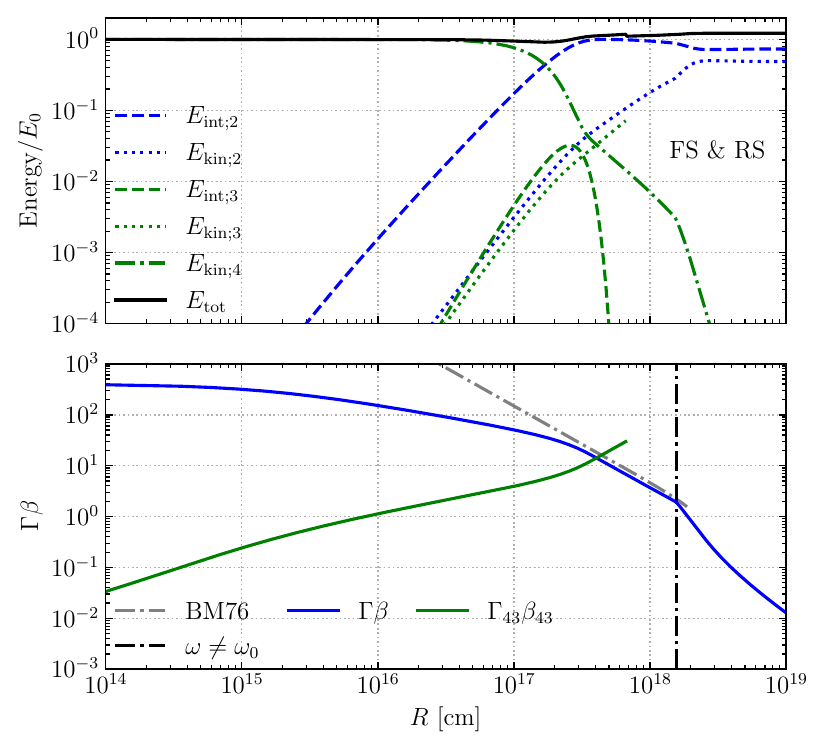}
	\caption{
		Energy (\textit{top panel}) and momentum (\textit{bottom panel})
		for top-hat-\ac{FS}-\ac{RS} simulation. 
		As in Fig.~\ref{fig:method:energy_fs}, 
		$\omega\neq\omega_0$ indicates the onset of the lateral spreading
		$\omega=\pi/2$ marks the end of spreading; 
		\ac{BM}76 is the analytic \ac{BM} solution. 
		When the \ac{RS} crosses the ejecta, its evolution terminates. This is indicated in the evolution of $E_{\rm kin;\,3}$ on the top panel and $\Gamma_{34}\beta_{34}$ in the bottom one. 
	}
	\label{fig:method:energy}
\end{figure}

\begin{figure}
	\centering 
	\includegraphics[width=0.49\textwidth]{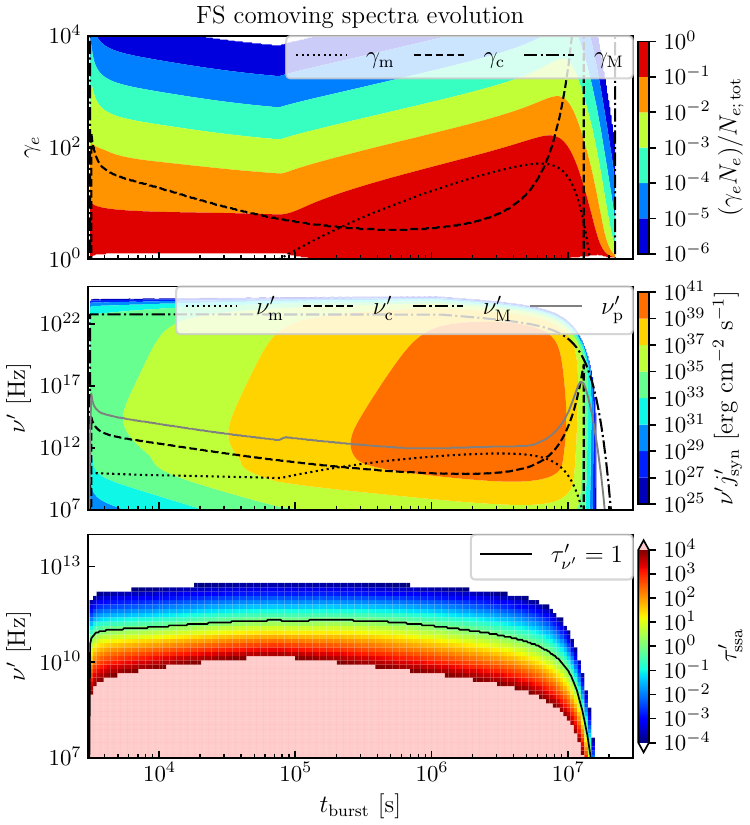}
	\caption{
		Same as Fig.~\ref{fig:method:comov_spectra_fs} but for the \ac{RS}. 
		Top, middle and bottom panels correspond to the electron, 
		synchrotron and \ac{SSA} spectra respectively. 
	}
	\label{fig:method:comov_spectra_rs}
\end{figure}

Fig.~\ref{fig:method:energy} shows the evolution of different energy components 
of the \ac{BW} (upper panel) as well as its bulk \ac{LF} (lower panel). 
The \ac{RS} shock starts with non-relativistic $\Gamma_{43} \sim 1$ and accelerates as 
it moves through the ejecta. Then its kinetic energy, defined as 
$E_{\rm kin;3} = (\Gamma - 1) m_{3} c^2$, increases. As the mass of the shocked 
material grows, the internal energy, $E_{\rm int;3} = \Gamma_{\rm eff;3} E_{\rm int;3}'$ 
also rises while the bulk \ac{LF} of the \ac{BW} steadily falls.
Shortly before the \ac{RS} crosses the ejecta, it becomes relativistic with 
$\Gamma_{43}\gg1$ and total energy  
$E_{\rm kin;3}+E_{\rm int;3}$ reaches ${\sim}10\%$ of the total 
\ac{BW} energy.

As in the case with the simulation where only \ac{FS} is included, we find that 
the total energy is conserved within ${\sim}1\,\%$ during the early evolution stages. 
However, when the \ac{RS} becomes relativistic, the total energy conservation violation reaches ${\sim}20\,\%$. 
This is due to our simplified treatment of ejecta profile in the upstream and \ac{EOS}. 

While in the top-hat-\ac{FS} simulation a large part of the \ac{BW} evolution -- after 
the onset of deceleration and before lateral spreading -- can be described with the 
\ac{BM} solution, in the top-hat-\ac{FS}-\ac{RS} simulation, only after the \ac{RS} 
crossing time does the \ac{BW} follow the self-similar solution.

\subsubsection{Comoving spectra evolution}

\begin{figure}
	\centering 
	\includegraphics[width=0.49\textwidth]{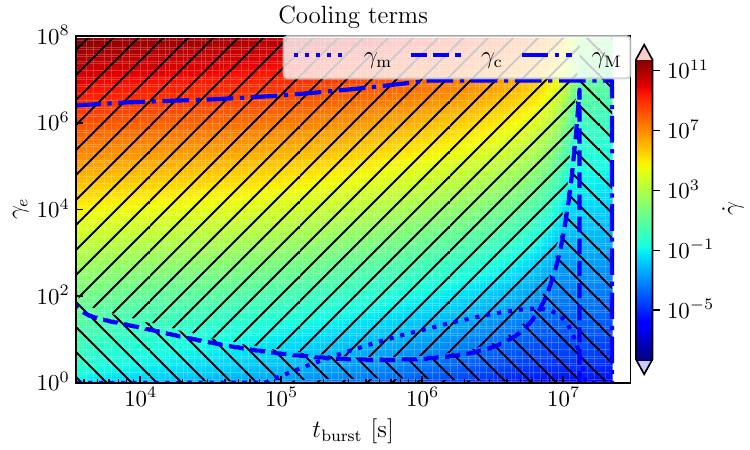}
	\caption{
		Same as Fig.~\ref{fig:method:gamm_dot_fs} but for the \ac{RS}.
	}
	\label{fig:method:ele_cool_rs}
\end{figure}

Comoving spectra from the \ac{RS} are shown in Fig.~\ref{fig:method:comov_spectra_rs}. 
Conditions at the \ac{RS} differ significantly from those at  \ac{FS}.   
Most importantly, while \ac{FS} starts highly relativistic and remains so (freely coasts) until 
the deceleration at late times, eventually reaching $\Gamma_{\rm fsh}\simeq1$ 
at very late times, the \ac{RS} starts with $\Gamma_{\rm rsh}\simeq1$ and then 
accelerates gaining maximum momentum shortly before it traverses the entire ejecta shell. 
Thus, at the beginning, the electron distribution in the \ac{RS} downstream is confined to 
$\gamma_{e}\lesssim 10$ as shown in the top panel 
of Fig.~\ref{fig:method:comov_spectra_rs}. 
Notably, only a fraction of particles crossing the shock is being injected, \ie, $\xi_{\rm DN} < 1$ (see Eq.~\eqref{eq:method:gm_lim}). 

When $\gamma_{e; \, \rm m}$ increases above $1$ the fraction $\xi_{\rm DN}$ becomes $1$
and the electron spectrum displays a sharp change as shown in the figure. 
This feature is largely a numerical artifact as electrons cannot have $\gamma_{e} < 1$ 
and thus only when $\gamma_{e; \, \rm m} > 1$ can electrons cool to 
$\gamma_{e} < \gamma_{e; \, \rm m}$. 
As \ac{RS} accelerates, $\gamma_{e; \, \rm m}$ grows rapidly while $\gamma_{e; \, \rm c}$, 
computed using Eq.~\eqref{eq:method:gc}, decreases and spectrum enters the fast cooling 
regime. The spectral regime changes again only at the end, when the upstream density 
for the \ac{RS} starts to decline exponentially (Eq.~\eqref{eq:method:rho4}). 
At this point internal energy density behind the \ac{RS} starts to decrease rapidly 
and so does $\gamma_{e; \, \rm m}$. 
As Fig.~\ref{fig:method:ele_cool_rs} shows, throughout most of the evolution, 
synchrotron cooling remains the dominant cooling process.

The synchrotron emissivity spectrum is shown on the second panel of 
Fig.~\ref{fig:method:comov_spectra_rs}. The spectral peak does not evolve 
as significantly. It remains between optical and radio bands till the shock crosses the ejecta.

As the upstream density for the \ac{RS} is significantly larger than the \ac{ISM} 
density and the shock itself is slower than the \ac{FS}. Thus, the \ac{SSA} 
plays a more significant role here. Specifically, radio emission at $\nu' < 10^{11}\,$Hz 
is self-absorbed. This is reflected in the bottom panel of 
Fig.~\ref{fig:method:comov_spectra_rs}. The self-absorbed radio spectrum from the \ac{RS} 
in \acp{GRB} is indeed expected theoretically \cite{Gao:2015lga} and there is observational 
support for it as well \cite{Laskar:2019xfo}. 

\begin{figure}
	\centering 
	\includegraphics[width=0.49\textwidth]{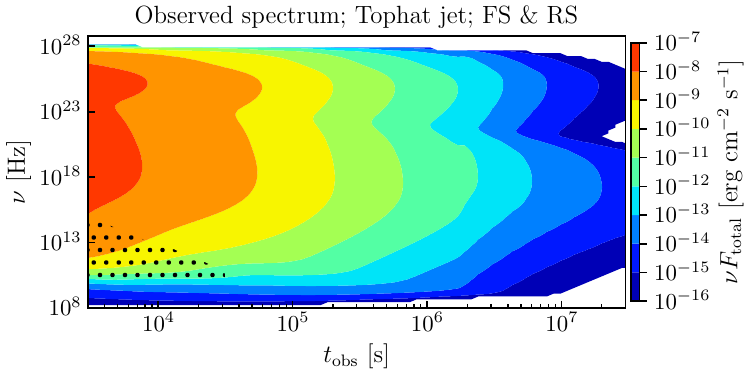}
	\caption{
		Time evolution of the observer spectrum from the top-hat-\ac{FS}-\ac{RS} 
		simulation with \ac{SSC}, \ac{PP} and \ac{EBL} absorption 
		turned on for the \ac{FS} emission and \ac{SSA} for both shocks. 
		The area where \ac{RS} emission is brighter than the \ac{FS} one 
		is indicated with $\cdots$ hatches.  
	}
	\label{fig:method:tophat_spec_fsrs}
\end{figure}

\subsubsection{Observed spectrum}

In Fig.~\ref{fig:method:tophat_spec_fsrs} the observed spectrum evolution, computed
via \ac{EATS} integration from the simulation is shown. Comparing this figure 
with Fig.~\ref{fig:method:tophat_spec_fs} we note that the \ac{RS} emission contributes 
primarily at the lower end of the spectrum. At the lowest frequencies, the \ac{SSA} process 
becomes important and the spectrum shows a steep decline.

\subsubsection{Sky map}

\begin{figure*}
	\centering 
	\includegraphics[width=0.99\textwidth]{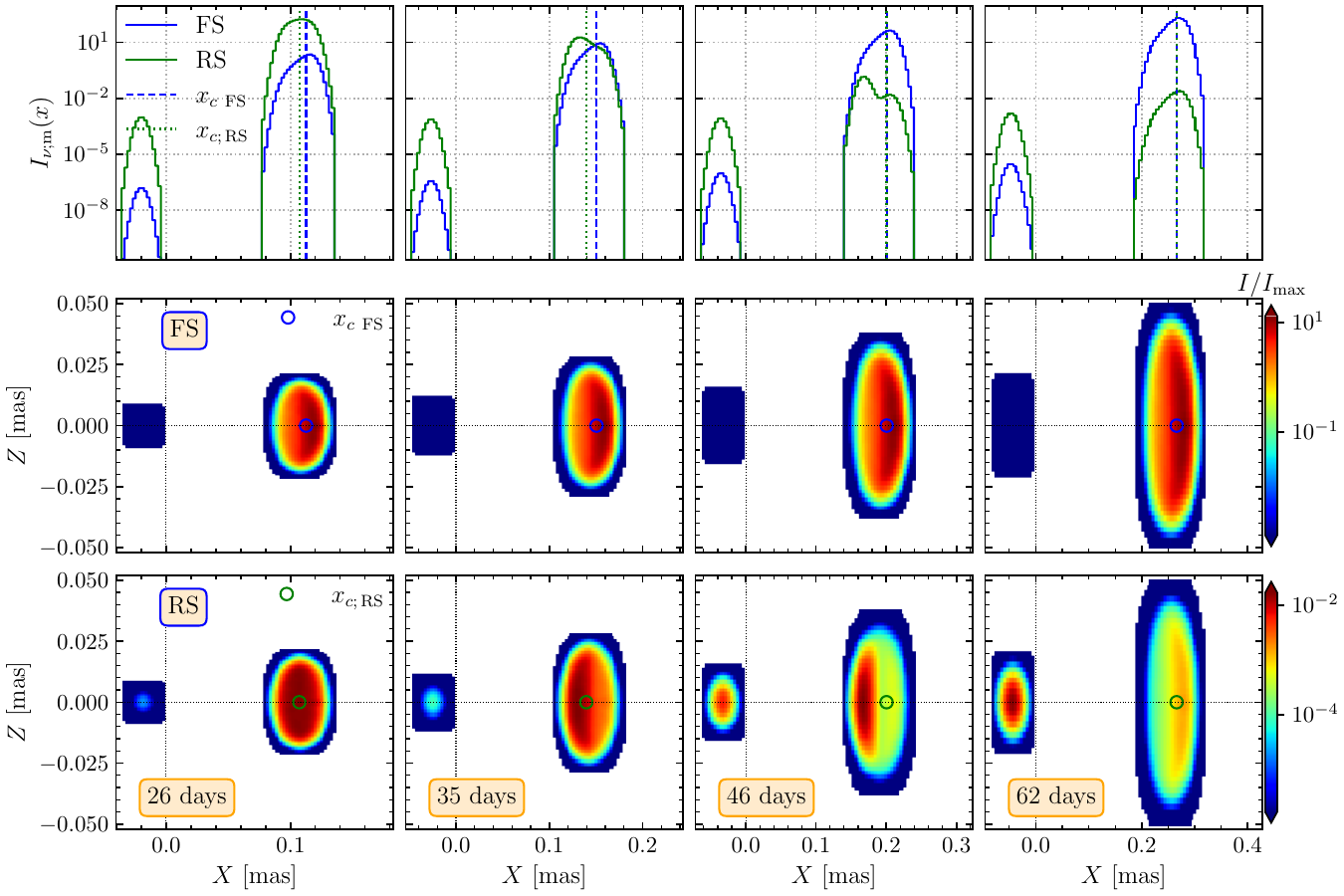}
	\caption{
		Radio sky maps of the top-hat-\ac{FS}-\ac{RS} simulation observed at 
		$\theta_{\rm obs}=\pi/4\,$rad. at different times (panel columns) as indicated 
		by the text at the bottom left of the last panel in a column.
		The top row of panels shows intensity integrated along the $z$
		axis, $I_{\nu;\,\rm m}(x)$. There the green (blue) line corresponds 
		to the sky map of associated with the \ac{RS} (\ac{FS}) emission only. 
		Second and third rows of panels show sky maps compute from \ac{FS} and 
		\ac{RS} emission respectively. 
		Moreover, each sky map is composed of the image associated with the 
		principle and counter jets that lie on the right and left sides with 
		respect to the origin, $X=0$, respectively. 
		Circular markers indicate the location of the 
		flux centroid, $x_c$. Location of the flux centroid for both sky maps is also 
		shown in the top row of panels with dashed and dotted vertical lines. 		 
	}
	\label{fig:method:top_hat_skymaps}
\end{figure*}

Sky maps for the top-hat-\ac{FS}-\ac{RS} simulation are shown in 
Fig.~\ref{fig:method:top_hat_skymaps} for several observer times and 
separately for the \ac{FS} and for the \ac{RS} (second and third panels,  respectively).
There we set $\nu_{\rm obs}=1\,$GHz and set $\theta_{\rm obs}=\pi/4\,$rad. 
The figure shows that as the \ac{RS} is significantly slower than the \ac{FS} its 
emission is less subjected to the Doppler beaming and thus the counter jet overtakes 
the principle jet in brightness much earlier in this case. 
The total sky map, however, is dominated by the emission from the \ac{RS} at very 
early times, when the \ac{FS} emission is beamed away form the \ac{LOS}. As the Doppler 
beaming decreases, the contributions from two shocks to the principle jet sky map 
becomes comparable. This transition is shown in the second column of panels in the figure.  
This is also the point where the position of the total sky map flux centroid is 
most affected by the emission from the \ac{RS}.

\subsection{Gaussian-\ac{FS}-\ac{RS} simulation}\label{sec:result:gauss}

\begin{figure}
	\centering 
	\includegraphics[width=0.49\textwidth]{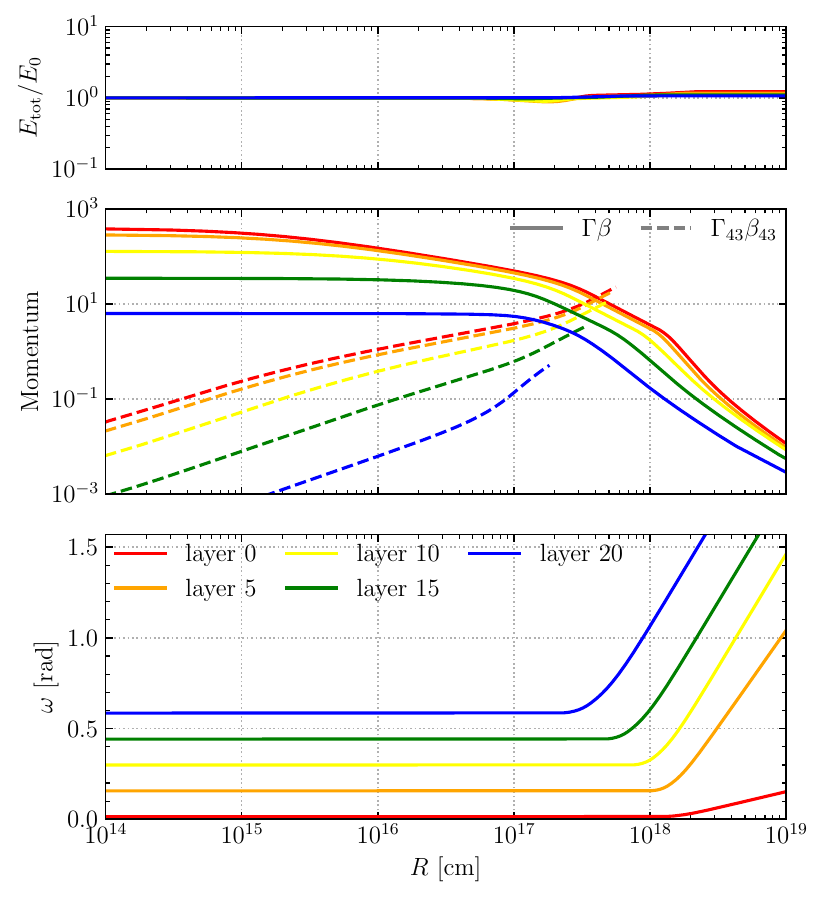}
	\caption{
		Dynamics of several \acp{BW} (layers) of a structured, Gaussian 
		jet, where layer=$0$ \ac{BW} corresponds to the core of the jet 
		(and thus it has the smallest initial half-opening angle and the 
		largest initial momentum). 
		\textit{Top panel} shows the total energy of each \ac{BW} divided by its 
		initial energy. If energy is conserved during the evolution 
		$E_{\rm tot}/E_0 = 1$. 
		\textit{Second panel} shows the evolution of the \ac{BW} momentum $\Gamma\beta$ 
		alongside the evolution of the momentum corresponding to the \ac{RS} 
		($\Gamma_{43}\beta_{43}$). 
		\textit{Third panel} shows the evolution of the \ac{BW} half-opening angle, 
		$\omega$, offset by half-opening angle assigned to the \ac{BW} within 
		a structured jet, $\omega_{0;\,l}$ (see Sec.~\ref{sec:method:discret}).  
	}
	\label{fig:method:energy_gauss}
\end{figure}

\begin{figure}
	\centering 
	\includegraphics[width=0.49\textwidth]{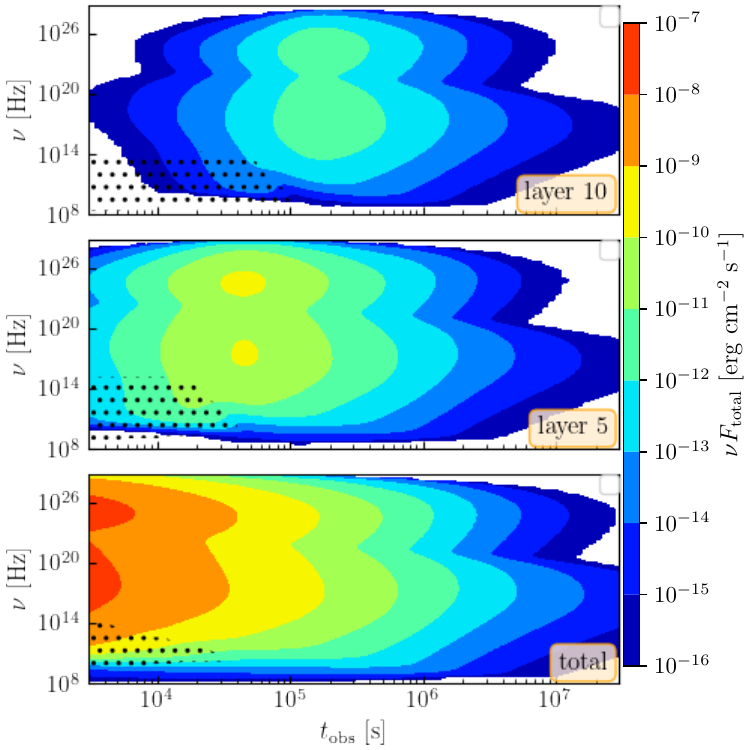}
	\caption{
		Observed spectrum evolution for Gaussian-\ac{FS}-\ac{RS} simulation. 
		First two panels show spectra from individual \ac{BW} within the jet, 
		while the last panel shows the total, summed spectrum. 
		Dotted hatched regions in each panel indicate where flux density 
		from \ac{RS} exceeds that from the \ac{FS} as in Fig.~\ref{fig:method:tophat_spec_fsrs}.  
	}
	\label{fig:method:obs_spec_gauss}
\end{figure}

\begin{figure}
	\centering 
	\includegraphics[width=0.49\textwidth]{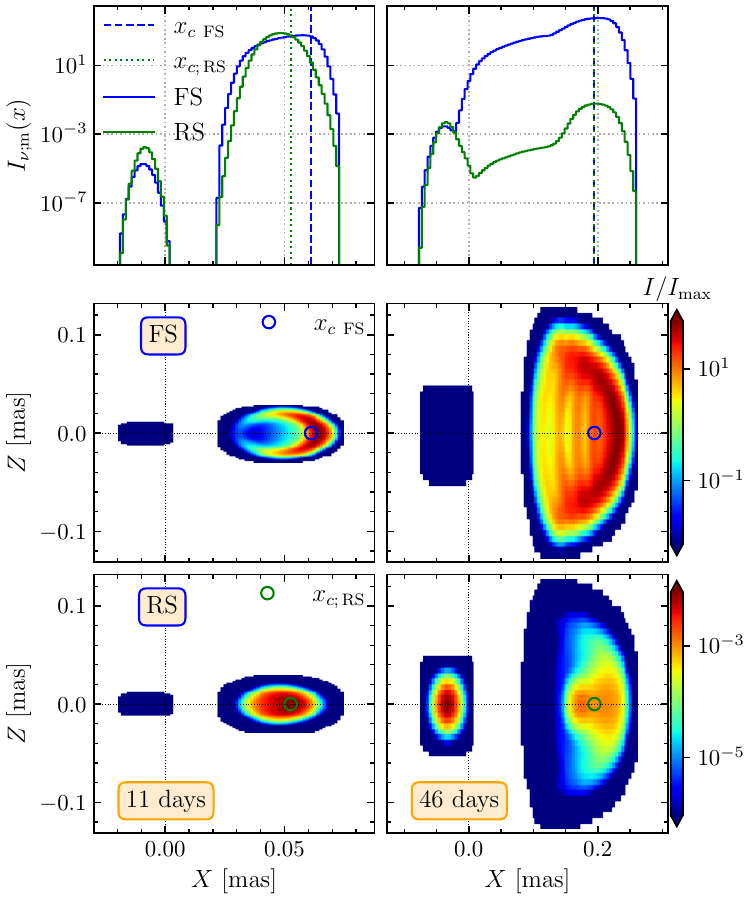}
	\caption{
		Same as Fig.~\ref{fig:method:top_hat_skymaps} but for the 
		Gaussian-\ac{FS}-\ac{RS} simulation, where the jet is observed at 
		$\theta_{\rm obs}=\pi/4\,$rad.
	}
	\label{fig:method:skymap_gaussian}
\end{figure}

In this subsection we discuss properties of the Gaussian-\ac{FS}-\ac{RS} simulation.  
The simulation parameters are set as follows: 
$E_{\rm iso;c}=10^{53}\,$ergs, $\Gamma_{0;c}=400$, 
$\theta_{\rm c}=0.1\,$rad. $\theta_{\rm w}=0.3\,$rad.
As before, we assume \ac{ISM} with constant density, $n_{\rm ISM} = 1\,\ccm$. 
Microphysics parameters are set as follows: 
$\epsilon_{e;\,\rm fs}=0.1$, $\epsilon_{b;\, \rm fs}=0.001$, $p_{\rm fs}=2.2$, 
$\epsilon_{e;\,\rm rs}=0.1$, $\epsilon_{b;\, \rm rs}=0.001$, $p_{\rm rs}=2.2$. 
For the \ac{RS}, we also set $t_{\rm prompt} = 10^3\,$sec.  
The jet is discretized into $21$ layers, each of which is represented by a \ac{BW}, 
(see Sec.~\ref{sec:method:numerics}).

The Gaussian jet structure implies that \ac{BW} initial conditions follow 
the Gaussian distribution, where \ac{BW} initial energy, 
\ac{LF} and mass are given as,
\begin{equation}\label{eq:method:gauss_id}
	\begin{aligned}
			E_{0;\, i} &= E_{\rm iso;c} \sin^2\Big(\frac{\theta_{i;\, \rm h}}{2}\Big) \exp\Big( -\frac{1}{2} \frac{\theta^2_{i;\, \rm c}}{\theta^2_{\rm c}} \Big)\, , \\
			\Gamma_{0;\, i} &= 1 + (\Gamma_{0;\, \rm c} - 1)  \frac{E_{0;\, i}}{\sin^2\big(\theta_{i;\, \rm h}/2\big)E_{\rm iso;c}}\, , \\
			M_{0;\, i} &= \frac{E_{0;\, i}}{(\Gamma_{0;\, i} - 1)c^2}\, , 
	\end{aligned}
\end{equation}
where $\theta_{i;\, \rm c}$ is the center of a ring corresponding to a 
given \ac{BW} and $\theta_{i;\, \rm h}$ is the outer boundary of 
the	ring (see Sec.~\ref{sec:method:discret}).

Dynamics of different \acp{BW} is shown in Fig.~\ref{fig:method:energy_gauss}. 
Each \ac{BW} goes through standard evolution stages: free-coasting, deceleration 
due to \ac{RS} crossing, \ac{BM}-like deceleration, lateral spreading enhanced 
deceleration and finally, \ac{ST}-like deceleration.
The \ac{RS} is weaker in slower \acp{BW} and thus does not significantly affect 
the \ac{BW} evolution, \ie, the free-coasting phase continues until the classical 
deceleration radius $R_{\rm d}$. Importantly, during the evolution of all \acp{BW} 
the total energy is reasonably well conserved. 

Evolution of the observed spectra is shown in Fig.~\ref{fig:method:obs_spec_gauss}, 
where spectra from two layers and the total spectrum from the entire jet are displayed 
in separate panels. As the observer angle is set to $\theta_{\rm obs}=0\,$rad, 
the \ac{LOS} passes through only one layer, layer=$0$ in the core of the jet. 
Other layers are seen off-axies and thus are visible only after \ac{BW} deceleration 
reduces the Doppler beaming. 
This is reflected in the spectra of individual layers reaching their respective maxima 
at later times depending on their position within the jet structure. 

At all times the spectra show two peaks corresponding to the synchrotron and 
the \ac{SSC} emissions. 
The \ac{RS} emission contribution is seen for longer and for more off-axis layers 
as Doppler beaming reduces the contribution from the \ac{FS} there. 
The \ac{SSA} plays an important role there and cuts off spectrum at early times and 
for slower \acp{BW}. 

The sky maps for this simulation, observed at an angle of $\theta_{\rm obs}=\pi/4\,$rad   
are shown in Fig.~\ref{fig:method:skymap_gaussian}. The \ac{RS} emission affects primarily 
part of the sky map that corresponds to the core of the jet as the \ac{RS} is stronger and 
faster in corresponding \acp{BW} (see Fig.~\ref{fig:method:energy_gauss}). 
However, because the jet energy is now distribution among fast and slow \acp{BW}, 
\ac{RS} stops significantly contributing to the total emission much earlier than 
in top-hat-\ac{FS}-\ac{RS} simulation.

\section{Comparison}\label{sec:comparison}

In this section we compare \PBA{} with two publicly available afterglow codes:  
\texttt{afterglowpy} \cite{Ryan:2019fhz,Ryan:2023pzk} and \texttt{jetsimpy} \cite{Wang:2024wbt}. 
Both codes can generate light curves and sky maps of \ac{GRB} afterglows. At the 
time of writing they only include synchrotron emission from the \ac{FS} from 
a top-hat or a structured jet.
Both codes assume similar \acp{EOS} for the transrelativistic ideal fluid 
and the same analytical prescription for computing the synchrotron emissivity, 
\ie, \ac{BPL} spectrum \cite{Sari:1997qe}.  

In these codes the observed radiation is computed though the \ac{EATS} integration, 
where the synchrotron emissivity in the observer frame is computed from the interpolated 
\ac{BW} properties. In other words, at each observer time and frequency, the 
radiation is evaluated from an interpolated state of the \ac{BW}. This is more 
computationally efficient than evaluating entire comoving spectra at each \ac{BW} 
evolution time step and then interpolating those spectra for a given observer time and 
frequency. However, by construction this approach does not allow for the continuous 
evolution of electron and photon spectra and thus has limited flexibility.    

While the two codes chosen for the comparison share a lot in common, there are 
significant differences between them. 
\texttt{afterglowpy} discretizes a \ac{BW} into a set of overlapping 
(in terms of angular extend) \acp{BW} with progressively larger initial half-opening 
angle each of which is evolved independently from each other under the standard 
$1$D thin-shell formulation. This is similar to what we employ in \PBA, especially to 
the setup when only \ac{FS} is evolved (see Sec.~\ref{sec:method:numerics}). 
The code assumes that each \ac{BW} starts in the \ac{BM} deceleration phase. 
It is worth noting, however, that the free-coasting phase can be enabled in \texttt{afterglowpy} 
source code. 
\texttt{jetsimpy} approximates a jet as a $2$D infinitesimally thin sheet of matter. 
In this formulation, internal pressure that drives the lateral expansion is included 
self-consistently. Thus, the code approximates the hydrodynamics of a laterally spreading 
jet more accurately. The code is further calibrated to hydrodynamic self-similar 
solutions wherever and includes the early free-coasting phase of the \ac{BW} evolution.
This, in turn, allows for more accurate modeling of early afterglow phase 
(see figure~6 in \citet{Wang:2024wbt}). 

We expect \PBA{} to agree with \texttt{jetsimpy} before the onset of lateral expansion, 
as the free-coasting phase is included in both codes, and with \texttt{afterglowpy} at 
late times due to similar treatment of lateral spreading. However, large differences 
in microphysics and radiation methods would make the comparison difficult. 
In order to have an intermediate point of comparison, we implemented an analytical 
synchrotron radiation prescription following \citet{Sari:1997qe,Wijers:1998st}. 
This configuration of our code is labeled as \PBA{}* in figures. 

\begin{figure}
	\centering 
	\includegraphics[width=0.49\textwidth]{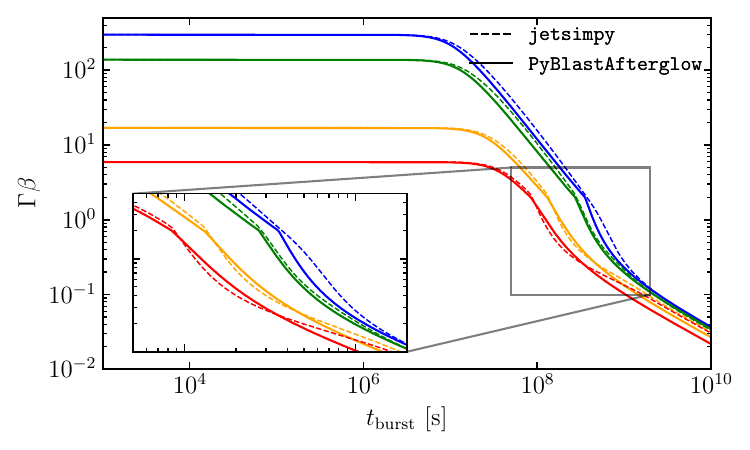}
	\caption{
		Comparison with \texttt{jetsimpy} in terms of time evolution 
		of the \ac{BW} momentum for several layers. 
		Simulation is performed with a Gaussian lateral structure 
		(Eq.~\eqref{eq:method:gauss_id}) and with \ac{FS} only 
		and the following model parameters: 
		$n_{\rm ism} = 0.00031,\ccm$, 
		$E_{\rm iso} = 10^{52}\,$ergs, $\Gamma_0 = 300$, 
		$\theta_{\rm c} = 0.085\,$rad and 
		$\theta_{\rm w} = 0.2618\,$rad. 
		Each color corresponds to one of the \acp{BW} that \PBA{} 
		evolves. \texttt{jetsimpy} values are taken by extracting 
		$\Gamma\beta$ at the times, $t_{\rm burst}$, and half-opening angle 
		$\omega$ (see text).
	}
	\label{fig:comp:dyn_jetsimpy}
\end{figure}

\subsection{Dynamics}

Since \PBA{} evolves separate \acp{BW} for different jet angular layers, while 
\texttt{jetsimpy} considers a single $2$D \ac{BW}, one-to-one comparison of the 
\ac{BW} dynamics is not trivial. We interpolate the \texttt{jetsimpy} \ac{BW} 
properties for each time, $t_{\rm burst}$, and half-opening angle, $\omega$, 
that a \PBA{} \ac{BW} attains during its evolution. We do this for several 
\acp{BW} of our code. The result is shown in Fig.~\ref{fig:comp:dyn_jetsimpy}. 

Predictably, the largest deviation in \ac{BW} dynamics with respect to 
occurs near the onset of lateral spreading, at late stages of \ac{BW} deceleration. 
At this point different elements of the laterally structured jet come into casual 
contact with each other and non-linear effects become important. 
An analytical prescription for the lateral spreading implemented in \PBA{}, \eg, Eq.~\eqref{eq:method:dthetadr}, cannot fully capture this effect. In part this can 
be mitigated by including a lateral spreading prescription that is ``aware'' of the jet 
structure at all times and naturally mimics the effects found in $2$D simulations.
Notably, since in \PBA{} all \acp{BW} are evolved simultaneously (see 
Sec.~\ref{sec:method:numerics}) such prescription can indeed be implemented. 
However, we leave this to afuture works. 
Meanwhile, we find that two codes agree the most when comparing dynamics of the inner 
layers of the jet (far from the core or the edge of the jet). 

\begin{figure}
	\centering 
	\includegraphics[width=0.49\textwidth]{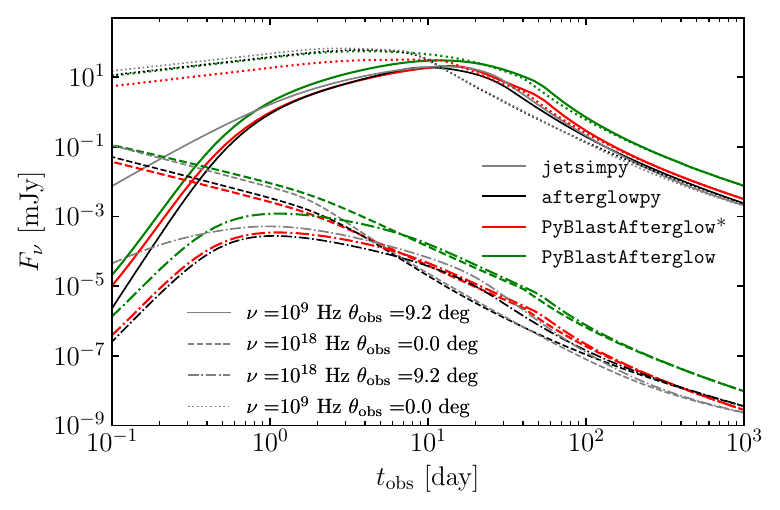}
	\includegraphics[width=0.49\textwidth]{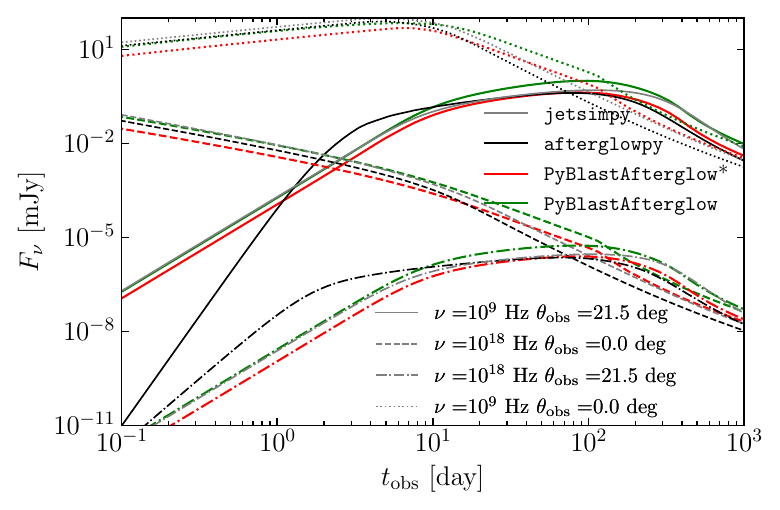}
	\caption{
		Comparison between \PBA{} (red and grid lines), 
		\texttt{afterglowpy} (black lines), and 
		\texttt{jetsimpy} (gray lines) in terms of light curves. 
		\textit{The top panel} shows comparison for the jet with top-hat 
		lateral structure (Eq.~\eqref{eq:method:tophat_id}) 
		and the following parameters:
		$d_{\rm L} = 3.09\times 10^{26}\,$cm, $Z = 0.028$, 
		$n_{\rm ism} = 10^{-2}\,\ccm$, 
		$E_{\rm iso}=10^{52}\,$ergs, $\Gamma_0 = 350$, 
		$\theta_{\rm c} = \theta_{\rm w} = 0.1\,$rad., 
		$\epsilon_e=0.1$, $\epsilon_b=0.1$ and $p=2.2$. 
		\textit{The bottom panel} corresponds to the 
		jet with the Gaussian lateral structure (Eq.~\eqref{eq:method:gauss_id}), 	
		with the following parameters:
		$d_{\rm L} = 1.27\times 10^{26}\,$cm, $Z = 0.0099$, 
		$n_{\rm ism} = 0.00031,\ccm$, 
		$E_{\rm iso} = 10^{52}\,$ergs, $\Gamma_0 = 300$, 
		$\theta_{\rm c} = 0.085\,$rad. and 
		$\theta_{\rm w} = 0.2618\,$rad, 
		$\epsilon_e = 0.0708$, $\epsilon_b = 0.0052$ and $p = 2.16$.
		In both cases only \ac{FS} is considered. 
		Different line styles indicate various observer angles and 
		observer frequencies. 
		\PBA{}* runs, depicted with red lines, were performed  
		using analytic synchrotron spectrum and not evolving electron 
		distribution. 
	}
	\label{fig:comp:lcs}
\end{figure}

\subsection{Light curves and sky maps}

Light curve comparison is shown in Fig.~\ref{fig:comp:lcs} for the top-hat jet 
(top panel) and the Gaussian jet (bottom panel), where parameters for the 
latter are chosen to be similar to those inferred for \GRB{} 
\cite{Fernandez:2021xce,Hajela:2019mjy}. 

For the top-hat jet, we observer a very good agreement between our, \PBA{}* 
generated light curves and those produced by \texttt{afterglowpy} at early and 
late times for both on-axis and off-axis observers. This is expected as we 
employ similar jet discretization and \ac{EATS} integrators. The remaining 
degree of difference, especially at $\nu_{\rm obs}=10^9\,$Hz, stems from 
the different treatment of $\gamma_{\rm m}$ and synchrotron critical frequencies.
At earlier times the disagreement between our light curves observed off-axis and 
those produced by \texttt{jetsimpy} arises from different treatments of the jet 
edge. In \PBA{} and in \texttt{afterglowpy}, a jet has a sharp edge determined by 
$\theta_{\rm w}$. Meanwhile, in \texttt{jetsimpy}, shallow wings start to emerge 
shortly after the beginning of deceleration (see Fig.~3 in \citet{Wang:2024wbt}). 

For the Gaussian jet the agreement between \texttt{jetsimpy} and \PBA{}* light 
curves is significantly better. However, the absence of the coasting phase 
in \texttt{afterglowpy} makes the light curve produced with that code qualitatively 
different from others. 

\begin{figure}
	\centering 
	\includegraphics[width=0.49\textwidth]{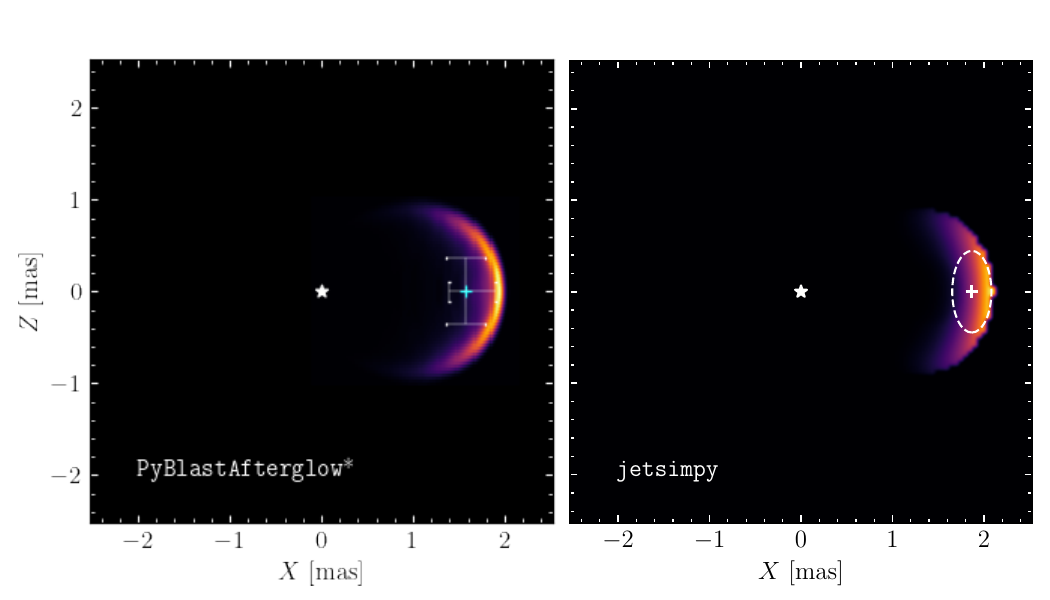}
	\caption{
		Comparison between the sky maps computed with 
		\PBA{}* (left), \texttt{jetsimpy} (right) for the  
		jet with Gaussian lateral structure, (same as in lower 
		panel of Fig.~\ref{fig:comp:lcs}), but observed at 
		an angle $\theta_{\rm obs}=0.38\,$rad., 
		$t_{\rm obs} = 75\,$days.   
		at the frequency, $\nu_{\rm obs}=10^{9}\,$Hz. 
		In both cases the specific intensity
		is shown in range $0.01-1$ of the maximum. 
		Cyan or white ``+'' marker indicate the location 
		of the image flux centroid, while white error bars 
		or white oval indicate the image size. 
	}
	\label{fig:comp:skymap}
\end{figure}

In addition to light curves, we also compare sky maps computed with our 
code and with \texttt{jetsimpy}. The simulation for comparison is performed 
using Gaussian structure and the result is shown in Fig.~\ref{fig:comp:skymap}. 
In both cases, additional Gaussian smoothing was applied to the sky map produced by 
afterglow codes. The figure shows that there is a small difference in sky map 
topography as well as in the position of the flux centroid. Notably, the latter is 
still small enough to be within the sky map image size, and thus may not affect 
inferences from a spatially unresolved source. 
These differences stem primarily from very different modeling of structured ejecta, 
-- $2$D \ac{BW} versus a set of independent $1$D \acp{BW} -- and are thus expected.

Overall, we conclude that while there are differences between all three \ac{GRB} 
afterglow codes considered here, arguably, they lie within expected ranges taking 
into account different physics input and numerical implementation of key physics inputs 
in these codes.

\subsection{\ac{VHE} spectrum comparison}

\begin{figure}
	\centering 
	\includegraphics[width=0.49\textwidth]{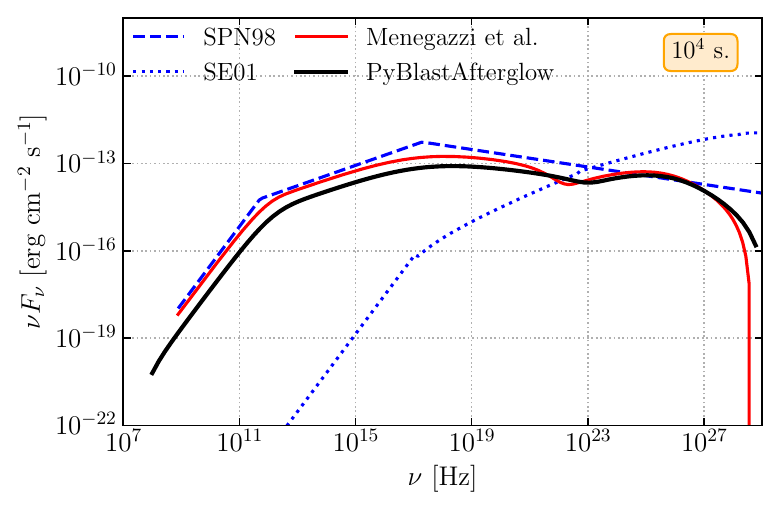}
	\caption{
		Comparison between spectrum produced by \PBA{} (black line) 
		spectrum produced by a simplified reference code (red line) 
		and analytic spectrum (blue lines).  
		The comparison is performed for a top-hat jet with the 
		following model parameters:
		$d_{\rm L} = 2.10\time10^{28}\,$cm, $Z = 1$, 
		$n_{\rm ism} = 1\,\ccm$, 
		$E_{\rm iso}=2 \times 10^{52}\,$ergs, $\Gamma_0 = 400$, 
		$\theta_{\rm c} = \theta_{\rm w} = \pi/2\,$rad., 
		$\epsilon_e=0.05$, $\epsilon_b=5\times10^{-4}$ and $p=2.3$. 
		The spectrum is computed for $\theta_{\rm obs}=0\,$rad. 
		and $t_{\rm obs}=10^4\,$s. 
	}
	\label{fig:comp:spec}
\end{figure}

As neither of the aforementioned codes includes \ac{SSC}, we turn to the 
literature for comparison of \ac{VHE} emission. Specifically, we consider 
a simplified spherically-symmetric, $1$-zone afterglow model developed and tested by 
Menegazzi et al. (in prep.), as well as analytical \acp{BPL} spectra derived by 
\citet{Sari:1997qe,Sari:2000zp}. 
The result of the comparison, for the top-hat jet spectrum computed at $10^4\,$s.
is shown in Fig.~\ref{fig:comp:spec}. The figure is made to be similar to Fig.~5 
from \citet{Miceli:2022efx} where the authors verified their \ac{SSC} model.

Menegazzi et al. (in prep.) code employs a simplified analytic formulation of the 
spherical \ac{BW} dynamics -- a smoothly-connected \ac{BPL} for the \ac{BW} \ac{LF}. 
Timesteps for the evolution are set according to Courant-Friedrichs-Lewy condition and thus are used directly in the 
implicit scheme for electron distribution evolution 
\cite{Chang:1970,Ghisellini:1999wu}. Note, in \PBA{}, by default, a sub-stepping procedure  
is used to increase computational efficiency (see Sec.~\ref{sec:method:numerics}). 
Numerically, the scheme implemented 
Synchrotron emission is computed using original formulation with Bessel functions, Eq.~\eqref{eq:method:synch_bessel}, while the \ac{SSC} emission is computed following 
\citet{Miceli:2022efx}. 
The code computes afterglow light curves and spectra without the \ac{EATS} integrator. 

As Fig.~\ref{fig:comp:spec} shows, there is an overall good agreement between our 
spectra and the one computed by Menegazzi et al. (in prep.). 
The difference in the synchrotron spectrum stems from the different treatments of synchrotron emissivity, \ac{BW} dynamics, and the calculation of the observed emission. Meanwhile, we observe a good agreement between our models in terms of high-energy \ac{SSC} emission as we use the same formulation for \ac{SSC} comoving emissivity. The large deviation between our \ac{SSC} spectra and SE01 is due to the Klein-Nishina effect suppressing \ac{SSC} emission at high energies.

\section{Application to observed \acp{GRB}}\label{sec:applications}

\begin{figure}
	\centering 
	\includegraphics[width=0.49\textwidth]{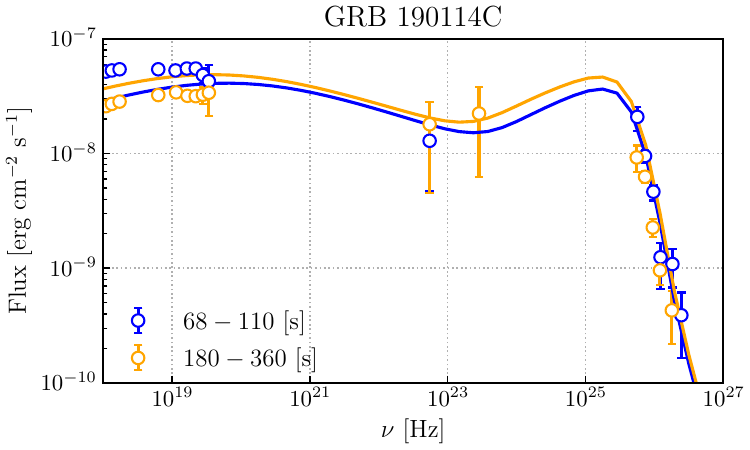}
	\caption{
		GRB~190114C \ac{SED} from soft X-rays to 1~TeV in two different time intervals. 
		Solid lines correspond to the model from the pre-computed grid that fits the 
		data best (lowest mean-squared-error; see text). 
		Observational data is adopted from \cite{MAGIC:2019irs}. 
	}
	\label{fig:method:GRB_190114C}
\end{figure}

In this work we opt not to perform any parameter inference studies and leave 
them to dedicated future works employing surrogate models that will make Bayesian 
inference feasible. However, for completeness, we provide examples on 
how the code can be applied to observed \acp{GRB}. 

\begin{table}
	\scalebox{1.05}{
		\begin{tabular}{l|l}
			\toprule
			Parameter & description \\
			\multicolumn{2}{c}{Microphysics} \\
			\midrule
			$\epsilon_{\rm e;\, fs}$ & fraction of internal energy deposited in electrons \\
			$\epsilon_{\rm b;\, fs}$ & fraction of internal energy deposited in magnetic fields \\
			$p_{\rm fs}$ & electron distribution slope \\
			\multicolumn{2}{c}{Environment} \\
			\hline
			$n_{\rm ISM}$ & \ac{ISM} number density \\
			$\theta_{\rm obs}$ & observer angle \\
			$d_L$ & luminosity distance to the source \\
			$Z$ & source redshift \\
			\multicolumn{2}{c}{Structure} \\
			\hline
			$E_{\rm iso;\,c}$ & ``isotropic equivalent'' kinetic energy \\
			$\Gamma_{0;\, \rm c}$ & \ac{LF} \\
			$\theta_{\rm w}$ & wings half-opening angle \\
			$\theta_{\rm c}$ & core half-opening angle \\
			\bottomrule
		\end{tabular}
	}
	\caption{
		List of parameters used in the paper and corresponding descriptions.  
	}\label{tbl:params}
\end{table}
 
In Table~\ref{tbl:params} model parameters required for modeling \ac{GRB} afterglows 
with \ac{FS} are shown. Overall, the are $11$ free parameters, $3$ for microphysics,
$4$ environmental and $4$ that describe \ac{GRB} jet structure. If the \ac{RS} is 
included, the number of microphysics parameters doubles, as each parameter for the 
\ac{FS} has a counterpart in \ac{RS}. In addition, $t_{\rm prompt}$ needs to be 
set that describes the duration of the \ac{GRB} shell ejection. 

In this section we focus on the \ac{GRB} afterglows considering \ac{FS} only. 
Observational data for \acp{GRB} with signatures of the \ac{RS} is more complex 
and requires a dedicated study. 

Furthermore, for simplicity, we focus on two jet structures: top-hat and the Gaussian, 
even though \PBA{} can be used with an arbitrary \ac{GRB} ejecta structure. Notably, 
for the top-hat jet, $\theta_{\rm w} = \theta_{\rm c}$, as there are no wings in the 
top-hat jet, and thus, the number of free parameters is reduced to $10$.

\subsection{\ac{VHE} of \ac{GRB}~190114C}

One of the key motivations behind developing \PBA{} is studying  the \ac{VHE} part of 
observed \ac{GRB} spectra. \ac{GRB}~190114C is a natural choice for us to apply our 
model to. \ac{VHE} emisison from this \ac{GRB} was observed by MAGIC \cite{MAGIC:2019lau}. 
Their figure~3 shows two time integrated spectra for $68-110\,$s and $110-180\,$s. 
covering the range from X-ray to gamma-rays. \ac{VHE} from this burst is generally 
attributed to \ac{SSC} emission generated during the afterglow phase \cite{Nava:2021}. 

\begin{table}
	\scalebox{1.05}{
		\begin{tabular}{l|l|l}
			\toprule
			Parameter & Values for the grid search & Result \\
			\multicolumn{3}{c}{Microphysics} \\
			\midrule
			$\epsilon_{\rm e;\, fs}$ & $[0.1,0.01,0.001]$ & $0.01$ \\
			$\epsilon_{\rm b;\, fs}$ & $[0.01,0.001,10^{-4},10^{-5}]$ & $10^{-5}$ \\
			$p_{\rm fs}$ & $[2.2,2.4,2.6,2.8]$ & $2.6$ \\
			\multicolumn{3}{c}{Environment} \\
			\hline
			$n_{\rm ISM}$ & $[1, 0.5, 0.1, 0.05, 10^{-2}, 10^{-3}]\,\ccm$ & $0.5\,\ccm$\\
			$\theta_{\rm obs}$ & $0\,$deg. \\
			$d_L$ & $2.3\times 10^9\,$pc \\
			$Z$ & $0.4245$ \\
			\multicolumn{3}{c}{Structure} \\
			\hline
			$E_{\rm iso;\,c}$ & $[10^{51},10^{52},10^{53},10^{54},10^{55}]\,$erg, & $10^{55}\,$erg \\
			$\Gamma_{0;\, \rm c}$ & $[100,300,600,1000]$ & $600$ \\
			$\theta_{\rm c}$ & $[5,10,15,20]\,$deg. & $15\,$deg. \\
			\bottomrule
		\end{tabular}
	}
	\caption{
		Values for the model parameters for the \ac{GRB}~190114C grid search. 
	}\label{tbl:params_grb190114c}
\end{table}

As we are interested in approximate modeling of this afterglow, we employ a grid 
search approach, where we perform a large number \ac{GRB} afterglow simulations and 
then select the one that fits the observational data the best. Values for the parameter 
grid are reported in Table~\ref{tbl:params_grb190114c}. For simplicity, in this case, 
we fixed the observational angle and the distance to the source using values from 
\citet{Melandri:2021rwu}. 

In total, we performed $23040$ runs using $1344$ CPU hours. For each simulation, 
we compute time-integrated spectra for $68-110\,$s and $110-180\,$s time windows 
and compare them with the observations using root-mean-square deviation as 
an error measure. In Fig~\ref{fig:method:GRB_190114C}, we show the spectra from 
the run with the lowest error. Parameter values are reported in the last column 
in Table~\ref{tbl:params_grb190114c}. Notably, they are broadly consistent with
what was previously inferred for this \ac{GRB}  \cite{Wang:2019zbs,Asano:2020grw,Joshi:2019opd,Derishev:2021ivd,Miceli:2022efx}.
For more details regarding different parameter inferences for this \ac{GRB}, see 
\cite{Derishev:2021ivd} and refs20240925 therein. 

As expected, a low $\epsilon_{\rm b;\, fs}$ is required to get sufficiently 
bright \ac{SSC} emission to explain MAGIC observations. Other parameters, however, 
are subjected to degeneracies and in order to get a more precise fit, a larger 
parameter grid or a Markov chain Monte Carlo simulation for Bayesian inference is
required. We leave this to future works.


\subsection{Structured jet of \GRB{}}

Another \ac{GRB} that we consider is \GRB{}. The unusually shallow slope 
of its afterglow's \ac{LC} prior to the peak is generally believed to be due to  
jet lateral structure and observational angle lying outside the code of the jet 
\citep[\eg][]{Hajela:2019mjy}. 

\begin{figure}
	\centering 
	\includegraphics[width=0.49\textwidth]{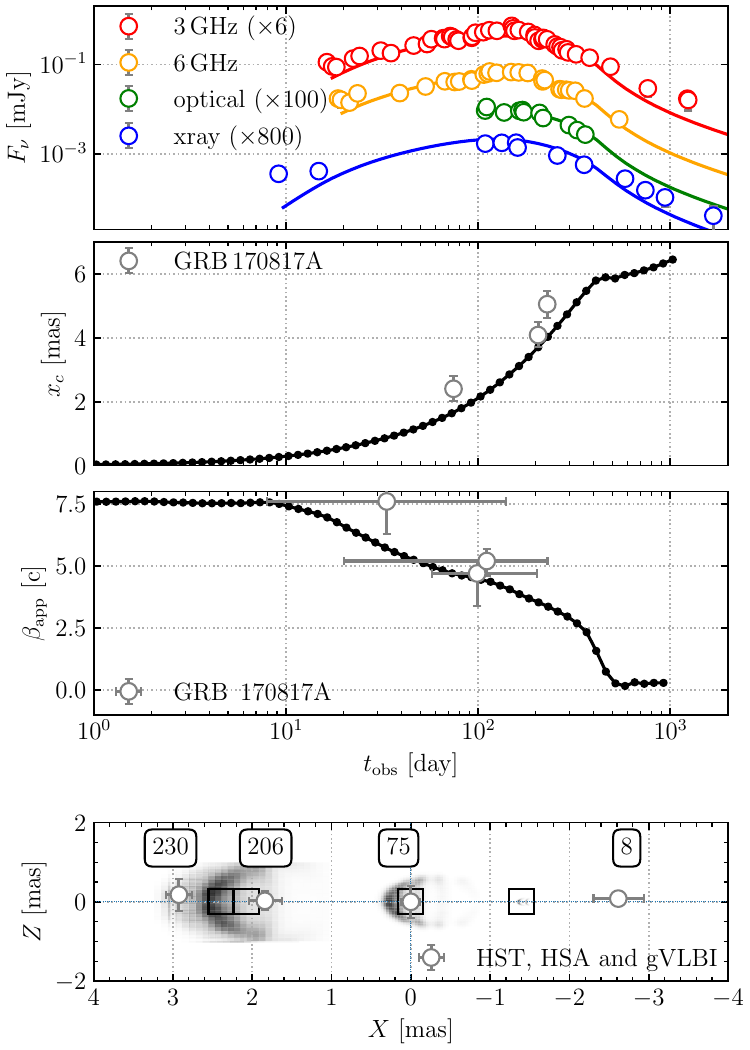}
	\caption{
		\GRB{} afterglow \acp{LC} (\textit{top panel}) and 
		sky map properties (\textit{last three panels}) 
		(see text for the details on the data used). 
		\textit{The second panel} shows the position of the 
		apparent proper motion of \GRB{} flux centroid, $x_c$,
		and the evolution of the $x_c$ from simulation with black line. 
		\textit{The third panel} shows the apparent velocity, $\beta_{\rm app}$ 
		for three time-intervals for \GRB{} and the evolution of 
		$\beta_{\rm app}$ from the simulation.
		\textit{The last panel} shows the positions of the flux 
		centroid at $4$ epochs labeled by the number of days since burst 
		and the sky maps computed at these epochs as well $x_c$ (displayed 
		with square markers). Positions are normalized using $75\,$ days 
		observation as a reference (having offset of $0$). 
		The \PBA{} simulation parameters are: 
		$n_{\rm ISM}=0.025\,\ccm$, 
		$E_{\rm iso;\,c}=1.26\times 10^{54}\,$erg,
		$\Gamma_{0;\, \rm c}=300$,
		$\theta_{\rm c}=3.5\,$deg.,
		$\theta_{\rm w}=25\,$deg.,
		$p_{\rm fs}=2.1$, 
		$\epsilon_{\rm e;\, fs}=3.8\times10^{-3}$,
		$\epsilon_{\rm b;\, fs}=9.5\times10^{-5}$,
		$\theta_{\rm obs}=20.8\,$deg. 
	}
	\label{fig:method:grb_170817a}
\end{figure}

\GRB{} has the largest amount of observational data collected for it 
including high cadence broad-band spectra and multi-epoch sky maps 
\cite{Hajela:2020data,Mooley:2022uqa}. Such a large amount of observational 
data should, in principle, allow for tight constraints on model parameters. However, 
a model of a structured jet observed off-axis has a large number of free parameters 
and significant degeneracies between them \cite{Ryan:2019fhz}. Specifically, 
modeling the afterglow of such a jet with \PBA{} requires setting all the 
Table~\ref{tbl:params} as free. At the same time, each simulation becomes significantly 
more computationally expensive as jet structure is modeled by discretizing the jet 
into a set (${\sim}20$) \acp{BW} each of which represents a top-hat jet that needs to 
be evolved. Moreover, computing multi-frequency \acp{LC} and sky maps adds to the 
simulation cost. With the present configuration of \PBA{} each run requires approximately 
$1$ CPU/hour. In light of this, as we aim to demonstrate the code applicability instead 
of performing parameter inference, we opt for a different strategy here. Instead 
of performing a grid parameter search, we start with parameter values derived by 
in a successful parameter inference run in \citet{Ryan:2019fhz} and then refine the 
values with a small number of fine-tuning runs. The result is shown in 
Fig.~\ref{fig:method:grb_170817a}. 

We use the following observational data to compare our model with. 
Radio band data was obtained by Karl G. Jansky Very Large Array \ac{VLA} \cite{Hallinan:2017woc,Alexander:2018dcl,Margutti:2017cjl,Mooley:2017enz}, 
Australia Telescope Compact Array \cite{Hallinan:2017woc,Dobie:2018zno,Mooley:2018clx,Mooley:2018dlz,Makhathini:2020ece},
the Giant Metrewave Radio Telescope \cite{Resmi:2018wuc,Mooley:2018dlz}, 
the enhanced Multi Element Remotely Linked Interferometer Network \cite{Makhathini:2020ece}, 
Very Long Baseline Array \cite{Ghirlanda:2018uyx}, 
and the MeerKAT telescope \cite{Mooley:2018dlz,Makhathini:2020ece}.
The optical data is from Hubble Space Telescope \cite{Lyman:2018qjg,Fong:2019vgn,Lamb:2018qfn}. 
The X-ray band data is from XMM-Newton \cite{DAvanzo:2018zyz,Piro:2018bpl} and 
Chandra \cite{Troja:2017nqp,Troja:2018ybt,Hajela:2019mjy,Troja:2020pzf,Troja:2021xsw,OConnor:2022}. 

For the sky map data we employ \ac{VLBI} observations at $8$, $75$, $206$ and $230\,$ 
days after the burst \cite{Mooley:2022uqa} comparing the positions of the flux centroid 
and the apparent proper motion (second and third panels in the figure, respectively). 
The latter is computed as $\beta_{\rm app} = \Delta x_{c} / \Delta t_{\rm obs} / c$ for 
three time intervals, $8-75$, $8-206$, and $8-230\,$days.
In the last panel of the figure, the position at $8\,$days is plotted relative to the 
radio \ac{VLBI} positions at $75$ and $230\,$ days measured with the High Sensitivity Array 
(HSA) and $206\,$days measurement made with $32$-telescope global \ac{VLBI} (g\ac{VLBI}) array. 
with $75\,$days \ac{VLBI} measurement used with offsets $0$ \cite{Mooley:2018dlz}. 
In the last panel of the figure we also display actual sky maps from our simulation using 
gray-scale colormap. 

Fig.~\ref{fig:method:grb_170817a} shows that \PBA{} is capable of reproducing the 
main features of \GRB{} including the shallow slope of the \ac{LC} before the peak, 
with overall good quantitative and qualitative agreements between the observations 
and the data with log-root-means-squared-error of $1.41$ that is given primarily 
due to our model underestimating fluxes in X-ray at early and late times. This can 
be improved by performing a dedicated parameter inference study. Similarly, 
sky map properties show qualitative and a reasonable quantitative agreement with 
observations when comparing apparent proper motion and apparent velocity. Slight 
underestimation in both that results in root-mean-squared-error of $0.77$ and $1.19$
respectively can be improved by consider a more narrow jet that is observed further 
off-axis. The last panel in the figure corroborates this observation.

\section{Discussion and conclusion}\label{sec:conclusion}

\acp{GRB} are ubiquitous in nature. They originate from the most energetic and 
the most complex events in the Universe: supernovae and mergers of 
compact object. Their understanding shapes our comprehension of cosmology, 
jet and plasma physics, properties of matter at extreme densities and gravity 
in the strong field regime. 

While prompt emission in \acp{GRB} is not yet sufficiently well understood to be 
employed in parameter inference studies, the afterglow phase is. There are numerous 
theoretical frameworks and numerical models that provide a connection between 
underlying physics and observables. 
Key physical processes include relativistic \ac{MHD}, shock microphysics, 
non-thermal radiation processes and radiation transport, while the observables 
include light curves, spectra and sky maps -- images of the brightest parts of the 
jet on the phrase plane of the sky (tangent to the celestial sphere).

From numerous \ac{GRB} afterglow studies it is deduced that this non-thermal, 
predominately synchrotron emission originates at collisionless shocks that form 
when relativistic ejecta collides with the surrounding matter, \eg, 
the \ac{ISM}. The success and relative simplicity of this picture has led to 
the development of a several analytical, semi-analytical and, to a lesser 
extend, numerical afterglow models capable of reproducing the key observables. 

Notably, due to a large number of observables (\eg, multi-epoch, spectra and sky maps), 
large number of free model parameters and unavoidable degeneracies 
between them analytical or semi-analytical models became the standard due to their 
relative simplicity and computational efficiency. Such models allow for Bayesian \
inference studies and parameter grid searches. 
Notably, certain physics components like \ac{RS} and \ac{SSC} emission are 
notoriously difficult to implement analytically and thus are rarely included in 
parameter inference studies. 

Meanwhile, extending existing models to account for the missing physics is highly 
non-trivial and generally requires re-derivation of the underlying theoretical 
framework and re-implementation of the numerics. On the other hand, models that 
are constructed from the ground up to include as much physics as possible, \eg, 
hydrodynamic or \ac{MHD} jet models with particle distribution evolution and 
radiation transport are prohibitively computationally expensive.

In this work we present a \ac{GRB} afterglow model \PBA{} that attempts to have a 
balance between the two approaches. The dynamics of the jet is treated under the 
piece-wise, thin-shell approximation where a jet is divided into non-interacting 
angular segments each of which is represented by a singular layer of matter 
-- a fluid element, a \ac{BW} -- that includes forward and reverse shocks and a 
contact discontinuity between them. Jet lateral spreading is accounted for with 
an analytical prescription based on local sound speed. This is conceptually 
similar to existing semi-analytical models such as \texttt{afterglowpy}.
However, contrary to most models, electron distribution evolution is treated numerically 
by solving continuity equation that includes injection of freshly shocked particles 
and cooling via synchrotron, adiabatic and \ac{SSC} processes. 
Radiation emission and absorption processes include synchrotron, \ac{SSC}, \ac{SSA}, 
\ac{PP} and \ac{EBL} absorption. Observables are computed via numerical \ac{EATS} 
integration from each angular layer and combining the result. 

Notably, such models have been presented in the literature before. However, to the 
best of our knowledge, there is no open-sourced or well-documented model. 
At present the code allows for \ac{GRB} and \ac{kN} afterglow modeling, with the 
latter discussed in \citet{Nedora:2022kjv,Nedora:2023hiz} and Nedora et al. (2024, in prep.). 
\PBA{}, is designed to be modular and easily extendable. For instance, \ac{BW} dynamics 
can be modified to include te effects of (i) non-uniform or pre-accelerated ambient medium 
\cite{Beloborodov:2005nd,Nava:2013}, 
(ii) late-time energy injection from a magnetar \cite{Metzger:2011,Gompertz:2013zga,Ren:2019nql}; 
spectrum of injected electrons can be further augmented to include more realistic profiles, 
motivated by \ac{PIC} simulations \cite{Vurm:2021dgo,Warren:2021whb}; 
proton distribution evolution \cite{Zhang:2023uei}, diffusion and escape 
terms can also be added to produce a spectrum of cosmic rays \cite{Das:2022gon,Biehl:2017zlw}; 
additional seed photons (\eg, from a kilonova or a supernova) can be included in 
\ac{IC} spectrum computation \cite{Linial:2018aep,Zhang:2019utn}.

Notably, the code is not fast enough to perform direct Bayesian inference runs, 
especially for laterally structured jets 
with \ac{FS}-\ac{RS} \acp{BW} numerical electron, synchrotron and \ac{SSC} spectra 
evolution. 
However, it is still possible to perform large set of simulation for a grid of values 
of model parameters and then train a surrogate model for fast inference. 
Modern machine learning techniques, such as conditional variational auto-encoders and 
generative adversarial neural networks can be trained on even marginally coarse grids of 
data due to their generative nature. Such surrogate models would allow for parameter 
inference runs that are significantly faster than any existing semi-analytic 
\ac{GRB} afterglow models and would have physics lacking in those. We aim to explore 
this possibility in future works.

\section*{Acknowledgements}

We thank Kohta Murase for providing us with comments regarding the manuscript.  
This work was partially funded by the European Union (ERC, SMArt, 101076369). Views and opinions expressed are those of the authors only and do not necessarily reflect those of the European Union or the European Research Council. Neither the European Union nor the granting authority can be held responsible for them. 
The simulations were performed on the national supercomputer HPE Apollo Hawk at the High Performance Computing (HPC) Center Stuttgart (HLRS) under the grant number GWanalysis/44189 and on the GCS Supercomputer SuperMUC at Leibniz Supercomputing Centre (LRZ) [project pn29ba]. 

We thank Professor Kohta Murase for his useful comments on the manuscript.  

\textit{Software:} We are grateful to the countless developers 
contributing to open-source projects that was used in the analysis 
of the simulation results of this 
work: \texttt{NumPy} \citep{numpy}, \texttt{Matplotlib} \cite{matplotlib}, and \texttt{SciPy} \cite{scipy}.

\section*{Data Availability:} 
The code presented and employed in this work is open source and 
can be found at \url{https://github.com/vsevolodnedora/PyBlastAfterglowMag}. 
The datasets generated during and/or analyzed during the current 
study are available from the corresponding author on reasonable request.

\appendix

\section{Lateral spreading prescriptions}

\begin{figure}
    \centering 
    \includegraphics[width=0.49\textwidth]{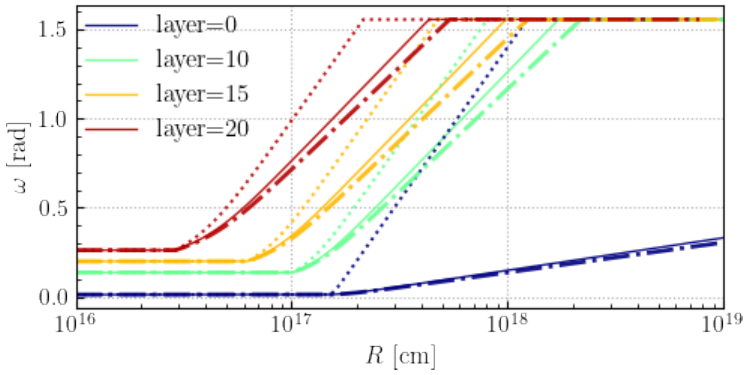}
    \caption{
			Comparison between different lateral spreading prescriptions 
			for several layers of a Gaussian jet. 
			Solid line corresponds to Eq.~\eqref{eq:method:dthetadr} 
			implemented in \PBA{} as a default option. Dashed-dotted 
			line corresponds to the prescription where ``TM'' \ac{EOS} 
			is assumed, and dotted line corresponds to the 
			prescription without structure-related terms (see text). 
        }
    \label{fig:app:spread}
\end{figure}

In this section we briefly compare \ac{BW} dynamics with different 
lateral spreading prescription. All simulations are performed using  
the structure and parameters of the Gaussian jet as discussed in the main 
text in Sec.~\ref{sec:result:gauss}. The result of the comparison is shown 
in Fig.~\ref{fig:app:spread}. 

Besides the prescription described in Sec.~\ref{sec:method:dynamics}, 
(Eq.~\eqref{eq:method:dthetadr})
we also consider here the original formulation of \cite{Ryan:2019fhz} 
where ``TM'' \ac{EOS} from \cite{Mignone:2005ns} was used. In this case 
velocity perpendicular to the norm of the shock surface reads, 
\begin{equation}
	\upsilon_{\perp} = \frac{1}{2} \frac{\beta_{\rm sh}}{\Gamma} \sqrt{\frac{2(\Gamma\beta)^2 + 3}{4(\Gamma\beta)^2 + 3}},
\end{equation}
where the shock velocity, 
$\beta_{\rm sh} = 4\Gamma^2 \beta / (4(\Gamma\beta)^2 + 3)$. 
The lateral spreading equation then becomes, 
\begin{equation}\label{eq:method:dthetadr_afgpy}
	\begin{aligned}
		\frac{d\omega}{dR} & = \frac{1}{2R\Gamma}\sqrt{\frac{2(\Gamma\beta)^2 + 3}{4(\Gamma\beta)^2 + 3}} 
		\\
		& \times 
		\begin{cases}
			0 & \text{ if } Q_0 \Gamma_{\rm sh}\beta_{\rm sh} \omega_c < 1 \\
			\frac{Q(1-Q_0\omega_c\Gamma_{\rm sh}\beta_{\rm sh})}{Q-Q_0} & \text{ if } Q \Gamma_{\rm sh}\beta_{\rm sh} \omega_c > 1 \\
			1 & \text{ otherwise } 
		\end{cases}
		\\
		& \times 
		\begin{cases}
			\tan(\omega_0/2) / \tan(\omega_c/2) & \text{ if } \omega_0 < \omega_c \\
			1 & \text{ otherwise } \\
		\end{cases}
	\end{aligned}
	\, ,
\end{equation} 
where $Q$, $Q_0$ and $\omega_c$ are set as before (see Eq.~\eqref{eq:method:dthetadr}).

The \ac{BW} dynamics computed with this equation is shown with dashed-dotted line in 
the figure. It deviates from the dynamics obtained with the default prescription 
only slightly and for \acp{BW} with $\Gamma_{0} \ll 100$. This degree of mismatch 
is expected as we employ different \acp{EOS}. 

Finally, we also compute spreading with a much simpler model, 
where the tangential velocity is still approximated with the sound speed but 
no additional constraints are applied (\eg, geometry and momentum constraints  
in Eq.~\eqref{eq:method:dthetadr}). The simplified equation then reads, 
\begin{equation} \label{eq:method:dthetadr_hang}
	\frac{d\omega}{dR} = \frac{c_s}{R \Gamma \beta c}\, .
\end{equation}
The \ac{BW} dynamics computed using Eq.~\eqref{eq:method:dthetadr_hang} corresponds to dotted 
line in the figure. As expected, the rate of lateral expansion for all layers of the jet,  
irrespective of their position with the jet, are similar as we fix the minimum 
$\Gamma\beta$ at which spreading can begin (see main text). This results in a large 
overestimation of the spreading rate for the core of the jet (layer=$0$). Overall this 
example highlights the importance of structure-aware lateral spreading prescriptions in 
semi-analytical modeling of jet dynamics.

\section{Sky Map comparison}

In \cite{Nedora:2022kjv} we presented a different method to compute \ac{GRB} 
afterglow properties. The main difference between that method and the one discussed 
in the main text is the jet discretization. 
While in the main text we discussed the model that approximates a jet as a series 
of overlapping \acp{BW} with progressively increasing half-opening angles 
(see Sec.~\ref{sec:method:discret}) 
in the aforementioned study we discretized a jet into a set of non-overlapping 
$(\theta{-}\phi)$-cells of equal solid angle, each of which had a \ac{BW} assigned 
to it. In that method, each hemisphere is split into $k=\{1,2,...n-1\}$ 
rings centered on the symmetry axis plus the single central spherical cap, $k=0$
with the opening angle $\theta_{l=1}$ \cite{Beckers:2012}. 
In order to achieve equal solid angle size per cell, each hemisphere was discretized 
uniformly in $\cos{(\theta_{l})}$ so that angle between two concentric circles on the 
sphere was $\theta_{l=i} = 2 \sin^{-1} \big( \sqrt{k/n} \sin( \theta_{\rm w} / 2 ) \big)$, 
where $\theta_{\rm w}$ is the initial opening angle of the jet.
Each layer consisted of $2 i + 1$ $(\theta-\phi)$-cells bounded by 
$\phi_{ij} = 2\pi j/ (2 i + 1)$ where $j=\{0,1,2...i\}$. 
In total the jet was descretized into $\sum_{i=0}^{i=n-1}(2 i+1)$ cells, for 
each of which an independent \ac{BW} was evolved. 
We label this jet discretization method as \texttt{PW} while the method discussed in 
the main text we label as \texttt{A} Since both of them are incorporated in \PBA{} it is 
natural to perform a comparison. 

Here we perform two afterglow simulations for top-hat jet and two for the Gaussian jet with 
the same structure and microphsycis properties as discussed in Sec.~\ref{sec:result}, but 
setting observer angle as $\theta_{\rm obs}=\pi/4\,$rad. Only \ac{FS} is considered here 
and synchrotron emission is computed analytically (\ie. using \PBA{}$^{*}$ configuration 
of the code discussed in Sec.~\ref{sec:comparison}). 

\begin{figure}
	\centering 
	\includegraphics[width=0.49\textwidth]{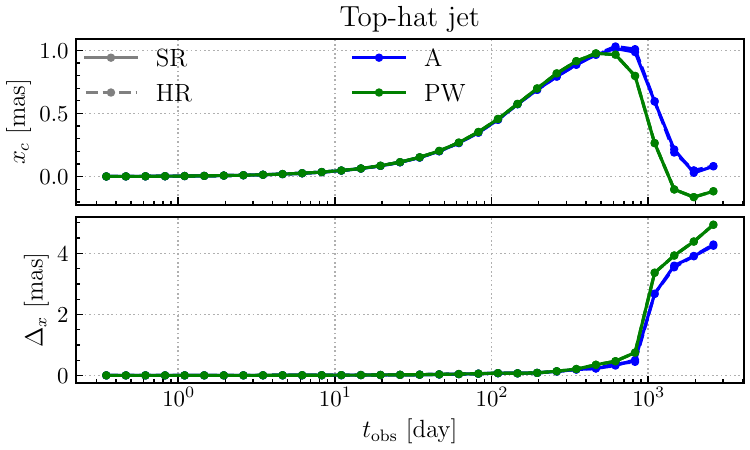}
	\includegraphics[width=0.49\textwidth]{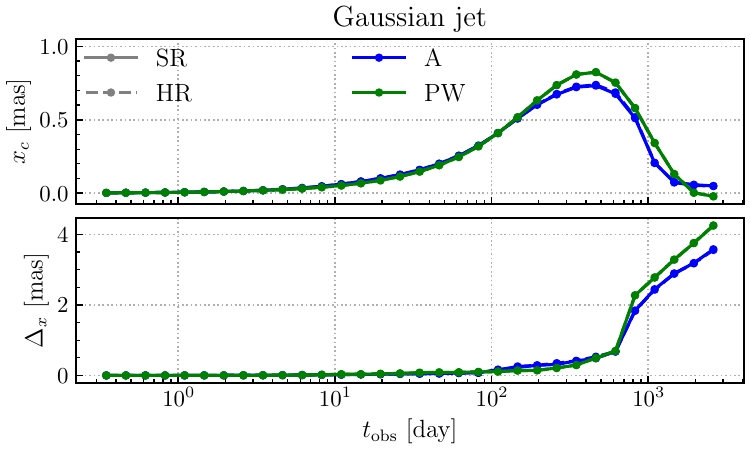}
	\caption{
		Comparison of sky-map properties evolution for top-hat jet 
		\textit{top pair of panels} and Gaussian jet \textit{bottom pair of panels}. 
		For both cases the evolution of sky map flux centroid position $x_c$ (\textit{top sub-panel}) 
		and image size, $\Delta_{x}$ (\textit{bottom sub-panel}) are shown. 
		\texttt{PW} and \texttt{A} stand for different jet discretization (see text).
	}
	\label{fig:app:skymap_method_comparison}
\end{figure}

In \texttt{A} method, a top-hat jet is approximated with a single \ac{BW} that has 
initial half-opening angle equal to that of the jet. In \texttt{PW} method, however, 
$n\gg1$ \acp{BW} are required so that the jet angular extend is properly covered with 
$(\theta-\phi)$-cells. Importantly, in \texttt{A} method we also need to have to 
discretized the hemisphere into cells to calculate sky map, but this is done after 
\ac{BW} evolution.  
There we adaptively resize the spherical grid until the required resolution is reached 
(See Sec.~\ref{sec:method:numerics:skymap}) and the grid resolution enters as a free 
parameter. In the case of a Gaussian jet, a certain number of angular layers 
(\ie, \acp{BW}) needs to be set for both methods, and thus, it is a natural parameter 
describing resolution. 

In Fig.~\ref{fig:app:skymap_method_comparison} we show a comparison of evolution of 
sky map properties computed with \texttt{A} method (blue color) and 
\texttt{PW} method (green color). We consider two resolution options labeled SR and HR, where 
SR stands for the resolution used in the main text and HR is two times higher resolution. 
Overall we observe a very good agreement in sky map flux centroid position $x_c$ and its size 
$\Delta_{x}$ between two simulations at early time.  At the point where $x_c$ reaches maximum,   
emission from the counter-jet becomes comparable to that from the principle jet 
(see Sec.~\ref{sec:result}).
Around this point, the \ac{BW} lateral spreading plays a key role in determining sky map properties. 
\acp{BW} in the simulation with \texttt{PW} method are less energetic and lateral spreading 
there starts much earlier (for the top-hat jet) and proceeds faster.
In the case of the Gausian jet, there are slow \ac{BW} in both cases, \texttt{A} and \texttt{PW} 
and the $x_c$ reaches maximum at the same time. Image size $\Delta_{x}$ grows faster and to 
larger values in the simulation with \texttt{PW} method as lateral spreading is enhanced. 
Overall, however, we highlight that sky map properties evolution is qualitatively the 
same in simulations with different discretization methods and quantitative agreement 
is good.

\bibliography{refs20240925}

\end{document}